\documentclass[12pt]{article}
\usepackage{amsmath}
\usepackage{amsfonts}
\usepackage{graphicx}
\usepackage{mathrsfs}

\newcommand{\R}{{\bf R}}
\newcommand{\Z}{{\bf Z}}
\newcommand{\J}{{\mbox{\boldmath $J$}}}

\begin{document}

%
\def\papertitlepage{\baselineskip 3.5ex \thispagestyle{empty}}
\def\preprinumber#1#2{\hfill \begin{minipage}{4.2cm}  #1
                 \par\noindent #2 \end{minipage}}
\renewcommand{\thefootnote}{\fnsymbol{footnote}}
\newcommand{\beq}{\begin{equation}}
\newcommand{\eeq}{\end{equation}}
\newcommand{\beqa}{\begin{eqnarray}}
\newcommand{\eeqa}{\end{eqnarray}}
\catcode`\@=11
\@addtoreset{equation}{section}
\def\theequation{\thesection.\arabic{equation}}
\catcode`@=12
\relax

\newcommand\km{k_{\rm min}}
\newcommand\cm{c_{\rm min}}
\newcommand\Jtot{J_0^{\rm tot}}
\newcommand\tildeJtot{\tilde J_0^{\rm tot}}

%
%
\papertitlepage
\setcounter{page}{0}
\preprinumber{KEK-TH-1280}{hep-th/yymmnnn}
\baselineskip 0.8cm
\vspace*{2.0cm}
\begin{center}
{\large\bf Localized Modes in Type II and Heterotic \\
Singular Calabi-Yau Conformal Field Theories}
\end{center}
\vskip 4ex
\baselineskip 1.0cm
\begin{center}
           {Shun'ya~ Mizoguchi} 
\\
\vskip 1em
       {\it High Energy Accelerator Research Organization (KEK)} \\
       \vskip -2ex {\it Tsukuba, Ibaraki 305-0801, Japan} \\
\end{center}
\vskip 5ex
%
\baselineskip=3.5ex
\begin{center} {\bf Abstract} \end{center}
We consider type\;II and heterotic string compactifications 
on an isolated singularity in the noncompact Gepner model approach. 
The conifold-type ADE noncompact Calabi-Yau threefolds, as well as the 
ALE twofolds, are modeled by a tensor product of the $SL(2,\R)/U(1)$ 
Kazama-Suzuki model and an $N=2$ minimal model. Based on the string 
partition functions on these internal Calabi-Yaus previously obtained
by Eguchi and Sugawara, we construct new modular invariant, 
space-time supersymmetric partition functions for both type\;II 
and heterotic string theories, where the GSO projection is performed 
before the continuous and discrete state contributions are separated. 
We investigate in detail the massless spectra of the localized 
modes. 
In particular, we propose an interesting three generation model, 
in which each flavor is in the ${\bf 27}\oplus{\bf 1}$ representation 
of $E_6$ and localized on a four-dimensional space-time 
residing at the tip of the cigar.

\vskip 2ex
\vspace*{\fill}
\noindent
August 2008
\newpage
\renewcommand{\thefootnote}{\arabic{footnote}}
\setcounter{footnote}{0}
\setcounter{section}{0}
\baselineskip = 0.6cm
\pagestyle{plain}

\section{Introduction}
Building phenomenologically realistic models in string theory is a 
challenging problem. Among others, one of the most serious obstacles 
to the construction is the issue of the moduli. 
Typically, we assume that the background is a product of a four-dimensional 
Minkowski space and some compact Calabi-Yau manifold. Various parameters 
characterizing the latter appear as scalar fields in the low-energy effective  
theory, which are massless until appropriate fluxes and quantum 
effects are taken into account. The basic question is whether or not, and if so how,
the moduli stabilization is realized dynamically.  This question is closely linked to 
the vacuum selection problem. With the recent recognition of the string landscape
\cite{Susskind}, 
one might be satisfied if any consistent ultra-violet completion of the 
Standard Model is obtained, but nothing can guarantee the uniqueness of 
the solution.

These difficulties stem from the complexity and diversity of compact 
Calabi-Yau manifolds. Let us 
suppose that we are given a Calabi-Yau which has only a few, 
say three, moduli. Then we would not need to worry about the moduli stabilization 
problem from the beginning. Although there are no such known {\em compact} 
Calabi-Yaus, there {\em are} such {\em noncompact} 
ones.  A typical example is the ADE series of the ALE manifolds.

After the discovery of D-branes, 
the use of noncompact local 
Calabi-Yau manifolds has been common --- geometric engineering
\cite{geometric_engineering},
topological string theory \cite{topological_string} and 
gauge theory \cite{gauge_theory} ---
in all these examples the central focus of the study
is the open string.
In this paper, in contrast, we use noncompact Calabi-Yaus as the internal sector of 
conventional closed string compactification in terms of conformal field 
theory \cite{Gepner},
for both type\;II and heterotic string theories.
We consider these superstrings in a four- (and also six-) dimensional Minkowski 
space with some internal noncompact conifold-like threefold (ALE twofold) 
of the ADE type, where  
the internal part is described \cite{GV,OV}
by a tensor product of 
an $N=2$ minimal model with level $\km=0,1,2\ldots$ and
the noncompact coset $SL(2,\R)/U(1)$ Kazama-Suzuki model 
with correlated level $\kappa$.

We present a compact expression for
space-time supersymmetric, modular invariant partition functions  
consisting not only of contributions from the continuous (principal unitary)
series representations 
of the mother $SL(2,\R)$ Lie algebra, but also of those from the discrete series 
representations. This new space-time supersymmetric partition function is an 
improvement of the earlier results in noncritical super strings or 
``noncompact" Gepner models 
\cite{Mizoguchi,ES1,Mizoguchi:Osaka,Murthy,
noncompact_Gepner_models}, and owes much to the recent construction 
of modular invariants for the internal noncompact Calabi-Yau 
CFTs by Eguchi and Sugawara \cite{ES}. 
We will show, by using the character decomposition 
technique \cite{MOS,HPT,ES}, there 
are massless matter supermultiplets coming from the discrete series, the 
number of which can be small depending on the value of the level $\km$.
In particular, if we consider the $E_8 \times E_8$
heterotic string compactification
\footnote{
We should note that in the heterotic case our construction is closely related 
to the ``heterotic coset models" \cite{heterotic_coset_models} studied earlier 
because an $N=2$ minimal model is realized \cite{N=2minimal} as 
an $SU(2)/U(1)$ coset theory.}    
 for $\km=3$, we will find precisely 
three generations of $N=1$ chiral multiplets in the 
${\bf 10}\oplus{\bf 16}\oplus {\bf 1}\oplus{\bf 1}$ of $SO(10)$
or ${\bf 27}\oplus{\bf 1}$ of $E_6$.
Since the discrete series representations in the $SL(2,\R)/U(1)$ 
gauged WZW model are known to be the modes localized \cite{DVV} near 
the tip of the ``cigar" \cite{Witten:2dBH}, 
these three flavors can move only in the four-dimensional Minkowski 
directions, and hence are trapped on some four-manifold at the tip of 
the cigar.
The schematic picture is shown in Figure 1.

\begin{figure}
\begin{center}
\includegraphics[width=150mm]{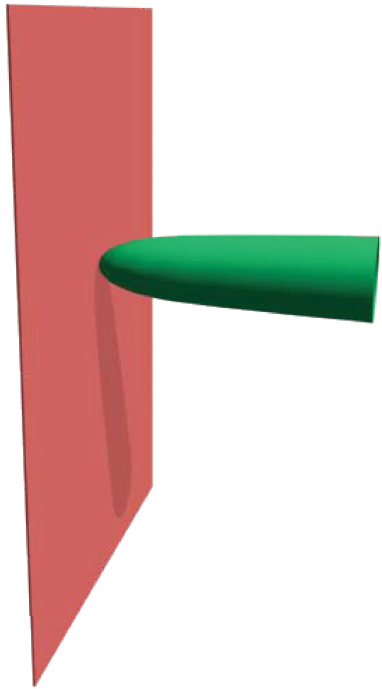}
\end{center}
\caption{The schematic picture. 
\label{schematic_picture}
}
\end{figure}

This ``brane" is not a D-brane; the localized modes are those of closed strings, 
which exist even in heterotic string theories. In fact, these modes can be regarded 
as the position moduli of NS5-branes.
Indeed, in the six-dimensional 
analysis with an ALE manifold, we will find \cite{ES} precisely as 
many massless supermultiplets in the discrete spectrum 
as the number of two-cycles, which are $D=6$ 
{\em nonchiral} $N=2$ multiplets (including vectors) in the type IIA case and 
{\em chiral} ones (including anti-selfdual tensors) in the type IIB case. 
They are opposite to the zero modes appearing on NS5-branes \cite{CHS2}
in agreement with the T-duality \cite{OV,Kutasov} between the NS5-brane and the 
ADE singularity. In the IIB case, the S-dual version was used in the past 
to explain \cite{BVS} 
the nonperturbative gauge symmetry enhancement near the singularity  \cite{Witten:Various_Dimensions} 
in terms of D-branes.\footnote{The nonperturbative ``W-bosons" cannot 
be seen in our closed string CFT partition functions.  They can, however, 
be analyzed \cite{ES:Modular_bootstrap,L,LLS} in the boundary Liouville 
CFT\cite{boundary_Liouville}, which we do not 
consider in this paper. We would like to thank Y.~Sugawara 
for discussion on this point.}  
In the four-dimensional case, relations between a deformed conifold 
and a system of intersecting NS5-branes are also known \cite{BVS,OY}.
We would like to emphasize that to even see 
geometric moduli of a noncompact Calabi-Yau as massless modes in a 
modular invariant CFT partition function has been a nontrivial problem.


%
The dynamics on NS5-branes in the framework of the CHS model \cite{CHS1,CHS2}
was much studied as ``Little String Theories" (LSTs) \cite{LST}. They are 
basically non-critical superstring theories \cite{KutasovSeiberg} coupled to 
some compact CFT, which is a supersymmetric $SU(2)$ WZW model 
for NS5-branes. In analogy to the AdS/CFT correspondence \cite{Maldacena}, it has 
been proposed that their vanishing string coupling ($g_s\rightarrow 0$) limit 
(and hence the decoupling gravity limit) has  
some holographic dual theory on the boundary at 
the weakly coupled linear dilaton region (the``mouth" of the throat). 
To avoid the strong-coupling singularity
far down the throat, we need a 
regularization in the bulk theory. There are two known ways: 
The first is the so-called double-scaled Little String Theory \cite{GK1,GK2},
that is, a particular limit of LST where the weak string coupling limit and 
the limit of collapsing areas of the homology cycles are taken in a correlated 
manner. In this limit, the physics depends only on a particular combination 
of the coupling constant and a deformation parameter, and the scaled theory 
can be weakly coupled.  
The second is to replace the linear-dilaton cylinder geometry with the cigar 
geometry \cite{GV,OV}. Later it was shown that these two are 
dual to each other \cite{GK1,HK}.

In fact, the link between the NS5-brane and the two-dimensional black hole 
goes back to the work by Gidding and Strominger (GM) \cite{GS} 
in 1991, where a similar double-scaling (and, at the same time, extremal) 
limit of a family of type II and heterotic 
{\em non-extremal} black five-brane solutions was 
considered to observe that the resulting geometry was a product of a
$(1+1)$-dimensional black hole, an $S^3$ and a five-dimensional 
Euclidean space. The CFT description of this geometry is very close to 
ours; although it is not exactly 
the same  (because, for instance, we consider a
Euclidean black hole), it is at least suggestive.  In this
GM's double-scaling limit, despite the $g_s\rightarrow 0$ limit taken 
there, the graviton, dilaton and other backgrounds do not disappear but are 
still present in the final geometry with nontrivial, though finite, profiles in 
the whole space-time.\footnote{With the standard embedding, 
this GM's double scaling 
also regularizes the small instanton singularity 
\cite{Witten:Small_Instantons, Witten:Heterotic_CFT} for heterotic five-branes.}

As we mentioned above, our new partition functions are constructed 
based on the ones for internal noncompact Calabi-Yaus obtained by 
\cite{ES}. Roughly speaking, what we do is to couple the noncompact 
Calabi-Yau CFT to that for the flat Minkowski space and perform a 
suitable GSO projection {\em before} the contributions from the 
continuous and discrete series 
representations are separated. The states in the latter class of $SL(2,\R)$ 
representations will be called the ``discrete states" in short. 
It turns out that the formulas are simple and similar in their form to 
those for the partition functions containing only the continuous 
representations obtained previously. One of the virtues of our formulas 
is that the couplings of the discrete states for the Calabi-Yau 
to states for the flat Minkowski space are automatically consistent with 
the modular invariance of the continuous sector.
Another advantage is that we can straightforwardly 
extend the type II analysis to heterotic strings by using the heterotic 
conversion procedure \cite{Gepner} of modular invariant partition functions. 
As we noted above, we can construct an interesting three generation model, 
in which each flavor consists of a {\bf 27}  and a singlet of $E_6$ and 
is localized on a four-dimensional space-time. 
Thus this ($\km=3$) model may offer a viable alternative string 
model for the $E_6$ unification \cite{E6}.

Gravity and gauge fields are not
localized; they are (apparently) massive due to the Liouville energy and 
propagate into the bulk. They correspond to the continuous series representations.
But still, we expect that the three generation model above will be 
useful for studying issues of flavors. 
While any particular phenomenological realization 
on a compact Calabi-Yau cannot be unique, singularities 
occur {\em universally} in the moduli space of any compact Calabi-Yau 
manifold. We hope we can capture some universal physics near 
the singularity by studying the localized modes in the conformal 
field theory.

This is a more detailed version of \cite{Mizoguchi:2007}, in which the summary of 
results presented here was already announced. The plan of this paper is as follows.
In Section 2, we review the basics of representation theory of the affine $SL(2,{\bf R})$
and the associated $N=2$ superconformal algebras. In Section 3, we also review
the previous constructions of modular invariant partition functions in the 
noncompact Gepner model approach, which consists of contributions from 
only the continuous (principal unitary) 
series representations. In Section 4, we construct new 
space-time supersymmetric, modular invariant partition functions on 
the ADE generalization of conifolds, for both type II and heterotic string 
theories. In Section 5, we describe the detail of how to separate the 
discrete series contributions from the new partition functions, and 
examine the spectrum. In particular, we propose the $\km=3$ three 
generation model mentioned above.  Section 6 is devoted to examples.
In Section 7, we briefly discuss the generalization to the six-dimensional 
space-time with the ordinary ALE manifolds. Finally, we conclude this paper
with a summary and discussion, which are given in Section 8.  
Appendix A contains  basic definitions of theta functions and characters,
and their identities. In Appendix B we collect useful formulas related the 
functions $F_{l,2r}(\tau,z)$ and  $\hat F_{l,2r}(\tau,z)$ we use in the text, which are 
important building blocks in the construction of the partition functions. 
Appendix C is a review of the heterotic conversion procedure of Gepner. 
Finally, in Appendix D we give a proof of the regularization formula of \cite{MOS}.

\section{$SL(2,{\R})$ paraferemions and $N=2$ superconformal algebra}
In this section we review the relation between the affine $SL(2,{\R})$ Kac-Moody 
and $N=2$ superconformal algebras based on the $SL(2,\R)$ parafermion 
construction of \cite{DLP}. This is relevant for our discussion because we construct 
a model by using the $N=2$ representations while the ``localization of modes" is 
a concept that has emerged in the $SL(2,{\R})$ ones.
\subsection{Free field realizations}
The $SL(2,{\R})$ Kac-Moody currents of level $\kappa$ are realized as follows:
\beqa
J^3(z)&=&i\sqrt{\frac\kappa 2}\partial\phi,\\
J^\pm(z)&=&i\left(
\sqrt{\frac \kappa 2} \partial\theta \pm i\sqrt{\frac{\kappa -2}2}\partial\rho
\right)
\exp\left(
\pm i \sqrt{\frac 2\kappa}(\theta -\phi)
\right),
\eeqa
where $\rho(z)$, $\theta(z)$ and $\phi(z)$ are 
free fields satisfying the following OPEs:
\beqa
\rho(z)\rho(w)&\sim&-\log(z-w),\\
\theta(z)\theta(w)&\sim&-\log(z-w),\\
\phi(z)\phi(w)&\sim&+\log(z-w).
\eeqa
The energy-momentum tensor $T^{SL(2,\R)}(z)$ is given 
by
\beqa
T^{SL(2,\R)}(z)
&=&-\frac 12( \partial\rho)^2+ \frac 1{\sqrt{2(\kappa -2)}}\partial^2\rho
-\frac 12( \partial\theta)^2
+\frac 12( \partial\phi)^2,
\eeqa
which has a central charge 
\beqa
c_{SL(2,\R)}&=&\frac{3\kappa}{\kappa -2}.
\eeqa

The $SL(2,{\R})$ parafermions $\psi^\pm(z)$ \cite{DLP} are fundamental fields 
in the  $SL(2,{\R})/U(1)$ coset conformal field theory. They are written in terms 
of the free fields as
\beqa
\psi^\pm(z)&=&i\left(
\sqrt{\frac 12} \partial\theta \pm i\sqrt{\frac{\kappa -2}{2\kappa}}\partial\rho
\right)
\exp\left(
\pm i \sqrt{\frac 2\kappa}\theta
\right).
\eeqa 
Using these fields with another free boson $\varphi$ \cite{GepnerQiu} satisfying
\beqa
\varphi(z)\varphi(w)&\sim&-\log(z-w),
\eeqa
a set of $N=2$ superconformal currents are 
realized as follows:
\beqa
T^{N=2}(z)&=&-\frac 12( \partial\rho)^2+ \frac 1{\sqrt{2(\kappa -2)}}\partial^2\rho
-\frac 12( \partial\theta)^2
-\frac 12( \partial\varphi)^2,\\
G^\pm(z)&=&\sqrt{\frac{2\kappa}{\kappa -2}} \psi^\pm(z)
\exp\left(
\pm i \sqrt{\frac{\kappa-2}\kappa \varphi}
\right),\\
J^{N=2}(z)&=&i\sqrt{\frac\kappa{\kappa -2}} \partial\varphi.
\eeqa
The central charge $c_{N=2}$ is the same as $c_{SL(2,\R)}$:
\beqa
c_{N=2}&=&\frac{3\kappa}{\kappa -2}.
\eeqa

\subsection{Unitary representations of the $SL(2,\R)$ and 
$N=2$ superconformal algebras}
A unitary module (``Fock space") of the affine $SL(2,\R)$ Kac-Moody algebra 
necessarily contains a unitary (non-affine) $SL(2,\R)$ algebra 
module at the lowest $L^{SL(2,\R)}_0$ level.
The states in the module are labeled by the eigenvalues of 
$J^3_0$ and
\beqa
\J&=&\frac12(J^+_0J^-_0+J^-_0J^+_0) -(J^3_0)^2.
\eeqa  
Let us denote\footnote{We consider the universal cover of $SL(2,\R)$.} 
such an eigenstate by 
\beqa
| l,m+\epsilon>~~~\left( 
m\in\Z,~~~ 0 \leq \epsilon <1\right),
\eeqa
where
\beqa
J^3_0 | l,m+\epsilon>&=&(m+\epsilon)| l,m+\epsilon>,\\
J^\pm_0 | l,m+\epsilon>&=&(m+\epsilon\pm l)| l,m+\epsilon\pm 1>,\\
\J | l,m+\epsilon>&=&-l(l-1)| l,m+\epsilon>.
\eeqa


A state $| l,m+\epsilon>$ corresponds to a vertex operator 
\beqa
| l,m+\epsilon>&\rightarrow& 
e^{\sqrt{\frac2{\kappa -2}}l\rho+i\sqrt{\frac 2 \kappa}(m+\epsilon)(\theta -\phi)}
\eeqa
in the free field realization of the affine $SL(2,\R)$ Kac-Moody algebra.
It has a conformal weight 
\beqa
L^{SL(2,\R)}_0
&=&-\frac{l^2 -l}{\kappa -2},\\
J^3_0&=&m+\epsilon.
\eeqa
The corresponding $N=2$ vertex operator is then given by 
\beqa
&\rightarrow& 
e^{\sqrt{\frac2{\kappa -2}}l\rho+i\sqrt{\frac 2 \kappa}(m+\epsilon)\theta
+i\sqrt{\frac2\kappa}(m+\epsilon)\varphi},
\eeqa
which is a primary field with eigenvalues
\beqa
L^{N=2}_0
&=&\frac{-(l^2 -l)+(m+\epsilon)^2}{\kappa -2}~~~(\equiv h),\\
J^{N=2}_0
&=&\frac{-2(m+\epsilon)}{\kappa -2}~~~(\equiv Q).
\eeqa
The point is that \cite{DLP} each individual state $| l,m+\epsilon>$ in a unitary representation of the 
non-affine $SL(2,\R)$ algebra corresponds to a unitary representation of the 
$N=2$ superconformal algebra.
We will consider each class of representations separately.
\begin{figure}
\begin{center}
\includegraphics[width=140mm]{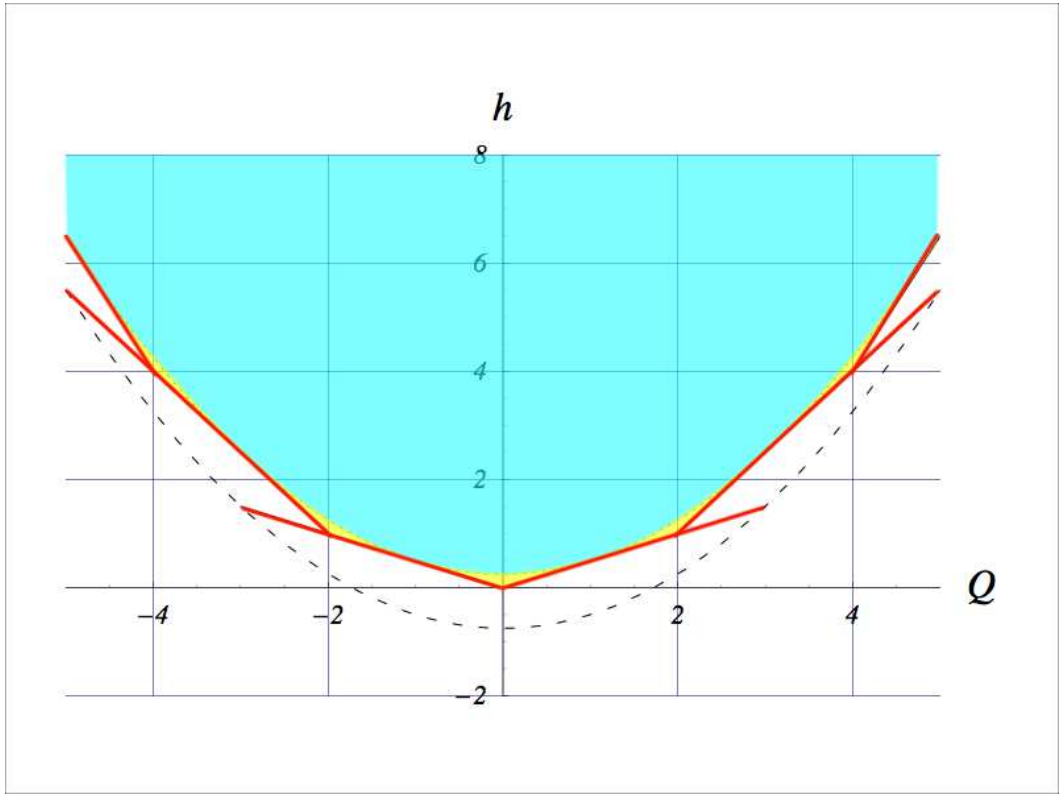}
\end{center}
\caption{The unitary region of the $N=2$ superconformal algebra \cite{BFK} ($c=9$, 
NS sector). 
\label{unitarity_region_conifold}
}
\end{figure}
\begin{itemize}
\item[(i)]{
The principal unitary series (The ``continuous series"). 
}
The representation space in this class consists of a set of states 
\beqa
\{| l,m+\epsilon>~|~ m\in\Z \}
\eeqa
for some 
$l=\frac12 + ip$, $p\in\R$ and  
$0 \leq \epsilon <1$.
There is neither upper nor lower $J^3_0$ bound in the states.
The corresponding $N=2$ representation has
\beqa
h&=&\frac 1{\kappa -2}\left(
p^2 +\frac14 + (m+\epsilon)^2
\right),\\
Q&=&-\frac{2(m+\epsilon)}{\kappa -2}.
\eeqa
Eliminating $m+\epsilon$, we obtain a family of parabola
\beqa
h&=&\frac{\kappa-2}4 Q^2 +\frac1{\kappa -2}\left(
p^2+\frac14
\right)
\label{p}
\eeqa
labeled by $p\in \R$ on the $(h,Q)$-plane.
They are shown in blue in Figure \ref{unitarity_region_conifold}.
Throughout this paper, the term ``continuous series"
will refer to this class of representations. 
\item[(ii)]{The discrete series ${\cal D}^+_n$ ($n=0,1,\ldots$).}
The representation space consists of states $| l,m+\epsilon>$ such that, 
for a given $n$, 
\beqa
l&=&n+\epsilon,\\
m+\epsilon&=&n+\epsilon+r~~~(r=0,1,2,\ldots).
\eeqa
The representation ${\cal D}^+_n$ has a lowest-$J^3_0$ state
\beqa
| l,m+\epsilon>&=&|n+\epsilon ,n+\epsilon>.
\eeqa
The values of $h$ and $Q$ of the corresponding $N=2$ 
representations are
\beqa
h&=&\frac 1{\kappa -2}\left(
(2r+1)(n+\epsilon)+ r^2
\right),\\
Q&=&-\frac 2{\kappa -2}(n+\epsilon+r).
\eeqa
Eliminating $n+\epsilon$ from above, we obtain
\beqa
h&=&-\left(
r+\frac12
\right)Q
-\frac1{\kappa -2}\left(
\left(r+\frac 12\right)^2-\frac14
\right)~~~(r=0,1,2,\ldots).
\eeqa
They are precisely the left half of the family of segments which 
bound the unitary region on the $(h,Q)$-plane.
They are shown in red lines in Figure \ref{unitarity_region_conifold}.

\item[(iii)]{The discrete series ${\cal D}^-_n$ ($n=1,2,\ldots$).}
The representation space of ${\cal D}^-_n$ consists of $| l,m+\epsilon>$ such that 
\beqa
l&=&n-\epsilon,\\
m+\epsilon&=&-n+\epsilon-r~~~(r=0,1,2,\ldots),
\eeqa
among which
\beqa
| l,m+\epsilon>&=&|n-\epsilon ,-n+\epsilon>
\eeqa
is the highest-$J^3_0$ state. The corresponding $h$ and $Q$ are 
similarly
\beqa
h&=&\frac 1{\kappa -2}\left(
(2r+1)(n-\epsilon)+ r^2
\right),\\
Q&=&+\frac 2{\kappa -2}(n-\epsilon+r),
\eeqa
and therefore
\beqa
h&=&+\left(
r+\frac12
\right)Q
-\frac1{\kappa -2}\left(
\left(r+\frac 12\right)^2-\frac14
\right)~~~(r=0,1,2,\ldots).
\eeqa
They are the right half of the family of the segments.
They are also shown in red lines.

\item[(iv)]{The complimentary series.}
The complimentary series is similar to the principal unitary series, but
in this case $l$ and $\epsilon$ satisfy ($0 \leq \epsilon <1$)
\beqa
-\epsilon(\epsilon -1)\leq -l(l-1) < \frac14.
\eeqa
The upper bound of $\J$
\beqa
-l(l-1)=\frac14
\eeqa
coincides with the principal unitary series with $p=0$, 
while the lower bound  
\beqa
-l(l-1)=-\epsilon(\epsilon -1)
\eeqa
the $N=2$ representations which arise from the states 
in  ${\cal D}^+_{n=0}\cup  {\cal D}^-_{n=1}$.
Therefore, the complimentary series fill the gap
between the paraboloid of the $p=0$ principal unitary series
and the polygonal boundary of the discrete series representations.
They are shown  in Figure \ref{unitarity_region_conifold} 
as the narrow yellow region between the blue area 
and the red segments.

Although this class of representations is much like the principal unitary 
series with continuous spectra, they do not arise in our model.

\item[(v)]{The trivial (identity) representation.}
It gives rise to the identity representation 
\beqa
h=0,~~~Q=0
\eeqa
of the $N=2$ superconformal algebra. 
\end{itemize}

\section{Noncompact Gepner Models} 
In this section we review the old noncompact 
Gepner model constructions and address their issues. 
\subsection{Modular invariant partition functions}
In usual Gepner models \cite{Gepner}, one uses a tensor product of 
$N=2$ minimal superconformal field theories so that their central charge 
add up to nine for compactification to four dimensions. 
They are subject to an orbifold projection, which is implemented by  
taking an alternating sum over shifted indices of theta functions in the 
minimal characters. This is called the ``$\beta$-method", with which 
one can achieve both an integral total $U(1)$ charge and modular 
invariance.  In modern terminology, it is equivalent to consider 
spectral flow orbits with respect to the $N=2$ $U(1)$ charge.

The $N=2$ minimal models are labeled by a nonnegative integer level $\km$
and have central charges
$\cm=\frac{3\km}{\km+2}$,
which do not exceed three. Therefore, we need at least four minimal models 
to have nine.
In \cite{Mizoguchi}, an attempt was made to construct a supersymmetric 
modular invariant partition function by using  $c=9$ characters directly, 
with no minimal models.
Such representations are necessarily nonminimal ones. A generic 
(and hence nonminimal) $N=2$ character 
of a representation with a highest weight $L_0^{N=2}=h$, $J_0^{N=2}=Q$ 
is given by \cite{N=2}
\beqa
{\rm Tr}_{\rm NS}~q^{L_0^{N=2}}y^{J_0^{N=2}}&=&
q^{h+\frac18}y^{Q} \frac{\vartheta_3(\tau,z)}{\eta^3(\tau)}
\eeqa 
for the NS sector, and 
\beqa
{\rm Tr}_{\rm R}~q^{L_0^{N=2}}y^{J_0^{N=2}}&=&
q^{h+\frac18}y^{Q} \frac{\vartheta_2(\tau,z)}{\eta^3(\tau)}
\eeqa 
for the R sector. The definitions of theta functions, as well as 
other functions used below, are summarized in Appendix A. To improve 
the modular property of these nonminimal characters and construct 
a modular invariant, the following two ideas were considered \cite{Mizoguchi}:
The first is to gather infinitely many generic representations with different 
$U(1)$ charges aligned on a lattice, so that the infinite sum produces 
another theta function. The second idea is to integrate over the  
continuous spectrum of generic characters  
with a Gaussian weight with respect to the Liouville momentum $p$ (\ref{p}). 
For reasons that will be explained below, 
we use level-1 theta functions for the first idea. Then, taking into account 
the Gaussian integration, we have roughly 
\beqa
\frac1{\sqrt{\tau_2}}\left|
\frac{\Theta_{*,1}\Theta_{*,2}}{\eta^3}
\right|^2,
\eeqa
which has modular weight $(0,0)$.

Next, to achieve spacetime supersymmetry, we need a GSO projection.  
We have two level-2 theta functions, one from the complex fermion for 
the flat two-dimensional transverse space and the other from the $N=2$ 
character above. The $\beta$-method tells us how to construct 
good combinations of theta functions. That is, we consider 
an alternating summation of the form
\beqa
\sum_{\nu}(-1)^\nu 
\Theta_{m+ \beta_0 \nu,k}
\Theta_{s_1+\beta_1 \nu,2}\Theta_{s_2+\beta_2 \nu,2}
\eeqa 
with the ``$\beta$-conditions" \cite{Gepner}:
\beqa
\frac{\beta_0^2}{2k}+\frac{\beta_1^2}4+\frac{\beta_2^2}4&=&1,\\
\frac{\beta_0 m}{2k}+\frac{\beta_1 s_1}4+\frac{\beta_2 s_2}4&=&\frac12.
\eeqa
The solution is  
$k=1$,
$(\beta_0,\beta_1,\beta_2)=(1,1,1)$ and 
$(m,s_1,s_2)=(1,0,0)$ or $(0,0,2)$.
Indeed, if we define
 \beqa
\Lambda_1(\tau,z)&\equiv&
2\sum_{\nu\in\Z_4}(-1)^\nu
\Theta_{1+\nu,1}(\tau,2z)
\Theta_{\nu,2}(\tau,z)\Theta_{\nu,2}(\tau,z)
\nonumber\\
&=&
\Theta_{1,1}(\tau,2z)
\Big(\vartheta_3^2(\tau,z)
+\vartheta_4^2(\tau,z)\Big)
-\Theta_{0,1}(\tau,2z)
\Big(\vartheta_2^2(\tau,z)+\tilde\vartheta_1^2(\tau,z)
\Big),
\label{Lambda1(z)}
\\
\Lambda_2(\tau,z)&\equiv&
2\sum_{\nu\in\Z_4}(-1)^\nu
\Theta_{\nu,1}(\tau,2z)
\Theta_{\nu,2}(\tau,z)\Theta_{2+\nu,2}(\tau,z)
\nonumber\\
&=&
\Theta_{0,1}(\tau,2z)
\Big(\vartheta_3^2(\tau,z)
-\vartheta_4^2(\tau,z)\Big)
-\Theta_{1,1}(\tau,2z)
\Big(\vartheta_2^2(\tau,z)-\tilde\vartheta_1^2(\tau,z)
\Big),
\label{Lambda2(z)}
\eeqa
then their modular transformations are \cite{Mizoguchi}
\beqa
\Lambda_1(\tau+1,0)&=&i~\Lambda_1(\tau,0),\nonumber\\
\Lambda_2(\tau+1,0)&=&-\Lambda_2(\tau,0),
\eeqa
and
%
\beqa
\Lambda_1\left(-\frac1\tau,0\right)&=&\frac{\tau^{3/2}e^{-\frac{3\pi i}4}}{\sqrt{2}}
\left(-\Lambda_1(\tau,0)+\Lambda_2(\tau,0)\right),\nonumber\\
\Lambda_2\left(-\frac1\tau,0\right)&=&\frac{\tau^{3/2}e^{-\frac{3\pi i}4}}{\sqrt{2}}
\left(+\Lambda_1(\tau,0)+\Lambda_2(\tau,0)\right).
\eeqa
Therefore
\beqa
\frac{
\left|\Lambda_1(\tau,0)\right|^2
+\left|\Lambda_2(\tau,0)\right|^2}{\left|\eta^3(\tau)\right|^2}
\label{conifold_invariant}
\eeqa
is modular invariant. 
In fact, the functions $\Lambda_1(\tau,z)$ and $\Lambda_2(\tau,z)$ 
vanishes identically for whatever value of $z$, and hence play the 
role of Jacobi's quartic identity in the ordinary 
ten-dimensional critical superstring theories. $\Lambda_1$ was used 
long time ago \cite{BG}, and $\Lambda_2$ was derived in \cite{Mizoguchi} 
by a modular transformation. It was clarified \cite{ES} that 
these (identically zero) functions, as well as more general combinations 
of theta functions and the $N=2$ minimal characters, were derived from 
Jacobi's identity through compositions of theta functions.  
Using (\ref{conifold_invariant}), we can write a modular invariant
\beqa
Z=\int\frac{d\tau d\overline{\tau}}{\mbox{Im}\tau}
	(\mbox{Im}\tau)^{-2}\left|\eta(\tau)\right|^{-4}
(\mbox{Im}\tau)^{-\frac12}\left|\eta(\tau)\right|^{-2}
%
\frac{
\left|\Lambda_1(\tau,0)\right|^2
+\left|\Lambda_2(\tau,0)\right|^2}{\left|\eta^3(\tau)\right|^2}
%
,
\label{partition_function}
\eeqa
%
where the factor $(\mbox{Im}\tau)^{-\frac12}$ comes 
from the Liouville momentum integration, and 
$\left|\eta(\tau)\right|^{-2}$ from the transverse fermions.
The transverse fermion thetas are contained in $\Lambda$'s.

\subsection{The link to singular Calabi-Yaus}
The modular invariant (\ref{partition_function}) 
is regarded as a partition function of type II strings ``compactified" 
on the conifold \cite{GV,OV}. The defining equation of the (deformed) 
conifold is \cite{conifold} 
\beqa
z_1^2+z_2^2+z_3^2+z_4^2&=&\mu
\label{deformed_conifold}
\eeqa
in ${\bf C}^4$ with a deformation constant $\mu$. If we 
view (\ref{deformed_conifold}) as an equation in 
inhomogeneous coordinates of some weighted projective space, 
we may recover the homogeneous expression
\beqa
-\mu z_0^{-1}+z_1^2+z_2^2+z_3^2+z_4^2&=&0,
\label{deformed_conifold2}
\eeqa
where the negative power of $z_0$ is determined by the Calabi-Yau 
condition. According to the well-known relation between the 
Landau-Ginzburg potential and the level of the minimal model \cite{Martinec,VW},
the first term suggests that the (deformed) conifold is described by 
the level-$(-1-2= -3)$ minimal model, which was 
interpreted\footnote{In these references 
the Kac-Moody level was denoted by $k$. Note that
we denote this by $\kappa$ while we differently use $k$ meaning 
$k=\kappa-2$ in this paper, following the notation of \cite{ES}. }
\cite{Witten:N_Matrix_Model,GV,OV}
as the level-(+3) $SL(2,\R)/U(1)$ Kazama-Suzuki model \cite{KazamaSuzuki} 
which has $c=9$. Prior to this, the circle of connections 
between the the topological string near the conifold limit,
twisted $SU(2)/U(1)$ coset at level-$(-3)$,
the $c=1$ string at the self-dual radius and matrix models had been noted 
\cite{BCOV1,BCOV2,Witten:N_Matrix_Model,KG,DV,MV}.

The ADE singularity of a Calabi-Yau twofold 
\cite{ALE}
can be considered similarly \cite{OV}.
The $X_n$ singularity ($X=A,D$ or $E$) is defined by 
an algebraic equation 
\beqa
W_{X_n}(z_1,z_2,z_3)&=&0
\label{W=0}
\eeqa
in ${\bf C}^3$,
where 
\beqa
W_{A_n}(z_1,z_2,z_3)&\equiv&z_1^{n+1}+z_2^2 +z_3^2,\\
W_{D_n}(z_1,z_2,z_3)&\equiv&z_1^{n-1}+z_1z_2^2 +z_3^2,\\
W_{E_6}(z_1,z_2,z_3)&\equiv&z_1^4+z_2^3 +z_3^2,\\
W_{E_7}(z_1,z_2,z_3)&\equiv&z_1^3z_2+z_2^3 +z_3^2,\\
W_{E_6}(z_1,z_2,z_3)&\equiv&z_1^5+z_2^3 +z_3^2.
\eeqa
The singularity equation (\ref{W=0}) is similarly deformed to 
\beqa
W_{X_n}(z_1,z_2,z_3)&=&\mu z_0^{-h^{\vee}(X_n)}, 
\label{W=mu}
\eeqa
where $h^{\vee}(X_n)$ is the (dual) Coxeter number of the Lie algebras 
$h^{\vee}=n+1,2(n-1),12,18$ and $30$ 
for $X_n=A_n$, $D_n$, $E_6$, $E_7$ and $E_8$, respectively. 
Again, (\ref{W=mu}) is understood as an equation in some weighted 
projective space specified by the weight of $z_0$, which is determined by 
the Calabi-Yau condition. (\ref{W=mu}) indicates that 
the deformed ADE singularities are
described by the $SL(2,\R)/U(1)$ Kazama-Suzuki 
model of level ($h^{\vee}+2$) coupled to the  level ($h^{\vee}-2$) $N=2$ minimal model of 
the corresponding modular invariant type.
It was also argued by using the character identity (\ref{character_identity}) 
\cite{Gepner} that the ADE singularity was T-dual to the NS5-brane.

The corresponding modular invariant partition functions for type II strings 
``compactified" on these noncompact manifolds, as well as on similar 
ADE generalizations of the conifold, were constructed 
in \cite{ES}
by using the generic noncompact $N=2$ characters, 
where it was revealed that the relevant spectral flow orbits which constituted 
the partition function were actually obtained in a unified way by composing the 
theta functions in Jacobi's identity. 
Namely, in the twofold case, the authors of \cite{ES} defined 
the functions
\beqa
F_l(\tau,z)&\equiv&
\frac12
\chi^{(\km)}_l (\tau,0)
\left(
\vartheta_3^4-\vartheta_4^4-\vartheta_2^4+\tilde\vartheta_1^4
\right)(\tau,z)
\nonumber\\
&=&
\sum_{\nu \in \Z_4} (-1)^\nu
\sum_{m\in\Z_{2(\km +2)}}
\chi^{l,\nu}_m(\tau,-z)
\sum_{
\mbox{\tiny $
							\begin{array}{c}
							\nu_0,\nu_1,\nu_2\in \Z_2\\
							\nu_0+\nu_1+\nu_2 \\
							\equiv 1({\rm mod}2)
							\end{array}$}
							}
							\Theta_{2\nu_0 +\nu,2}(\tau,z)
							\Theta_{2\nu_1 +\nu,2}(\tau,z)
							\nonumber\\
							&&~~~~~~~~~~~~~~~~~~~~~~
							~~~~~~~~~~~
							~~~~\cdot
							\Theta_{2\nu_2 +\nu,2}(\tau,z)
							\Theta_{m,\km+2}\left(\tau,\frac{2z}{\km +2}\right)
							\label{F_l}
\eeqa
for $l=0,\ldots,\km$,
where  the familiar character identity 
(\ref{character_identity}) \cite{Gepner,OV} was used.
The second line enables us to identify $F_l(\tau,z)$ as a spectral flow orbit of 
a system consisting of the level-$\km$, $N=2$ minimal model, two complex 
fermions and a noncompact $N=2$ CFT with some appropriate $U(1)$-charge 
lattice. Therefore, we can write
\beqa
Z&=&\int\frac{d\tau d\overline{\tau}}{\mbox{Im}\tau}
(\mbox{Im}\tau)^{-3}\left|\eta^{-4}(\tau)\right|^2
(\mbox{Im}\tau)^{-\frac12}\left|\eta^{-2}(\tau)\right|^2
\sum_{l,\tilde l}
N_{l,\tilde l}
\frac{
F_{l}(\tau,0)
(F_{{\tilde l}}(\tau,0))^*
}{\left|\eta^3(\tau)\right|^2}\nonumber\\
&=&
\int\frac{d\tau d\overline{\tau}}{(\mbox{Im}\tau)^2}
(\mbox{Im}\tau)^{-\frac52}\left|\eta^{-5}(\tau)\right|^2
\sum_{l,\tilde l}
N_{l,\tilde l}
\frac{
F_{l}(\tau,0)
(F_{{\tilde l}}(\tau,0))^*
}{\left|\eta^4(\tau)\right|^2},
\label{ADE_partition_function}
\eeqa
which is clearly modular invariant.
This was regarded as a supersymmetric partition function 
modeling a deformed ADE singularity (\ref{W=mu})
with a six-dimensional Minkowski space.

In the threefold case, the relevant building blocks are
\begin{eqnarray}
F_{l,2r}(\tau,z)&\equiv&\frac14 
\sum_{m\in {\bf Z}_{4(\km+2)}} 
\left(
(\vartheta_3(\tau,z))^2
\mbox{ch}^{\mbox{\scriptsize NS}}_{l,m}(\tau,z)
-(-1)^{r-\frac m2}
(\vartheta_4(\tau,z))^2
\mbox{ch}^{\widetilde{\mbox{\scriptsize NS}}}_{l,m}(\tau,z)
\right.
\nonumber\\
&&\left.~~~~~~~~~~~~~~~\rule{0ex}{1em}
-(\vartheta_2(\tau,z))^2
\mbox{ch}^{\mbox{\scriptsize R}}_{l,m}(\tau,z)
+(-1)^{r-\frac m2 +\frac12}
(\tilde\vartheta_1(\tau,z))^2
\mbox{ch}^{\widetilde{\mbox{\scriptsize R}}}_{l,m}(\tau,z)
\right)
\nonumber\\
&&
~~~~~~~~~~~~~~~~
\cdot
\Theta_{
(k_{\mbox{\scriptsize min}}+2)2r-(k_{\mbox{\scriptsize min}}+4)m,
2(k_{\mbox{\scriptsize min}}+2)(k_{\mbox{\scriptsize min}}+4)
}
\left({\textstyle
\tau,\frac z{k_{\mbox{\scriptsize min}}+2}
}\right)
\label{F}
\end{eqnarray}
for $r\in\Z_{\km+4}+\frac l2$.
As we explain in Appendix, this $F_{l,2r}(\tau,z)$ is derived from 
Jacobi's identity and satisfies
\beqa
\frac14
\chi^{(\km)}_l(\tau,0)
(\vartheta_3^4-\vartheta_4^4-\vartheta_2^4-\tilde\vartheta_1^4)(\tau,z)
&=&
\sum_{r\in\Z_{\km+4}+\frac l2}F_{l,2r}(\tau,z)\Theta_{2r,\km+4}(\tau,0).
\eeqa
The modular properties of $F_{l,2r}(\tau,z)$ can be read off from this equation.
We can similarly write a modular invariant \cite{ES}\footnote{For simplicity, 
we take the coefficients of the modular invariant theta system 
$M^{\km}_{r,r'}$\cite{GepnerQiu} to be diagonal.}
\beqa
Z=\int\frac{d\tau d\overline{\tau}}{(\mbox{Im}\tau)^2}
(\mbox{Im}\tau)^{-\frac32}\left|\eta(\tau)\right|^{-6}
\sum_{l,\tilde l}
\sum_{r\in \Z_{\km +4}+\frac l2}
N_{l,\tilde l}
\frac{
F_{l,2r}(\tau,0)
(F_{{\tilde l},2r}(\tau,0))^*
}{\left|\eta^3(\tau)\right|^2},
\label{conifold_type_partition_function}
\eeqa
which can be regarded as the type II partition function for the ADE 
generalization of the conifold
\beqa
W_{X_n}(z_1,z_2,z_3)+z_4^2&=&0, 
\label{W=0_threefold}
\eeqa
which is deformed to
\beqa
W_{X_n}(z_1,z_2,z_3)+z_4^2&=&\mu z_0^{-\frac{2 h(X_n)}{h(X_n)+2}}.
\label{W=mu_threefold}
\eeqa
Since the level of the $N=2$ minimal model is $\km=h(X_n)-2$, 
the level of the noncompact $N=2$ CFT is 
\beqa
\kappa&=&\frac{2 h(X_n)}{h(X_n)+2}+2\nonumber\\
&=&\frac{2(\km+2)}{\km+4}+2.
\eeqa

Heterotic partition functions were also 
constructed in \cite{Murthy}. 


%

\subsection{Absence  of localized modes}
\begin{figure}
\begin{center}
\includegraphics[width=150mm]{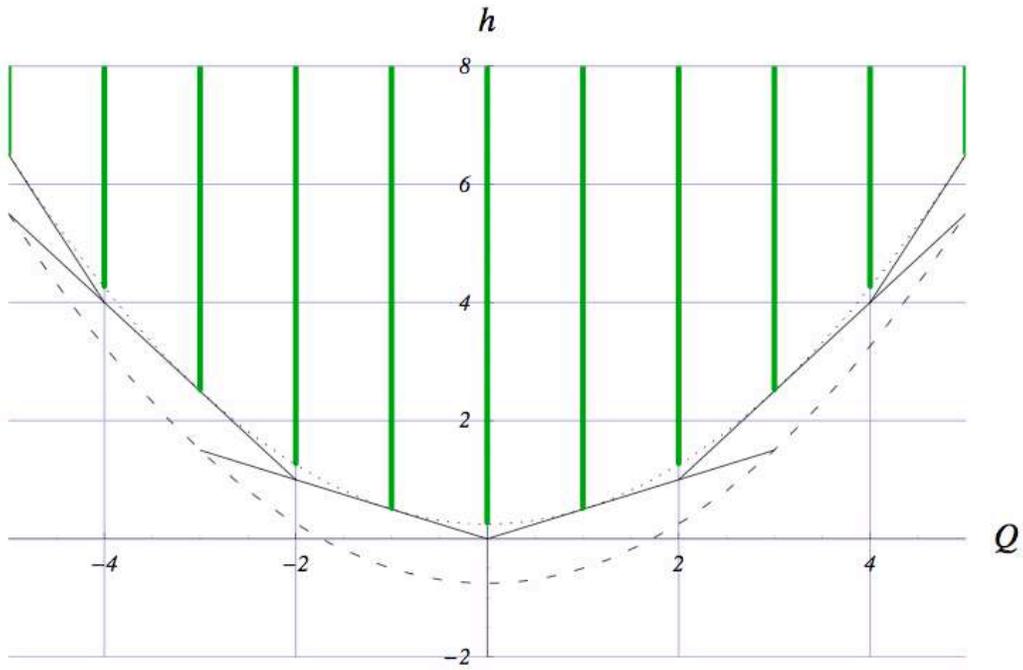}
\end{center}
\caption{The spectrum of the $c=9$, $N=2$ representations 
in the old conifold partition function (NS sector). 
The green lines at odd (even) $Q$ are the spectrum of representations 
contributing to $\Lambda_1$ ($\Lambda_2$).
\label{conifd_spectrum}
}
\end{figure}

The spectrum of the $c=9$, $N=2$ representations used 
in the partition function (\ref{partition_function}) is shown in 
Figure \ref{conifd_spectrum}. What is nontrivial here is that \cite{Mizoguchi} 
the level-1 $U(1)$ theta functions determined by modular invariance 
and supersymmetry are consistent with unitarity. That is, 
the envelope of the lowest ends of the continuous spectra, which 
are set by the level-1 theta functions, {\em coincides exactly} with 
the lowest $L_0$ bound of possible $N=2$ representations corresponding 
to the continuous series of $SL(2,\R)$.

By construction, there are only the representations 
coming from the continuous series of $SL(2,\R)$.
The graviton is massive due to the Liouville energy \cite{Seiberg}.
All the modes have continuous Liouville momenta 
and propagate into the extra dimension (that is, the Liouville 
direction).
There are no localized massless 
modes.\footnote{
There is a subtlety associated with the gapless spectrum. 
The appearance of this is the common feature of the spectrum 
for even $\km$, the level of the  $N=2$ minimal model coupled
in the generalized models. See section \ref{evenkm}.}
This is also the case for the partition functions for the ADE singularity 
obtained in \cite{ES,Murthy}; they do not reflect the geometry 
in their massless spectrum \cite{Mizoguchi:Osaka}.

In \cite{ES}, new modular invariant partition functions including 
contributions from both the continuous and discrete series representations 
have been constructed for noncompact Calabi-Yau manifolds with an 
isolated singularity. (See also \cite{recent} for more recent related works.) 
They obtained them via the path-integral approach. 
They used the character decomposition technique developed in different 
but similar models \cite{MOS,HPT} to show the existence of the localized 
modes. In particular, they found \cite{ES} the correct chiral ring structures 
expected from the geometry of the ALE manifolds.

In the next section, based on this result, we construct spacetime supersymmetric
partition functions (that is, the ones in which the fermions for  
the four-dimensional Minkowski space are coupled and GSO-projected 
before the continuous and discrete representations are separated) 
on the conifold-type threefolds for type II strings, and also for heterotic strings.

\section{Partition functions of superstrings on noncompact 
singular Calabi-Yau threefolds}
\label{Partition_functions}
We start with the toroidal partition function of the $SL(2,\R)/U(1)$ 
Kazama-Suzuki model \cite{ES}
\begin{eqnarray}
Z_{CY}^{\rm (NS)}(\tau)&=&C
\int_0^1 ds_1 \int_0^1 ds_2
\frac{|\vartheta_3(\tau,s_1\tau-s_2)|^2}{|\vartheta_1(\tau,s_1\tau-s_2)|^2}
\sum_{v,w\in {\bf Z}}e^{-\frac{k\pi}{\tau_2}|(w+s_1)\tau-(v+s_2)|^2}
\label{Z_superSL2RoverU1}.
\end{eqnarray}
This expression was obtained by a path integration 
\cite{Gawedzki,GawedzkiKupiainen,MOS,HPT,IKP}
in the $H^+_3/\R$ gauged 
WZW model coupled to fermions. The detail of the derivation of (\ref{Z_superSL2RoverU1})
can be found in Appendix C of ref.\cite{ES}.
The partition functions for other spin structures $Z_{CY}^{(\widetilde{\rm NS})}(\tau)$,
$Z_{CY}^{(\rm R)}(\tau)$ and $Z_{CY}^{(\widetilde{\rm R})}(\tau)$ are given by similar 
expressions 
with $\vartheta_3$ replaced by $\vartheta_4$, $\vartheta_2$ and $\vartheta_1$, 
respectively. The overall constant $C$ is arbitrary at this point but later, after 
the discrete series contributions are separated, it is chosen to be $4k$ 
so that the discrete states partition function becomes a polynomial of $q$ with 
integer coefficients. Here we set $C=1$ for simpler notation.

By a Poisson resummation we may write
\begin{eqnarray}
\sum_{v,w\in {\bf Z}}e^{-\frac{k\pi}{\tau_2}|(w+s_1)\tau-(v+s_2)|^2}
&=&
\sum_{n,w\in {\bf Z}}
e^{
-\pi\tau_2\big(\frac{n^2}k+k(s_1+w)^2 \big)
-2\pi i n ((s_1+w)\tau_1-s_2 )
}\nonumber
\\
&=&\sqrt{\frac{\tau_2}k}
\sum_{m,\tilde m}
e^{-k \pi \tau_2 s_1^2}
q^{\frac{m^2}k}e^{-2\pi i m (s_1\tau -s_2)}
\bar{q}^{\frac{{\tilde m}^2}k}e^{+2\pi i {\tilde m} (s_1\bar\tau -s_2)},
\label{lattice_decomposition}
\end{eqnarray}
where $m=\frac{n-kw}2$, ${\tilde m}=-\frac{n+kw}2$. They run over an appropriate 
direct sum of orthogonal lattices determined by $n,w\in {\bf Z}$.
\beqa
k&=&\frac{2(\km+2)}{\km+4} ~~~(\km=0,1,2,\ldots).
\label{k}
\eeqa

\subsection{$\km=0$ : The conifold}
To get insight into how the GSO projection is accomplished before separating 
the continuous and discrete series representations, we consider the 
$\km=0$ ($k=1$) case first.

If $k=1$,  (\ref{lattice_decomposition}) becomes
\beqa
(\ref{lattice_decomposition}) 
&=&\sqrt{\tau_2}
\sum_{\mbox{\scriptsize$\begin{array}{c} m,\tilde{m}\in\Z\\ m=\tilde{m}~{\rm mod}~2\end{array}$}}
e^{- \pi \tau_2 s_1^2}
q^{m^2}e^{-2\pi i m (s_1\tau -s_2)}
\bar{q}^{{\tilde m}^2}e^{+2\pi i {\tilde m} (s_1\bar\tau -s_2)}
\\
&=&\sqrt{\tau_2}
\sum_{\nu\in {\bf Z_2}}
e^{- \pi \tau_2 s_1^2}
\Theta_{\nu,1}(\tau,s_2-s_1\tau)
\left(
\Theta_{\nu,1}(\tau,s_2-s_1\tau)
\right)^*
.
\label{thetatheta}
\eeqa
The level-1 theta functions are precisely the ones which are used to construct 
a modular invariant partition function on the conifold consisting of only 
continuous series representations. 
This leads us to define, generalizing the continuous series result, 
the new functions
 \beqa
\hat\Lambda_1(\tau,z)&\equiv&\Theta_{1,1}(\tau,z)
\Big(\vartheta_3(\tau,z)\vartheta_3(\tau,0)
+\vartheta_4(\tau,z)\vartheta_4(\tau,0)\Big)
-\Theta_{0,1}(\tau,z)
\;\vartheta_2(\tau,z)\vartheta_2(\tau,0),
\label{hatLambda1}
\nonumber\\
\\
\hat\Lambda_2(\tau,z)&\equiv&\Theta_{0,1}(\tau,z)
\Big(\vartheta_3(\tau,z)\vartheta_3(\tau,0)
-\vartheta_4(\tau,z)\vartheta_4(\tau,0)\Big)
-\Theta_{1,1}(\tau,z)
\;\vartheta_2(\tau,z)\vartheta_2(\tau,0)
\nonumber\\
\label{hatLambda2}
\eeqa
and write
\beqa
Z_{{\cal M}_4\times {\rm conifold}}(\tau)&=&
\int_0^1 ds_1 \int_0^1ds_2 \sqrt{\tau_2} (q\overline{q})^{\frac{s_1^2}4}
\frac{\left|\hat\Lambda_1(\tau,s_1\tau -s_2) \right|^2 + \left|\hat\Lambda_2(\tau,s_1\tau -s_2) \right|^2}
{|\eta(\tau)|^2 |\vartheta_1(\tau,s_1\tau-s_2)|^2},
\label{Z_conifold_new(tau)}
\eeqa
where 
$\tau=\tau_1 + i\tau_2$.

By definition of $\hat\Lambda_1$ and $\hat\Lambda_2$,
we see that 
$Z_{{\cal M}_4\times {\rm conifold}}(\tau)$ is a partition function 
for the $N=2$ CFT for the conifold coupled to a complex 
fermion for the transverse space, with a GSO projection 
performed  before the discrete series representations 
are separated.

Including the four-dimensional boson contributions, we obtain 
the full modular invariant partition function of type\;II strings on the conifold 
\begin{eqnarray}
Z_{{\cal M}_4\times {\rm conifold}}^{\rm full}
&=&\int\frac{d\tau d\bar\tau}{\tau_2}\frac1{\tau_2^2|\eta^2(\tau)|^2} 
Z_{{\cal M}_4\times {\rm conifold}}(\tau).
\label{totalZ}
\end{eqnarray}

In the following we will show that $Z_{{\cal M}_4\times {\rm conifold}}^{\rm full}$:
\begin{itemize}
\item[(i)]{ 
is modular invariant.} 
\item[(ii)]{ 
reduces to 
(\ref{partition_function}) if, 
after a certain regularization, 
divided by a divergent volume factor.}
\item[(iii)]{
 also contains contributions from the discrete series of $SL(2,\R)$,
 which transform as four-dimensional ${\cal N}=2$ hyper/vector multiplets in type\;IIA/IIB 
 string compactifications.}
\end{itemize}

Here we note that, in going from (\ref{thetatheta}) to (\ref{Z_conifold_new(tau)}),
we have extended the summation region of $(n,w)$ from $(\Z , \Z)$ to 
$(\Z ,  \Z) \oplus (\Z+\frac12 ,  \Z+\frac12)$ to have a supersymmetric 
partition function. This is because we need 
$\Theta_{\nu,1}(\tau,s_2-s_1\tau)
(\Theta_{\tilde\nu,1}(\tau,s_2-s_1\tau))^*
$ with $\nu\tilde{\nu}=$ odd 
in order to 
contain spacetime fermions.
Therefore, we assume that $n$ and $w$ are allowed to 
take values in ${\bf Z}+\frac12$ as well as in ${\bf Z}$.
We also note that the particular $z$-dependence of the functions 
$\hat\Lambda_1$ and $\hat\Lambda_2$ is crucial to the construction, and 
is different from (\ref{Lambda1(z)}),(\ref{Lambda2(z)})
which are obtained by a composition of level-2 theta functions 
in Jacobi's quartic identity. (See Appendix.)

\subsection{Modular invariance of the type\;II conifold partition function}
We first prove the modular $S$-invariance of (\ref{Z_conifold_new(tau)}).
The following modular $S$-properties are well-known:
\beqa
\tau_2&\rightarrow&\frac{\tau_2}{|\tau|^2},\\
\eta(\tau)&\rightarrow&\eta\left(
-\frac1\tau
\right)=\sqrt{-i\tau}~\eta(\tau).
\eeqa
Also it is easy to see that 
\beqa
(q\bar q)^{\frac{s_1^2}4}&\rightarrow&(q\bar q)^{\frac{(1-s_2)^2}4}
\left|~ 
e^{-\frac{\pi i}2 \left(
\tau (1-s_2)^2+\frac{s_1^2}\tau
\right)}
 \right|^2.
\eeqa
On the other hand,  we have the following relation in general: 
\begin{eqnarray}
\Theta_{M,K}\left(
\tau,\frac{s_1\tau-s_2}{-a}
\right)
&
\stackrel{\tau\rightarrow -\frac1\tau}{\rightarrow}
&
e^{\frac{Ki\pi}{2a^2} \left(
\frac{s_1^2}\tau +\tau(1-s_2)^2 +2(s_1s_2-s_1)
\right)}
\nonumber\\
&&
\cdot
\sqrt{\frac\tau{2Ki}}
\sum_{M'\in{\bf Z}_{2K}} e^{-\frac{MM'}K \pi i}
\Theta_{M'+\frac  K a ,K}\left(\tau,\frac{(1-s_2)\tau-s_1}{-a}
\right)\nonumber\\
\label{S-transformed_Theta}
\end{eqnarray}
for any divisor $a$ of a positive integer $K$.
Comparing (\ref{S-transformed_Theta}) with 
\begin{eqnarray}
\Theta_{M,K}\left(
\tau,0
\right)
&
\stackrel{\tau\rightarrow -\frac1\tau}{\rightarrow}
&
\sqrt{\frac\tau{2Ki}}
\sum_{M'\in{\bf Z}_{2K}} e^{-\frac{MM'}K \pi i}
\Theta_{M' ,K}\left(\tau,0
\right),
\label{S-transformed_ordinaryTheta}
\end{eqnarray}
we see that
$\Theta_{M,K}\left(
\tau,\frac{s_1\tau-s_2}{-a}
\right)$ undergoes the following additional changes:
\begin{itemize}
\item{The exponential factor.}
\item{The replacement $(s_1,s_2)\rightarrow(1-s_2,s_1)$.}
\item{The shift in the first subscript of the theta function (``spectral flow").}
\end{itemize} 
 
Using (\ref{S-transformed_Theta}), we find
\beqa
\vartheta_1(\tau,s_1\tau-s_2)&\rightarrow&
-e^{\pi i \left(
\frac{s_1^2}\tau +\tau(1-s_2)^2 +2(s_1s_2-s_1)
\right)}
\vartheta_1\left(\tau,(1-s_2)\tau-s_1\right),
\\
\Theta_{m,1}(\tau,s_1\tau -s_2) &\rightarrow&
e^{\frac12\pi i \left(
\frac{s_1^2}\tau +\tau(1-s_2)^2 +2(s_1s_2-s_1)
\right)}
\sum_{m'\in\Z_2}e^{-mm' \pi i}
\Theta_{m'-1,1}\left(\tau,(1-s_2)\tau-s_1\right),
\nonumber\\
\\
\Theta_{s,2}(\tau,s_1\tau -s_2) &\rightarrow&
e^{\pi i \left(
\frac{s_1^2}\tau +\tau(1-s_2)^2 +2(s_1s_2-s_1)
\right)}
\sum_{s'\in\Z_4}e^{-\frac{ss'}4 \pi i}
\Theta_{s-2,2}\left(\tau,(1-s_2)\tau-s_1\right).
\eeqa
Since $2\pi i (s_1s_2-s_1)$ is pure imaginary, it is irrelevant if the absolute value is 
taken. Then the exponential factors of 
$e^{\mbox{\scriptsize const.}\times\pi i \left(
\frac{s_1^2}\tau +\tau(1-s_2)^2 \right)}$ arising from various 
factors of 
(\ref{Z_conifold_new(tau)}) cancel out.
The replacement  $(s_1,s_2)\rightarrow(1-s_2,s_1)$ acts trivially on $\int_0^1 s_1
\int_0^1 s_2 $.
Therefore, 
since 
\beqa
\frac{\left|\Lambda_1(\tau,0) \right|^2 + \left|\Lambda_2(\tau,0) \right|^2}
{|\eta^3(\tau)|^2 }
\eeqa
is modular invariant, we have only to worry about the shift of $m'$ and $s'$
in the theta functions in $\hat\Lambda_1$ and $\hat\Lambda_2$.
It turns out that they 
simply amount 
to the permutation
\beqa
\hat\Lambda_1&\rightarrow&\hat\Lambda_2,\\
\hat\Lambda_2&\rightarrow&\hat\Lambda_1,
\eeqa
which obviously preserves (\ref{Z_conifold_new(tau)}). Thus we have 
proved that  
$Z_{{\cal M}_4\times {\rm conifold}}(\tau)$
(\ref{Z_conifold_new(tau)})
is modular $S$-invariant.

The proof of the modular $T$-invariance is easier. Since 
\begin{eqnarray}
\hat\Lambda_1(\tau+1,z)&=&i\hat\Lambda_1(\tau,z),\\
\hat\Lambda_1(\tau+1,z)&=&-\hat\Lambda_2(\tau,z)
\end{eqnarray}
hold independently of $z$, all we need to do is 
to examine the effect of the change of $s_1\tau-s_2$, 
which amounts to a change of variables $(s_1, s_2)\rightarrow (s_1, s_2-s_1)$.
In fact, the integrand of 
$Z_{{\cal M}_4\times {\rm conifold}}(\tau)$
(\ref{Z_conifold_new(tau)}) is periodic (with a period of 1) in $s_2$, so the integral is invariant 
under the change of variables. Thus 
$Z_{{\cal M}_4\times {\rm conifold}}(\tau)$
is also T-invariant.


\subsection{Type II string partition functions for general $k$}
In this section, we extend the discussion for the conifold  
to more general singularities  
in which a nontrivial $N=2$ minimal model with a
general non-negative integer level $\km$ is coupled to the 
noncompact $N=2$ coset theory. 

The expression of the new partition function for the conifold 
$Z_{{\cal M}_4\times {\rm conifold}}(\tau)$ (\ref{Z_conifold_new(tau)}) is very suggestive;
it is similar in form to the old partition function (\ref{conifold_invariant}). 
In particular, the alternating 
sum is realized in similar functions $\Lambda_i(\tau,z)$ $(i=1,2)$ and 
$\hat\Lambda_i(\tau,z)$ $(i=1,2)$, which differ only in the $z$-dependences.
This motivates us to define
\begin{eqnarray}
\hat{F}_{l,2r}(\tau,z)&=&\frac14 
\sum_{m\in {\bf Z}_{4(\km+2)}} 
\left(
\vartheta_3(\tau,0)\vartheta_3(\tau,z)\mbox{ch}^{\mbox{\scriptsize NS}}_{l,m}(\tau,0)
-(-1)^{r-\frac m2}
\vartheta_4(\tau,0)\vartheta_4(\tau,z)\mbox{ch}^{\widetilde{\mbox{\scriptsize NS}}}_{l,m}(\tau,0)
\right.
\nonumber\\
&&\left.~~~~~~~~~~~~~~~\rule{0ex}{1em}
-\vartheta_2(\tau,0)\vartheta_2(\tau,z)\mbox{ch}^{\mbox{\scriptsize R}}_{l,m}(\tau,0)
\right)
\nonumber\\
&&
~~~~~~~~~~~~~~~~
\cdot
\Theta_{
(k_{\mbox{\scriptsize min}}+2)2r-(k_{\mbox{\scriptsize min}}+4)m,
2(k_{\mbox{\scriptsize min}}+2)(k_{\mbox{\scriptsize min}}+4)
}
\left({\textstyle
\tau,\frac z{k_{\mbox{\scriptsize min}}+4}
}\right).
\label{hatF}
\end{eqnarray}
Again, this $\hat{F}_{l,2r}(\tau,z)$  is obtained from
$F_{l,2r}(\tau,z)$ (\ref{F}),
which was defined in \cite{ES1} to construct the modular invariant 
partition function for the ADE type conifold-like singularities with 
only the continuous series representations. 
Remarkably, the level-$2(\km+2)(\km+4)$ theta functions 
are precisely the ones which appear in the $U(1)$-charge lattice decomposition 
\beqa
(\ref{lattice_decomposition}) 
&=&\sqrt{\frac{\tau_2}k}
\sum_{j_1,j_2\in\Z} 
~\sum_{m\in\Z_{4(\km +2)}}
~\sum_{r\in\Z_{2(\km +4)}}
e^{- k\pi \tau_2 s_1^2}\nonumber\\
&&\cdot
q^{\frac{\km +4}{2(\km +2)}\left(
2(\km+2)(j_1-j_2)
+\frac m2
-\frac{(\km +2)r}{\km +4}
\right)^2}
e^{-2\pi i (s_1\tau -s_2)
\left(
2(\km+2)(j_1-j_2)
+\frac m2
-\frac{(\km +2)r}{\km +4}
\right)}
\nonumber\\
&&\cdot
\bar q^{\frac{\km +4}{2(\km +2)}\left(
2(\km+2)(-j_1-j_2)
-\frac m2
-\frac{(\km +2)r}{\km +4}
\right)^2}
e^{+2\pi i (s_1\bar\tau -s_2)
\left(
2(\km+2)(-j_1-j_2)
-\frac m2
-\frac{(\km +2)r}{\km +4}
\right)}\nonumber\\
\\
&=&\sqrt{\frac{\tau_2}k}
~\sum_{m\in\Z_{4(\km +2)}}
~\sum_{r\in\Z_{2(\km +4)}}
e^{- k\pi \tau_2 s_1^2}\frac12
\nonumber\\
&&
\cdot
\Theta_{(\km+4)m-2(\km+2)r,
2(\km +2)(\km +4)}
\left(\tau,\frac{s_2-s_1\tau}{\km +4}\right)
\nonumber\\
&&
\cdot
\left(
\Theta_{-(\km+4)m-2(\km+2)r,
2(\km +2)(\km +4)}
\left(\tau,\frac{s_2-s_1\tau}{\km +4}\right)
\right)^*,
\label{thetatheta_generalk}
\eeqa
where $*$ denotes the complex conjugate. 
Therefore, $\hat{F}_{l,2r}(\tau,z)$ specifies a particular way of GSO projection in 
the $N=2$ minimal model, the transverse fermion and the noncompact 
$N=2$ coset Hilbert spaces.
Using $\hat{F}_{l,2r}(\tau,z)$, 
we can similarly write a modular invariant expression 
\begin{eqnarray}
Z_{{\cal M}_4\times CY(X_n)}(\tau)
& =&
\int_0^1  ds_1  \int_0^1 ds_2 \sqrt{\frac{\tau_2}k} (q\overline{q})^{\frac{ks_1^2}4}
\nonumber\\
&&
\cdot 
\sum_{l,\tilde{l}}N_{l,\tilde{l}}\sum_{r\in{\bf Z}_{\km+4}+\frac l2}
\frac{
\hat{F}_{l,2r}(\tau,s_1\tau -s_2)
\left(
\hat{F}_{\tilde{l},2r}(\tau,s_1\tau -s_2)
\right)^*
}
{|\eta(\tau)|^2 |\vartheta_1(\tau,s_1\tau-s_2)|^2}
\label{Z_ADE_new(tau)}
\end{eqnarray}
for general $\km$. 
$N_{l,\tilde l}$ is the coefficients of the $X_n$ $(X=A,D,$ or $E)$ modular invariant \cite{CIZ,Kato}.
Since $N_{l,\tilde l}$ vanishes if $l-\tilde l=1$ (mod 2) for any modular 
invariant, the summation over $r\in{\bf Z}_{\km+4}+\frac l2$ is 
equivalent to the one over $r\in{\bf Z}_{\km+4}+\frac{\tilde l}2$.
If $\km =0$, 
$Z_{{\cal M}_4\times CY(A_1)}(\tau)$
is reduced to $Z_{{\cal M}_4\times {\rm conifold}}(\tau)$
(\ref{Z_conifold_new(tau)}). 

The proof of the modular invariance of 
$Z_{{\cal M}_4\times CY(X_n)}(\tau)$
is parallel to 
the conifold case. 
Again,
the only nontrivial point is the $\tau$-dependence 
through the $z$-argument. In the present case the modular $S$-transformation 
simply permutes $\hat F$'s cyclically, and  
$Z_{{\cal M}_4\times CY(X_n)}(\tau)$
as a whole remains invariant. The proof of the modular $T$-invariance is also 
similar.

\subsection{Heterotic string partition functions for general $k$}
Once we have a modular invariant partition function for type\;II strings, we can 
easily convert it to one for heterotic strings by a standard procedure \cite{Gepner}, 
as we review in Appendix. All we need to do is replace the holomorphic 
$\hat F_{l,2r}(\tau,s_1\tau-s_2)$ in (\ref{Z_ADE_new(tau)}) with
$\hat F^{E_8\times E_8}_{l,2r}(\tau,s_1\tau-s_2)/\eta^{12}(\tau)$ 
(\ref{F^E_8xE_8})-(\ref{Lambda2^E_8xE_8}) for the $E_8\times E_8$ theory, 
and with
$\hat F^{SO(32)}(\tau,s_1\tau-s_2)/\eta^{12}(\tau)$ 
(\ref{F^SO(32)})-(\ref{Lambda2^SO(32)}) for the $SO(32)$ theory. 
The anti-holomorphic $(\hat F_{{\tilde l},2r}(\tau,s_1\tau-s_2))^*$ is left 
unchanged. 
Since $\hat F^{E_8\times E_8}_{l,2r}(\tau,s_1\tau-s_2)/\eta^{12}(\tau)$ or 
$\hat F^{SO(32)}(\tau,s_1\tau-s_2)/\eta^{12}(\tau)$ transforms in the same way as 
$\hat F_{l,2r}(\tau,s_1\tau-s_2)$ does, the resulting heterotic partition function 
is automatically modular invariant. Their massless spectra will be investigated 
in the next section.

\section{Separation of the  discrete series contributions}
We now separate the contributions from the discrete series representations 
from the partition functions obtained in the previous section. In section \ref{modules}, 
we first define modules of various algebras and describe relevant spectral 
flow operations in them, which are needed later. Then we consider the separation 
for type II strings from section \ref{flow-orbit} through section \ref{evenkm},
and for heterotic strings in section \ref{separation_heterotic}.

\subsection{Modules and  spectral flows}
\label{modules}
\begin{itemize}
\item{\it The SL(2,R) Kac-Moody algebra module ${\cal H}^{SL(2,{\bf R})}_{\pm,(h,l_0)} $.}

The affine $SL(2,{\bf R})$ algebra relations are \cite{DLP}
\beqa
{[}J^3_n, ~~J^3_m{]}&=&-\frac\kappa 2 n\delta_{n,-m},\\
{[}J^3_n, ~~J^\pm_m{]}&=&\pm J^\pm_{n+m},\\
{[}J^+_n, ~~J^-_m{]}&=&-\kappa  n\delta_{n,-m} +2 J^3_{n+m}
\eeqa
for $n$, $m\in \Z$, where 
\beqa
\kappa&=&k+2.
\eeqa
The Virasoro generators are
\newcommand{\La}{L^{SL(2,{\bf R})}}
\beqa
\La_0&=&\frac1{2(\kappa -2)}
\left(\rule{0em}{4ex}
J^+_0 J^-_0 + J^-_0 J^+_0  -2(J^3_0)^2 \right.\nonumber\\
&&\left.
+2 \sum_{m=1}^\infty  \left(J^+_{-m} J^-_m + J^-_{-m} J^+_m  -2 J^3_{-m} J^3_m 
\right)
\right),
\nonumber\\
\La_n&=&\frac1{2(\kappa -2)}
\sum_{-\infty}^\infty  \left(J^+_{-m} J^-_m + J^-_{-m} J^+_m  -2 J^3_{-m} J^3_m \right).
\eeqa
We define ${\cal H}^{SL(2,{\bf R})}_{\pm,(h,l_0)} $ as an $SL(2,{\bf R})$ Kac-Moody 
algebra module generated from a state $|h,l_0\rangle$ such that 
\begin{eqnarray}
L^{SL(2,{\bf R})}_0|h,l_0\rangle&=& h|h,l_0\rangle,\\
J_0^3|h,l_0\rangle&=& l_0|h,l_0\rangle,\\
L^{SL(2,{\bf R})}_n|h,l_0\rangle&=&J_n^3|h,l_0\rangle=J_n^+|h,l_0\rangle=J_n^-|h,l_0
\rangle=0,
~~~(n>0)~~~\\
J_0^\mp|h,l_0\rangle&=&0.
\end{eqnarray}
The character for a {\it generic} representation is given by
\begin{eqnarray}
\mbox{Tr}_{{\cal H}^{SL(2,{\bf R})}_{\pm,(h,l_0)}}q^{L_0^{SL(2,{\bf R})}}y^{J_0^3}=
\frac{\pm i q^{\frac18 + h} y^{\mp\frac12 + l_0}}{\vartheta_1(\tau,z)}.
\end{eqnarray}

Let us define the spectral-flow operation 
\beqa
J^\pm_n&=&\tilde J^\pm_{n\mp w},\\
J^3_n&=&\tilde J^3_n + \frac{\kappa w}2 \delta_{n,0},\\
\La_n&=&\tilde{L}^{SL(2,R)}_n- w\tilde{J}^3_n-\frac{\kappa w^2}4 \delta_{n,0},
\eeqa
then the tilde generators also satisfy the same algebra relations as those without 
tildes, so it is an isomorphism. In particular, if we set $w=1$, then
\beqa
J^-_0&=&\tilde{J}^-_1,\\
J^+_1&=&\tilde{J}^+_0,
\eeqa
and 
${\cal H}^{SL(2,{\bf R})}_{+,(h,l_0)}$
as a module generated by $J_n$'s can be identified to be 
${\cal H}^{SL(2,{\bf R})}_{-,(h+l_0-\frac\kappa 4,l_0-\frac\kappa 2)}$ 
as a module generated by $\tilde{J}_n$'s. 
Therefore, for any function $f(\La_0,J^3_0)$, the following equation
holds true:
\beqa
{\rm Tr}_{{\cal H}^{SL(2,{\bf R})}_{+,(h,l_0)}}
f\left(\La_0,J^3_0\right)
&=&
{\rm Tr}_{{\cal H}^{SL(2,{\bf R})}_{-,(h+l_0-\frac\kappa 4,l_0-\frac\kappa 2)}}
f\left(\tilde{L}^{SL(2,R)}_0- \tilde{J}^3_0-\frac{\kappa}4,~
\tilde J^3_0 + \frac{\kappa}2 \right).
\eeqa

\item{\it The free fermion module ${\cal H}_{\nu,2}$ ($\nu\in{\bf Z}_4$).}

The complex fermion algebra  is generated by $\psi^{\pm}_r$, where 
$r\in\Z+\frac12$ in the NS sector and $r\in\Z$ in the Ramond sector, 
with the relations
\beqa
&&\{\psi^+_r,~\psi^+_s\}=\{\psi^-_r,~\psi^-_s\}=0,\\
&&\{\psi^+_r,~\psi^-_s\}=\{\psi^-_r,~\psi^+_s\}=\delta_{r,-s}.
\eeqa
The $L_0$ and fermion number operators are 
\beqa
L_0^{(\rm NS)}&=&\sum_{r\in\Z+\frac12,>0}
\frac r2\left(
\psi^+_{-r}\psi^-_r+\psi^-_{-r}\psi^+_r
\right),\\
L_0^{(\rm R)}&=&\sum_{r\in\Z,>0}
\frac r2\left(
\psi^+_{-r}\psi^-_r+\psi^-_{-r}\psi^+_r
\right)+\frac18,\\
F^{(\rm NS)}&=&\frac12\sum_{r\in\Z+\frac12,>0}
\left(
\psi^+_{-r}\psi^-_r-\psi^-_{-r}\psi^+_r
\right),\\
F^{(\rm R)}&=&\frac12\psi^+_0\psi^-_0
+\frac12\sum_{r\in\Z,>0}
\left(
\psi^+_{-r}\psi^-_r-\psi^-_{-r}\psi^+_r
\right)
-\frac12.
\eeqa
As usual, we introduce the NS ground state $|0\rangle_{\rm NS}$ 
such that
\beqa
\psi^\pm_r|0\rangle_{\rm NS}&=&0
\eeqa
for $r=\frac12, \frac32,\ldots$,
and the Ramond ground state $|0\rangle_{\rm R}$
such that  
\beqa
\psi^+_r|0\rangle_{\rm R}&=&0
\eeqa
for $r=1,2,\ldots$, while 
\beqa
\psi^-_r|0\rangle_{\rm R}&=&0
\eeqa
for $r=0,1,2,\ldots$. Then
\beqa
L_0^{(NS)}|0\rangle_{\rm NS}&=&0,\\
F^{(NS)}|0\rangle_{\rm NS}&=&0,\\
L_0^{(R)}|0\rangle_{\rm R}&=&\frac18 |0\rangle_{\rm R},\\
F^{(R)}|0\rangle_{\rm R}&=&-\frac12 |0\rangle_{\rm R}.
\eeqa
Let us call the free fermion modules generated from these ground states 
${\cal H}^{(NS)}$ and ${\cal H}^{(R)}$, respectively. 
We also define 
${\cal H}_{\nu,2}$ ($\nu\in{\bf Z}_4$)
to be a free fermion module such that 
\begin{eqnarray}
\mbox{Tr}_{{\cal H}_{\nu,2}}q^{L_0^{(\nu)}}y^{F^{(\nu)}}=
q^{\frac1{24}}\frac{\Theta_{\nu,2}(\tau,z)}{\eta(\tau)},
\end{eqnarray}
where 
\beqa
L_0^{(\nu)}&=&L_0^{(\rm NS)}~~~\mbox{if $\nu=0,2$},\\
&=&L_0^{(\rm R)}~~~~~\mbox{if $\nu=\pm1$},\\
F^{(\nu)}&=&F^{(\rm NS)}~~~\mbox{if $\nu=0,2$},\\
&=&F^{(\rm R)}~~~~~\mbox{if $\nu=\pm1$}.
\eeqa
Clearly, ${\cal H}_{0,2}$ (${\cal H}_{2,2}$) consists of even (odd) 
$F^{(\rm NS)}$ states in  ${\cal H}^{(NS)}$, and similarly 
${\cal H}_{1,2}$ (${\cal H}_{-1,2}$) consists of states with 
$F^{(R)}=+\frac12 +$ even- (odd-) integer in  ${\cal H}^{(\rm R)}$. 
Also
\beqa
{\cal H}^{(\rm NS)}&=&{\cal H}_{0,2}\oplus{\cal H}_{2,2},\\
{\cal H}^{(\rm R)}&=&{\cal H}_{1,2}\oplus{\cal H}_{-1,2}.
\eeqa

For both the NS and the Ramond sectors, 
\beqa
\psi^{\pm(\rm NS,R)}_r&=&\tilde\psi^{\pm(\rm NS,R)}_{r+1}
\eeqa
is an isomorphism and maps ${\cal H}^{(\rm NS)}$ to ${\cal H}^{(\rm NS)}$,
and ${\cal H}^{(\rm R)}$ to ${\cal H}^{(\rm R)}$.
We also have, any function $f(L^{(\nu)}_0,F^{(\nu)})$, the following equation
\beqa
{\rm Tr}_{{\cal H}_{\nu,2}}
f\left(L^{(\nu)}_0,F^{(\nu)}\right)
&=&
{\rm Tr}_{{\cal H}_{\nu+2,2}}
f\left(\tilde{L}^{(\nu+2)}_0-\tilde{F}^{(\nu+2)}+\frac12,~\tilde{F}^{(\nu+2)}-1\right).
\eeqa

\item{\it The free boson module ${\cal H}_{m,K}$ ($K=2(\km+2)(\km+4)$, $m\in{\bf Z}_{2K}$).}

Let us consider a free scalar field $\phi(z)$ with the OPE
\beqa
\phi(z)\phi(w)&\sim&-\log(z-w)
\eeqa
with the energy-momentum tensor and the $U(1)$ current 
\newcommand{\Lb}{L^{U(1)}}
\newcommand{\Jb}{J^{U(1)}}
\beqa
T^{U(1)}(z)&=&-\frac12(\partial\phi(z))^2,\\
J^{U(1)}(z)&=&i\sqrt{\frac K2}\partial\phi(z).
\eeqa
for some integer $K=2(\km+2)(\km+4)$. 
Let $\Lb_0$ and $\Jb_0$ be their zeromodes. Then ${\cal H}_{m,K}$ ($m\in{\bf Z}_{2K}$) 
is defined to be a (reducible)
free boson module such that
 \begin{eqnarray}
\mbox{Tr}_{{\cal H}_{m,K}}q^{L_0^{U(1)}}y^{J_0^{U(1)}}&=&
q^{\frac1{24}}\frac{\Theta_{m,K}(\tau,z)}{\eta(\tau)},
\end{eqnarray}
and hence
 \begin{eqnarray}
\mbox{Tr}_{{\cal H}_{m,K}}q^{L_0^{U(1)}}y^{\frac{J_0^{U(1)}}{\km +4}}&=&
q^{\frac1{24}}\frac{\Theta_{m,K}(\tau,\frac z{\km +4})}{\eta(\tau)}.
\end{eqnarray}
${\cal H}_{m,K}$ is a direct product of free boson modules 
generated from the ground states
\beqa
e^{i\sqrt{2K}(n+\frac m{2K})\phi(0)}|0\rangle~~~(n\in\Z),
\eeqa
where $|0\rangle$ is the $\Jb_0=0$ ground state.

The replacement
 \beqa
 \Lb_0&=&\tilde{L}_0^{U(1)} -\frac{\tilde{J}_0^{U(1)}}{\km +4} +\frac{\km +2}{2(\km+4)},\\
  \Jb_0&=&\tilde{J}_0^{U(1)} -(\km +2)
 \eeqa
is a spectral flow by $2(\km+2)$ units, and hence is an isomorphism.
As before, we have a relation
\beqa
&&{\rm Tr}_{{\cal H}_{m,K}}
f\left(\Lb_0,\Jb_0\right)\nonumber\\
&=&
{\rm Tr}_{{\cal H}_{m+2(\km +2),K}}
f\left(\tilde{L}_0^{U(1)} -\frac{\tilde{J}_0^{U(1)}}{\km +4} +\frac{\km +2}{2(\km+4)},
~\tilde{J}_0^{U(1)} -(\km +2)\right)
\eeqa
for any  $f(\Lb_0,\Jb_0)$.

\item{\it The \mbox{$N=2$} minimal superconformal algebra
 module ${\cal H}^{(k_{\mbox{\rm\scriptsize min}})l,s}_{m}$. }

Finally, we define the \mbox{$N=2$} minimal superconformal algebra
 module ${\cal H}^{(\km)l,s}_{m}$ such that
\beqa
\mbox{Tr}_{{\cal H}^{(\km)l,s}_{m}}q^{L_0^{N=2}}y^{J_0^{N=2}} &= &
q^{\frac{c_{\mbox{\tiny min}}}{24}}\chi^{(\km)l,s}_m(\tau,z).
\eeqa 
In Appendix we collect useful formulas of the $N=2$ minimal characters 
$\chi^{(\km)l,s}_m(\tau,z)\equiv\chi^{l,s}_m(\tau,z)$, where the 
$\km$-dependence is suppressed for notational simplicity. 
We do not need spectral flow formulas for them because they do not have 
the denominator $U(1)$ charge of the gauged WZW model. 
\end{itemize}

\subsection{The flow-orbit representation of the partition functions}
\label{flow-orbit}
To extract the discrete series contributions from
(\ref{Z_ADE_new(tau)}), we will write it as a trace of some operator 
over appropriate modules defined in the previous subsection. 
First, we note that the function $F_{l,2r}(\tau,z)$, introduced in \cite{ES1}
to construct modular invariants containing only the continuous series, 
can be written as 
a spectral flow orbit with respect to the $N=2$ $U(1)$ charge (which is not 
the same as the denominator $U(1)$ charge of the $SL(2,\R)/U(1)$ coset 
counted by $J_0^{\mbox{\scriptsize tot}}$ below,  
as emphasized in \cite{ES}), as shown in 
Appendix. Since $F_{l,2r}(\tau,z)$ and $\hat F_{l,2r}(\tau,z)$ differ
only in the $z$-dependences of theta functions,  $\hat F_{l,2r}(\tau,z)$ can 
also be expressed as a similar alternating sum (see (\ref{Fl2r}))
\beqa
\hat F_{l,2r}(\tau,z)&=&\frac12\sum_{\nu\in\Z_{4(\km +2)}}
\sum_{
\mbox{\tiny $
							\begin{array}{c}
							\nu_0,\nu_1,\nu_2\in Z_2\\
							\nu_0+\nu_1+\nu_2 \\
							\equiv 1({\rm mod}2)
							\end{array}$}
							}
(-1)^{\nu}
							\chi^{l,l-2r+2\nu_0+\nu}_{l+\nu}(\tau,0)
							\Theta_{2\nu_1+\nu,2}(\tau,0)
							\nonumber\\
&&~~~~~~~~~
\cdot\Theta_{2\nu_2+\nu,2}(\tau,z)
\Theta_{(\km +2)2r-(\km+4)(l+\nu),2(\km+2)(\km +4)}
\left(
\tau,\frac z{\km +4}
\right).
\nonumber\\
\eeqa
It motivates us to define 
\cite{Mizoguchi:2007}
\begin{eqnarray}
{\cal H}^{(\nu)}_{F_{l,2r}}&\equiv&
							{\oplus}_{\!\!\!\!\!\!\!\!\!\!\!\!\!\!\!\!\!\!\!\!\mbox{\tiny $
							\begin{array}{c}\\
							\nu_0,\nu_1,\nu_2\in Z_2\\
							\nu_0+\nu_1+\nu_2 \\
							\equiv 1({\rm mod}2)
							\end{array}$}
							}
											\!\!\!\!\!\!\left(
					{\cal H}_{l+\nu}^{(\km)l, l-2r+ 2\nu_0+\nu} \otimes
					{\cal H}_{2\nu_1+\nu,2}
					\otimes {\cal H}_{2\nu_2+\nu,2}
				\right)
				\nonumber\\
&&				\otimes ~
				{\cal H}
				_{(\km+2)2r
				-(\km+4)(l+\nu),
									2(\km+2)
									(\km+4)
				}.
												\end{eqnarray}
We can then write 
\begin{eqnarray}
&&\int_0^1 ds_1 \int_0^1 ds_2 
\sqrt{\frac{\tau_2}k} (q\bar q)^{\frac{ks_1^2}4}
\frac{
\hat F_{l,2r}(\tau, s_1\tau - s_2)
\left(\hat F_{\tilde l,2 r}(\tau, s_1\tau - s_2)\right)^*
}
{\left|
\eta(\tau) \vartheta_1(\tau, s_1\tau - s_2)\right|^2}
\nonumber\\
&=&
\int_0^1 ds_1 \int_0^1 ds_2 
\sqrt{\frac{\tau_2}k} (q\bar q)^{\frac{ks_1^2}4}
\frac{|\eta^2(\tau)|^{2}}4 \sum_{\nu,\tilde{\nu}\in{\bf Z}_{4(\km+2)}}(-1)^{\nu+\tilde{\nu}}
\nonumber\\
&&\cdot 
\mbox{Tr}_{\left( {\cal H}^{SL(2,{\bf R})}_{+,(0,0)} 
	\otimes{\cal H}_{F_{l,2r}}^{(\nu)}
	\right)
	\otimes
	 \left({\cal H}^{SL(2,{\bf R})}_{+,(0,0)} 
	\otimes{\cal H}_{F_{\tilde l,2r}}^{(\tilde\nu)}
	\right)}
\nonumber\\
&&~~~~~~~~~~~~~~~~~~~~~~~~~~~~~~~~~~
\cdot 
q^{L_0^{SL(2,{\bf R})}+L_0^{N=2}+L_0^{(\nu)}+L_0^{(\nu)}+L_0^{U(1)}
	\!\!\!-\frac14-\frac{c_{\mbox{\tiny min}}}{24}
	+s_1\left(J_0^3+F^{(\nu)}+\frac{J_0^{U(1)}}{\km+4}+\frac12\right)}	
	\nonumber\\
&&
~~~~~~~~~~~~~~~~~~~~~~~~~~~~~~~~~~
\cdot 
\bar{q}^{\tilde{L}_0^{SL(2,{\bf R})}+\tilde{L}_0^{N=2}+\tilde{L}_0^{(\tilde{\nu})}+\tilde{L}_0^{(\tilde{\nu})}+
 	\tilde{L}_0^{U(1)}
	\!\!\!-\frac14-\frac{c_{\mbox{\tiny min}}}{24}
	+s_1\left(\tilde{J}_0^3+\tilde{F}^{(\tilde{\nu})}+\frac{\tilde{J}_0^{U(1)}}{\km+4}
	+\frac12\right)}	\nonumber\\
&&~~~~~~~~~~~~~~~~~~~~~~~~~~~~~~~~~~
\cdot e^{-2\pi i s_2 \left(J_0^3+F^{(\nu)}+\frac{J_0^{U(1)}}{\km+4}
-\tilde{J}_0^3-\tilde{F}^{(\tilde{\nu})}-\frac{\tilde{J}_0^{U(1)}}{\km+4}\right)}. 
\label{to_do_s1s2integrals}
\end{eqnarray}
%
The $s_2$ integral yields a constraint 
\beqa
(J_0^{\mbox{\scriptsize tot}}\equiv)~
J^3_0+F^{(\nu)}
+\frac{J_0^{U(1)}}{\km+4}
&=&\tilde{J}^3_0+\tilde{F}^{(\tilde{\nu})}
+\frac{\tilde{J}_0^{U(1)}}{\km+4}
~(\equiv \tilde J_0^{\mbox{\scriptsize tot}}).
\label{J0totcondition}
\eeqa  
Using the Fourier transformation
\begin{eqnarray}
\sqrt{\frac{\tau_2}k}(q\bar q)^{\frac{ks_1^2}4}
&=&\frac1k\int_{-\infty}^\infty dc ~e^{-\frac \pi{k\tau_2} c^2 -2 \pi i c s_1},
\end{eqnarray}
we can perform the $s_1$ integral as
\beqa
&=&\frac{|\eta^2(\tau)|^2}{4k}
\int_{-\infty}^\infty dp
\sum_{\nu,\tilde\nu\in\Z_{4(\km +2)}}
(-1)^{\nu +\tilde\nu}
\mbox{Tr}_{\left(
{\cal H}^{SL(2,{\bf  R})}_{+,(0,0)}
\otimes
{\cal H}^{(\nu)}_{F_{l,2r}}
\right)
	\times
\left(
{\cal H}^{SL(2,{\bf  R})}_{+,(0,0)}
\otimes
{\cal H}^{(\tilde\nu)}_{F_{\tilde l,2\tilde r}}
\right)
	}
	\nonumber\\
	&&
\cdot	q^{L_0^{SL(2,{\bf R})}+L_0^{N=2}+L_0^{(\nu)}+L_0^{(\nu)}+L_0^{U(1)}
	\!\!\!-\frac14-\frac{c_{\mbox{\tiny min}}}{24}
	}
	\bar q^{\tilde L_0^{SL(2,{\bf R})}+\tilde L_0^{N=2}
	+\tilde L_0^{(\tilde\nu)}+\tilde L_0^{(\tilde\nu)}+\tilde L_0^{U(1)}
	\!\!\!-\frac14-\frac{c_{\mbox{\tiny min}}}{24}
	}\nonumber\\
	&&
\cdot
\frac{(q\bar q)^{\frac1k\left(
p+\frac{ik}2
\right)^2
+J^{\mbox{\tiny tot}}_0 +\frac{k+2}4}
-(q\bar q)^{\frac{p^2}k}
}
{-2\pi(ip+ J^{\mbox{\scriptsize tot}}_0 +\frac12)},
\eeqa
where we set  $c=2\tau_2 p$.

The first term of the numerator contains $J^{\mbox{\scriptsize tot}}_0$
in its exponent. If we use the isomorphisms of various modules 
in the previous section, we can eliminate this  
$J_0^{\mbox{\scriptsize tot}}$ dependence, and also
the module over which the first trace is taken changes from 
$
{\cal H}^{SL(2,{\bf R})}_{+,(0,0)} 
	\otimes 
	{\cal H}^{(\nu)}_{F_{l,2r}}
$
to
$
{\cal H}^{SL(2,{\bf R})}_{-,(-\frac\kappa 4,-\frac\kappa 2)} 
	\otimes 
	{\cal H}^{(\nu)}_{F_{l,2(r+1)}}
$ ($\kappa=k+2$), and similarly in the anti-holomorphic sector:
\begin{eqnarray}
(\ref{to_do_s1s2integrals})
&=&\frac{|\eta^2(\tau)|^{2}}{4k} 
 \sum_{\nu,\tilde{\nu}\in{\bf Z}_{
4(\km+2)
}}(-1)^{\nu+\tilde{\nu}}
\nonumber\\
&&\cdot
 \left(
 \mbox{Tr}_{\left( {\cal H}^{SL(2,{\bf R})}_{-,(-\frac\kappa 4,-\frac\kappa 2)} 
	\otimes 
	{\cal H}^{(\nu)}_{F_{l,2(r+1)}}
	\right)
	\otimes
	 \left({\cal H}^{SL(2,{\bf R})}_{-,(-\frac\kappa 4,-\frac\kappa 2)} 
	 {\cal H}^{(\tilde\nu)}_{F_{l,2(r+1)}}
	\right)}
	\right.
\nonumber\\
&&
\cdot \int_{-\infty}^\infty \frac{dp}{-2\pi(ip+J_0^{\mbox{\scriptsize tot}}+\frac12)} 
q^{\frac1k(p+\frac{ik}2)^2-\frac14 -\frac{c_{\mbox{\tiny min}}}{24}
+L_0^{SL(2,{\bf R})}+L_0^{N=2}+L_0^{(\nu)}+L_0^{(\nu)}
+L_0^{U(1)}+\frac\kappa 4
}
\nonumber\\ &&
~~~~~~~~~~~~~~~~~~~~~~~~~~~~~~~
\cdot
\bar{q}^{\frac1k(p+\frac{ik}2)^2-\frac14 -\frac{c_{\mbox{\tiny min}}}{24}
+\tilde{L}_0^{SL(2,{\bf R})}+\tilde{L}_0^{N=2}
+\tilde{L}_0^{(\tilde{\nu})}
+\tilde{L}_0^{(\tilde{\nu})}
+\tilde{L}_0^{U(1)}+\frac\kappa 4
}
\nonumber\\
&&
-\mbox{Tr}_{\left( {\cal H}^{SL(2,{\bf R})}_{+,(0,0)} 
	\otimes 
	{\cal H}^{(\nu)}_{F_{l,2r}}
	\right)
	\otimes
	 \left({\cal H}^{SL(2,{\bf R})}_{+,(0,0)} 
	\otimes 
	{\cal H}^{(\tilde\nu)}_{F_{l,2r}}
	\right)}
\nonumber\\
&&
\cdot 
\int_{-\infty}^\infty \frac{dp}{-2\pi(ip+J_0^{\mbox{\scriptsize tot}}+\frac12)} 
q^{\frac{p^2}k-\frac14 -\frac{c_{\mbox{\tiny min}}}{24}
+L_0^{SL(2,{\bf R})}+L_0^{N=2}+L_0^{(\nu)}+L_0^{(\nu)}+L_0^{U(1)}
}
\nonumber\\ &&\left.
~~~~~~~~~~~~~~~~~~~~~~~~~~~~~~~
\cdot
\bar{q}^{\frac{p^2}k-\frac14 -\frac{c_{\mbox{\tiny min}}}{24}
+\tilde{L}_0^{SL(2,{\bf R})}+\tilde{L}_0^{N=2}
+\tilde{L}_0^{(\tilde{\nu})}+\tilde{L}_0^{(\tilde{\nu})}+\tilde{L}_0^{U(1)}
}
\right).
\label{1sttrace-2ndtrace}
\end{eqnarray}
The first trace is simplified by replacing 
${\cal H}^{SL(2,{\bf R})}_{-,(-\frac\kappa 4,-\frac\kappa 2)} $'s with
${\cal H}^{SL(2,{\bf R})}_{-,(0,-\frac\kappa 2)} $'s, and at the same time 
removing $\frac\kappa4$ from the exponents of $q$ and $\bar q$.

\begin{figure}
\includegraphics[width=160mm]{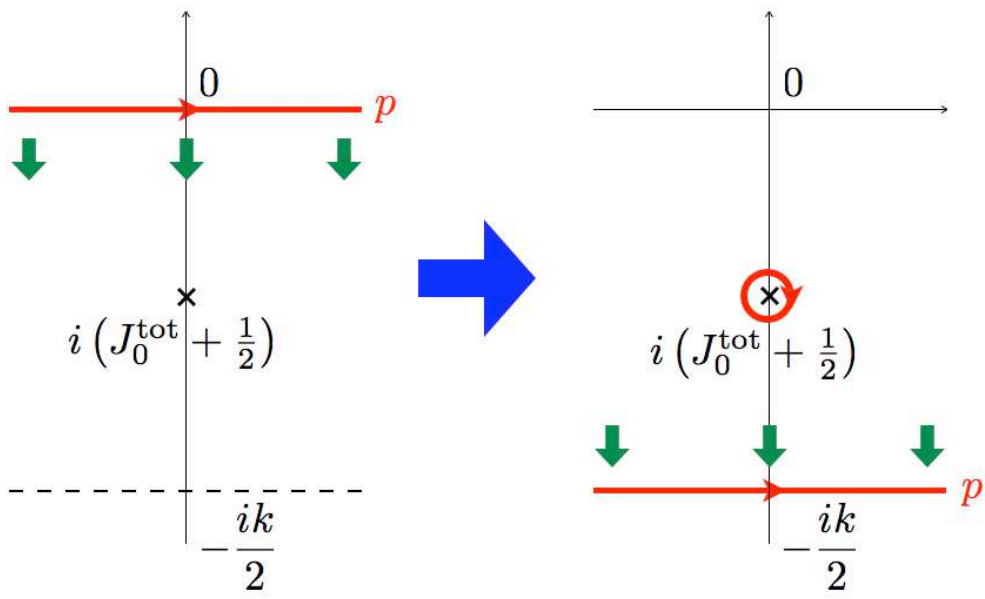}
\caption{The contour deformation of the $p$ integration. 
\label{contr_deformation}
}
\end{figure}

As was done in \cite{MOS,HPT,ES}, we will now change the integration contour 
of the first trace from $p'\equiv p+\frac {ik}2 \in {\bf R}+\frac {ik}2$ to ${\bf R}$.
Then it picks up a residue of the pole at $p=i(J_0^{\mbox{\scriptsize tot}}+\frac12)$  for the states satisfying 
$-\frac{k+1}2<J_0^{\mbox{\scriptsize tot}}<-\frac12$ (Figure \ref{contr_deformation}). 
We will show that in sect.\ref{oddkm} that 
these imaginary-momentum 
states reside below the lower bound of the continuous spectrum, and precisely on 
the boundary of the unitary region \cite{BFK,DLP}. That is, they are the discrete series representations. 

\subsection{The continuous spectrum}
As in \cite{MOS,HPT,ES}, we consider the continuous and 
discrete spectra separately.

The continuous spectrum arises from the first trace of (\ref{1sttrace-2ndtrace}) 
with the $p$-integration 
contour deformed, and the second trace for which we do not need any 
deformation.
Since
\beqa
&&q^{-\frac14 -\frac{c_{\mbox{\tiny min}}}{24}
+L_0^{SL(2,{\bf R})}+L_0^{N=2}+L_0^{(\nu)}+L_0^{(\nu)}+L_0^{U(1)}}
\nonumber\\
&=&
q^{\left(L_0^{SL(2,{\bf R})}-\frac18\right)
+\left(L_0^{N=2}-\frac{c_{\mbox{\tiny min}}}{24}\right)
+\left(L_0^{(\nu)}-\frac1{24}\right)
+\left(L_0^{(\nu)}-\frac1{24}\right)
+\left(L_0^{U(1)}-\frac1{24}\right)},
\eeqa
if the denominator $ip+J_0^{\mbox{\scriptsize tot}}+\frac12$ were absent,
we would formally obtain
\beqa
\sum_{\nu}
(-1)^\nu
 \mbox{Tr}_{
 {\cal H}^{SL(2,{\bf R})}_{\pm,(0,-\frac\kappa 2)} 
	\otimes 
	{\cal H}^{(\nu)}_{F_{l,2r}}
	}
	q^{-\frac14 -\frac{c_{\mbox{\tiny min}}}{24}
+L_0^{SL(2,{\bf R})}+L_0^{N=2}+L_0^{(\nu)}+L_0^{(\nu)}+L_0^{U(1)}}
&=&
\frac{\pm i}{\vartheta_1(\tau,0)}\cdot
\frac{F_{l,2r}(\tau,0)}{\eta^3(\tau)},
\nonumber\\
\eeqa
which contains a divergent factor $\frac{\pm i}{\vartheta_1(\tau,0)}$.
This divergence comes from the zero mode contributions in the $SL(2,\R)$ 
module. In reality, the traces in the holomorphic and anti-holomorphic sectors 
are not independent but are constrained by the condition (\ref{J0totcondition}),
but still the trace is divergent because, for a given pair of holomorphic and 
anti-holomorphic states with fixed values of $L_0$, $\tilde L_0$ and 
$J_0^{\mbox{\scriptsize tot}}(=\tilde J_0^{\mbox{\scriptsize tot}})$, there are 
infinitely many states having the same $L_0$, $\tilde L_0$ but different 
$J_0^{\mbox{\scriptsize tot}}(=\tilde J_0^{\mbox{\scriptsize tot}})$, and 
the sum of the form
\beqa
-\sum_{n=0}^\infty \frac{1}{z-n}
\eeqa
does not converge. Following \cite{MOS}, we use the formula
\beqa
-\sum_{n=0}^\infty\frac{e^{-n\epsilon}}{z-n}
&=&-\log\epsilon+\frac\partial{\partial z}\log\Gamma(-z)
-{\cal C}+O(\epsilon)+O(\epsilon\log\epsilon),
\label{regularization_formula}
\eeqa
to regularize this divergence to obtain a finite answer. 
We give a proof for (\ref{regularization_formula}) in Appendix D, 
thereby correcting (irrelevant)
typos (the minus sign in front of $\log$ and Euler's constant) in \cite{MOS}.
Then the contribution to  
$-\frac{1}{2\pi(ip+J_0^{\mbox{\scriptsize tot}}+\frac12)}$ 
from an arbitrary number of $J_0^-$ multiplications
in ${\cal H}^{SL(2,{\bf R})}_{-,(0,-\frac\kappa 2)} $
(in the first trace of (\ref{1sttrace-2ndtrace})) 
is
$-\frac{\log\epsilon}{2\pi}$ times 
\begin{eqnarray}
\mbox{Tr}
_{{\cal H}^{SL(2,{\bf R})}_{-,(0,0)} /\{J_0^-\}}
q^{L_0^{SL(2,{\bf R})}
}y^{J_0^3}
&=&\frac{- i q^{\frac18} y^{+\frac12}}{\vartheta_1(\tau,z)}
\Big/ \frac1{1-y^{-1}}
\nonumber\\
&\stackrel{z\rightarrow 0}{\rightarrow}&
\frac{q^{\frac18}}{\eta^3(\tau)}
\end{eqnarray}
to leading order.
Here we denote by ${\cal H}^{SL(2,{\bf R})}_{-,(h,j_0)} /\{J_0^-\}$ 
the coset of the module ${\cal H}^{SL(2,{\bf R})}_{-,(h,l_0)}$ 
obtained by modding out the $J_0^-$ multiplication. 
Similar equations hold for 
the second trace.
Therefore, (\ref{1sttrace-2ndtrace}) becomes
\beqa
(\ref{1sttrace-2ndtrace}) 
&=&\frac{-\log\epsilon}{2\pi}
\cdot
\frac1{4k}
\int_{-\infty}^\infty dp (q\bar q)^{\frac{p^2}k}
\frac1{|\eta(\tau)|^2}
\left(
\frac{F_{l,2(r+1)}(\tau,0)(F_{\tilde l,2(r+1)}(\tau,0))^*}{|\eta^3(\tau)|^2}
\right.
\nonumber\\
&&~~~~~~~~~~~~~~~~~~~~~~~~~~~~~~~~~~~~~~~~~~~
+
\left.
\frac{F_{l,2r}(\tau,0)(F_{\tilde l,2r}(\tau,0))^*}{|\eta^3(\tau)|^2}
\right)+O(\epsilon^0). 
\label{regularized1sttrace-2ndtrace}
\eeqa
$Z_{{\cal M}_4\times CY(X_n)}(\tau)$ 
(\ref{Z_ADE_new(tau)}) is obtained by summing (\ref{regularized1sttrace-2ndtrace})
over $r\in\Z_{\km +4}+\frac l2$ and $l,\tilde l$ with a weight $N_{l,\tilde l}$, and hence
\beqa
Z_{{\cal M}_4\times CY(X_n)}(\tau)&=&
\frac{-\log\epsilon}{8\pi k}
\sum_{l,\tilde l}N_{l,\tilde l}\sum_{r\in\Z_{\km +4}+\frac l2}
\int_{-\infty}^\infty dp (q\bar q)^{\frac{p^2}k}
\frac1{|\eta(\tau)|^2}
\nonumber\\&&
~~~
\cdot
\left(
\frac{F_{l,2(r+1)}(\tau,0)(F_{\tilde l,2(r+1)}(\tau,0))^*}{|\eta^3(\tau)|^2}
+
\frac{F_{l,2r}(\tau,0)(F_{\tilde l,2r}(\tau,0))^*}{|\eta^3(\tau)|^2}
\right)+O(\epsilon^0)
\nonumber\\
&=&
\frac{-\log\epsilon}{8\pi k}
\sum_{l,\tilde l}N_{l,\tilde l}\sum_{r\in\Z_{\km +4}+\frac l2}
\sqrt{\frac k{\tau_2}}~
\frac1{|\eta(\tau)|^2}
\cdot
\frac{F_{l,2r}(\tau,0)(F_{\tilde l,2r}(\tau,0))^*}{|\eta^3(\tau)|^2}
+O(\epsilon^0).
\nonumber\\
\label{regularizedZ_ADE_new}
\eeqa
This
shows that the coefficient of 
the  $\log\epsilon$ divergence of 
$Z_{{\cal M}_4\times CY(X_n)}(\tau)$
is precisely the integrand of the old partition function (\ref{conifold_type_partition_function})
(without the transverse boson factor)
consisting of only 
the continuous series representations.
 
\subsection{The discrete spectrum}
\label{The_discrete_spectrum}
Let us now consider the discrete spectrum, which is the main focus of this paper.
In section \ref{flow-orbit}, we have deformed the $p$-integration contour of the 
first trace in (\ref{1sttrace-2ndtrace}), the summation of which over $l$, $\tilde l$ 
(with a weight $N_{l,\tilde l}$) 
and $r$ is equal to 
$Z_{{\cal M}_4\times CY(X_n)}(\tau)$. 
Then any state in 
$
{\left( {\cal H}^{SL(2,{\bf R})}_{-,(-\frac\kappa 4,-\frac\kappa 2)} 
	\otimes 
	{\cal H}^{(\nu)}_{F_{l,2(r+1)}}
	\right)
	\otimes
	 \left({\cal H}^{SL(2,{\bf R})}_{-,(-\frac\kappa 4,-\frac\kappa 2)} 
	\otimes
	 {\cal H}^{(\tilde\nu)}_{F_{l,2(r+1)}}
	\right)}
	$
such that the eigenvalue of $\Jtot$ is between $-\frac{k+1}2$ and $-\frac12$ 
gives rise to a pole in the integrand.
The resulting small contour around the pole is clock-wise, 
and the residue integral just cancels the $-2\pi i$ factor of the denominator. 
The residue contributions to 
$Z_{{\cal M}_4\times CY(X_n)}(\tau)$
are therefore
\beqa
\mbox{Residues}&=&
\sum_{l,\tilde l} N_{l,\tilde l}
\sum_{r\in \Z_{\km +4} +\frac l 2}
\frac{|\eta^2(\tau)|^2}{4k}\nonumber\\
&&\cdot
\sum_{\nu,\tilde\nu\in \Z_{4(\km +2)}}
(-1)^{\nu + \tilde \nu}
\mbox{Tr}_
{\left.\left( {\cal H}^{SL(2,{\bf R})}_{-,(
       0,-\frac\kappa 2)} 
	\otimes 
	{\cal H}^{(\nu)}_{F_{l,2r}}
	\right)
	\otimes
	 \left({\cal H}^{SL(2,{\bf R})}_{-,(
	 0,-\frac\kappa 2)} 
       \otimes
       {\cal H}^{(\tilde\nu)}_{F_{l,2r}}
	\right)\right|_{-\frac{k+1}2\leq\Jtot\leq\frac12,~\Jtot=\tildeJtot}}
	\nonumber\\
	&&~~~~~~~~~~~~~~~~\cdot
	(q\bar q)^{\frac 1k\left(
	i\left(
	\Jtot + \frac12
	\right) +\frac{ik}2
	\right)^2}
	\nonumber\\
	&&~~~~~~~~~~~~~~~~\cdot
q^{
\left(L^{SL(2,{\bf R})}_0 -\frac18\right)
+\left(L^{N=2}_0 -\frac{c_{\mbox{\tiny min}}}{24}\right)
+\left(L^{(\nu)}_0 -\frac1{24}\right)
+\left(L^{(\nu)}_0 -\frac1{24}\right)
+\left(L^{U(1)}_0 -\frac1{24}\right)
}
\nonumber\\
&&~~~~~~~~~~~~~~~~\cdot
\bar q^{
\left(\tilde L^{SL(2,{\bf R})}_0 -\frac18\right)
+\left(\tilde L^{N=2}_0 -\frac{c_{\mbox{\tiny min}}}{24}\right)
+\left(\tilde L^{(\nu)}_0 -\frac1{24}\right)
+\left(\tilde L^{(\nu)}_0 -\frac1{24}\right)
+\left(\tilde L^{U(1)}_0 -\frac1{24}\right)
},
\nonumber\\
\label{residues}
\eeqa
where we have shifted $r+1$ to $r$ in the $r$-summation, and also  
$L^{SL(2,{\bf R})}_0$ (and $\tilde L^{SL(2,{\bf R})}_0$) by $\frac\kappa 4$ 
as we mentioned below eq.(\ref{1sttrace-2ndtrace}).

As we noted at the beginning of sect.\ref{Partition_functions}, 
(\ref{residues}) would become a polynomial with integer coefficients 
if we had started from the partition function (\ref{Z_superSL2RoverU1}) 
with an extra overall factor of $C=4k$.


To obtain the discrete spectrum, we first relax the conditions for 
$\Jtot$ and $\tildeJtot$ and consider 
\beqa
&&\sum_{l,\tilde l} N_{l,\tilde l}
\sum_{r\in \Z_{\km +4} +\frac l 2}
\frac{|\eta^2(\tau)|^2}{4k}
\nonumber\\
&&\cdot
\sum_{\nu,\tilde\nu\in \Z_{4(\km +2)}}
(-1)^{\nu + \tilde \nu}
\mbox{Tr}_
{
\left( 
{\cal H}^{SL(2,{\bf R})}_{-,(0,-\frac\kappa 2)}
	\otimes 
	{\cal H}^{(\nu)}_{F_{l,2r}}
	\right)
	\otimes
	 \left(
{\cal H}^{SL(2,{\bf R})}_{-,(0,-\frac\kappa 2)} 
       \otimes
       {\cal H}^{(\tilde\nu)}_{F_{l,2r}}
	\right)
	}
	\nonumber\\
	&&~~~~~~~~~~~~~~~~\cdot
q^{
\left(L^{SL(2,{\bf R})}_0 -\frac18\right)
+\left(L^{N=2}_0 -\frac{c_{\mbox{\tiny min}}}{24}\right)
+\left(L^{(\nu)}_0 -\frac1{24}\right)
+\left(L^{(\nu)}_0 -\frac1{24}\right)
+\left(L^{U(1)}_0 -\frac1{24}\right)
}y^{\Jtot}
\nonumber\\
&&~~~~~~~~~~~~~~~~\cdot
\bar q^{
\left(\tilde L^{SL(2,{\bf R})}_0 -\frac18\right)
+\left(\tilde L^{N=2}_0 -\frac{c_{\mbox{\tiny min}}}{24}\right)
+\left(\tilde L^{(\nu)}_0 -\frac1{24}\right)
+\left(\tilde L^{(\nu)}_0 -\frac1{24}\right)
+\left(\tilde L^{U(1)}_0 -\frac1{24}\right)
}\bar y^{\tildeJtot}
\label{instead}
\eeqa
instead of (\ref{residues}). Next we 
find the states which satisfy the conditions 
$-\frac{k+1}2\leq \Jtot \leq-\frac12$ and $\Jtot=\tildeJtot$, and then 
we take into account the ``drop" of $L_0$  due to 
the imaginary momentum factor
\beqa
(q\bar q)^{\frac 1k\left(
	i\left(
	\Jtot + \frac12
	\right) +\frac{ik}2
	\right)^2}
\label{imaginary_momentum}
\eeqa
 in (\ref{residues}). 
Without the factor (\ref{imaginary_momentum}), 
we can easily evaluate (\ref{instead}):
\footnote{ 
$\hat F_{l,2r}(\tau,z)$ is so defined that $\hat F_{l,2r}(\tau,z)$ 
coincides with $F_{l,2r}(\tau,0)$, where 
the latter was defined in \cite{ES}. It is more natural to consider 
$2\hat F_{l,2r}$ here because it is a polynomial of $q$ with integer coefficients.} 
\beqa
(\ref{instead})&=&\frac 1{4k} 
\sum_{l,\tilde l} N_{l,\tilde l}
\sum_{r\in \Z_{\km +4} +\frac l 2}
\frac{
2\hat F_{l,2r}(\tau,z)(2\hat F_{\tilde l,2r}(\tau,z))^*}
{
\left|
y^{\frac{\kappa-1}2}
\tilde\vartheta_1(\tau,z)\eta(\tau)
\right|^2}.
\label{J0tot_dependence}
\eeqa

\subsection{Massless spectra for odd $\km$}
\label{oddkm}
\begin{figure}
\hskip -15ex
\includegraphics[width=200mm]{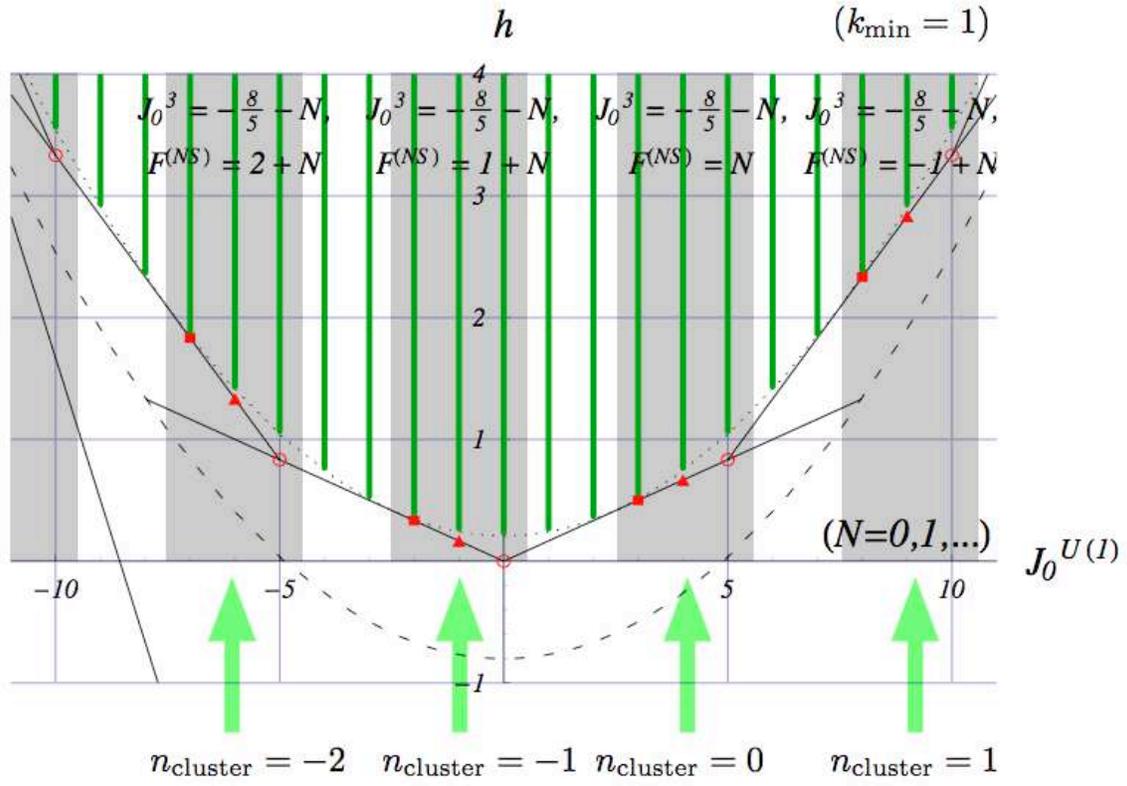}
\caption{The clusters and the discrete states ($\km=1$, NS sector). 
The circle at $(J^{U(1)}_0,h)=(0,0)$ in the $n_{cluster}=-1$ cluster 
and the square at $(3,\frac12)$ in the $n_{cluster}=0$ cluster 
correspond to two massless supermultiplets for type\;II compactifications. 
\label{clusters}
}
\end{figure}

We consider the cases $\km$ odd and $\km$ even separately.
We first assume that $\km$ is odd.
Massless states in type\;II string theories come from those with the total 
conformal weight $\frac12$. Therefore, they must lie at the lowest $L^{SL(2,\R)}_0$ 
level.
Since $J^3_0$ takes values 
\beqa
J^3_0&=&-\frac\kappa 2,~-\frac\kappa 2 -1,~-\frac\kappa 2 -2,\ldots~~~(\kappa =k+2)
\label{J^3_0}
\eeqa
at the lowest $L^{SL(2,\R)}_0$ 
level
in ${\cal H}^{SL(2,{\bf R})}_{-,(0,-\frac\kappa 2)}$, the condition 
\beqa
-\frac{k+1}2<\Jtot<-\frac12
\eeqa
for the existence of a pole implies that 
a noncompact $N=2$ representation
can contribute to the discrete series spectrum only if 
it carries a $J^{U(1)}_0$ charge in the ranges
\beqa
\frac{\km+4}2 +(\km+4)n_{cluster}<&J^{U(1)}_0&<
\left(\frac{\km+4}2 +(\km+4)n_{cluster}\right)+\km +2,
\nonumber
\label{J^U(1)_0range}
\eeqa
\beqa
n_{cluster}&\equiv&-\frac\kappa2-J^3_0-F^{(\nu)}~\in 
\left\{ 
\begin{array}{ll}
\Z&\mbox{(NS sector),}\\
\Z+\frac12&\mbox{(R sector),}\\
\end{array}
\right.
\label{ncluster}
\eeqa
where we have introduced 
a label $n_{cluster}$
to distinguish different ``clusters" of relevant noncompact $N=2$ 
representations (Figure \ref{clusters}).

Let us consider a continuous family of 
noncompact $N=2$ representations with a definite $J^{U(1)}_0$ charge
in the range (\ref{J^U(1)_0range}), which is drawn as a semi-infinite 
line in Figure \ref{clusters}.
As we discussed, such a family in the partition function 
is accompanied by a residue contribution, which has a 
conformal weight yet lower than the lower bound of the continuous spectrum 
by an amount equal to the exponent of (\ref{imaginary_momentum}).
For the holomorphic part, it is found to be
\beqa
-\frac{\left(
J^{U(1)}_0-(\km +4)(n_{cluster}+\frac12)
\right)^2}
{2(\km+2)(\km+4)}.
\label{drop}
\eeqa
The lower bound of the continuous spectrum (which can be read off 
from the level-$2(\km+2)(\km+4)$ theta function) is
\beqa
\frac14 - \frac{\cm}{24}
+2(\km+2)(\km+4)\left(
\frac{J^{U(1)}_0}{2(\km+2)(\km+4)}
\right)^2.
\label{lower_bound}
\eeqa
Adding (\ref{drop}) to (\ref{lower_bound}), we find the conformal weight 
of the residue contribution
\beqa
 (\ref{lower_bound})+(\ref{drop})
 &=&
\frac{\left(
n_{cluster}+\frac12
\right)
 J^{U(1)}_0}{\km +2}
-\frac{{\hat c}_{KS}-1}2
\left(
\left(
n_{cluster}+\frac12
\right)
^2 -\frac14
\right),
\label{h_discrete}
\eeqa 
where 
\beqa
{\hat c}_{KS}&=&\frac\kappa{\kappa-2}
\eeqa
is $\frac13$ of the central charge of the  
noncompact $N=2$ CFT.
(\ref{h_discrete}) is precisely the series of equations of the boundary lines 
surrounding the polygonal region of the $N=2$ unitary representations \cite{BFK}
(Figure {\ref{clusters}}).
Thus we have shown that they are indeed the $N=2$ representations coming 
from the discrete series of $SL(2,{\bf R})$.

It turns out that the states with the total conformal weight $\frac12$
exist only in the $n_{cluster}=0$ and
$n_{cluster}=-1$ clusters.
If $n_{cluster}=0$, (\ref{J^3_0}) implies that the ``Liouville fermion number"
(that is, the number of the fermion oscillators
of the noncompact $N=2$ CFT) 
in the NS sector
$F^{(\nu)}$ ($=F^{(\rm NS)}$)
takes values $0,1,\dots$. (The R sector can be analyzed in the same 
way as done below; anyway the supersymmetry ensures that the massless spectra 
must be identical.) For massless states $F^{(\rm NS)}$ must be $0$ or $1$ because 
otherwise the conformal weight $h$ exceeds $\frac12$.
The noncompact $N=2$ CFT representations in the $n_{cluster}=0$ cluster
have $J^{U(1)}_0$ charges in the range
\beqa
\frac{\km}2 +2 &<~J^{U(1)}_0~<& \frac{3\km}2 +4.
\eeqa
The upper limit for massless states is much stronger than 
this:
\beqa
\frac{\km}2 +2 &<~J^{U(1)}_0~\leq& \km+2
\eeqa
because otherwise the straight-line boundary, on which 
the discrete series resides, already goes above $h=\frac12$. 
Therefore, there are $\frac{\km +1}2$ different possible 
$J^{U(1)}_0$ charges
\beqa
J^{U(1)}_0=\km+2-j~~~\left(j=0,\ldots,\frac{\km-1}2\right).
\eeqa
Since it does not contain $J^{U(1)}_0=0$, $F^{(\rm NS)}$ needs to be $0$ 
and we look for $h=\frac12$ combinations of states of the noncompact coset and 
$N=2$ minimal CFT sectors.
It turns out that for every $j$ above, there exists precisely one $\hat F_{j,2r}$ 
which contains an $h=\frac12$ combination; this is $\hat F_{j,j+2}$.
Indeed, it contains NS-sector terms such as (See Appendix.)
\beqa
\hat F^{(+)\rm NS}_{j,j+2}(\tau,z)\cdot q^{\frac14
}
&=&q^{\frac14}\chi^{j,0}_j(\tau,0)(\Theta_{0,2}(\tau,0)\Theta_{0,2}(\tau,z)+
\Theta_{2,2}(\tau,0)\Theta_{2,2}(\tau,z))
\nonumber\\
&&~~~~~~~~~~~\cdot
\Theta_{-2j+2(\km+2),2(\km+2)(\km +4)}
\left(
\tau,\frac z{\km +4}
\right)+\cdots\nonumber\\
&=&
q^{\frac14-\frac{\cm}{24}
+\frac j{2(\km +2)}
+\frac{\left(
-j+\km+2
\right)^2}{2(\km +2)(\km +4)}
}
y^{\frac{-j+\km+2}{\km +4}}
+
\cdots,
\eeqa
where we have taken into account 
the extra factor of $q^\frac14$
because of the shortage of 
the eta (or theta) functions\footnote{To read off the conformal weights of 
the internal CFT representations from (\ref{J0tot_dependence}), 
we write the denominator as 
an integer power series of $q$ with an overall ghost ground state factor 
(in the spherical worldsheet coordinates) of 
$q^{\frac12}$.
In the usual critical strings, this factor may be thought of as 
provided by the $12$ eta functions coming from the $8$ transverse 
bosons and the $4$ complex fermions, that is, the normal  ordering constant 
factor in the cylindrical worldsheet coordinates. In the present case, we have only one 
$\vartheta_1$ and three $\eta$'s (one from (\ref{J0tot_dependence}) 
and two from the transverse bosons (\ref{totalZ})), so we need to multiply both 
the denominator and the numerator by $q^\frac14$. This is the Liouville energy,
which makes the tachyon be massless in two-dimensional string theory.}
in the denominator 
of the partition function.
If we include the imaginary-momentum contribution (\ref{drop}) with $n_{cluster}=0$, 
we have the conformal weight
\beqa
\frac14-\frac{\cm}{24}
+\frac j{2(\km +2)}
+\frac{\left(
-j+\km+2
\right)^2}{2(\km +2)(\km +4)}
-\frac{\left(
-j+\km+2-\frac{\km +4}2
\right)^2}
{2(\km+2)(\km+4)}=
\frac12.
\label{1/2}
\eeqa
Since $\hat F_{j,j+2}=\hat F_{\km-j,\km-j+2}$ (\ref{Fhatsymmetry}),   
$\hat F^{(+)\rm NS}_{l,l+2}$ give rise to an $h=\frac12$ 
state for every $l=0,1,\ldots,\km$. 
The character of the $N=2$ minimal model is  $\chi^{j,0}_j$.
The anti-holomorphic sector is similar. 
So for the $A_{\km +1}$ modular invariant there are
$\frac{\km +1}2$ complex scalars in the NS-NS sector.

Taking also the Ramond sector 
into account, each 
NS-NS complex scalar becomes a part of 
a single hyper-multiplet for type\;IIA
(since the two massless Ramond-Ramond states are spacetime scalars), 
and a single vector multiplet for typeIIB strings (since 
the Ramond-Ramond states become the two helicity states of a massless 
vector).


Next we turn to the $n_{cluster}=-1$ cluster. In this case 
(\ref{J^3_0}) and (\ref{ncluster}) require that the NS-sector 
Liouville fermion number $F^{(\rm NS)}$ takes values $\geq 1$.
This means that the Liouville fermion 
already ``spends" the maximal conformal weight for massless states 
(that is, $h=\frac12$) so that 
the remaining CFT sectors can only associate an $h=0$ state 
with it.
In this $n_{cluster}=-1$ cluster,
$J^{U(1)}_0$ takes 
integral values in the range 
\beqa
-\frac{\km}2 -2 &<~J^{U(1)}_0~<& \frac\km2,
\eeqa
and indeed contains $J^{U(1)}_0=0$. The family of noncompact 
$N=2$ representations with $J^{U(1)}_0=0$ are contained
in several $\hat F_{l,2r}$'s, among which 
\beqa
\hat F^{(-)\rm NS}_{0,0}(\tau,z)\cdot q^{\frac14}
&=&q^{\frac14}\chi^{0,0}_0(\tau,0)
(\Theta_{0,2}(\tau,0)\Theta_{2,2}(\tau,z)+
\Theta_{2,2}(\tau,0)\Theta_{0,2}(\tau,z))
\nonumber\\
&&~~~~~~~~~~~\cdot
\Theta_{0,2(\km+2)(\km +4)}
\left(
\tau,\frac z{\km +4}
\right)+\cdots
\eeqa
only gives rise to an $h=0$ combination of states of the noncompact 
and minimal $N=2$ CFTs, if the effect (\ref{drop}) of the imaginary-momentum factor 
is taken into account.
Since $J^{U(1)}_0=0$ for the $n_{cluster}=-1$ cluster of $\hat F^{\rm NS}_{0,0}(\tau,z)$,
(\ref{drop}) becomes
\beqa
(\ref{drop})&=&-\frac{\km+4}{8(\km +2)},
\eeqa
which cancels the extra conformal weight of the continuous series 
$\frac14-\frac\cm{24}$. 
The first level-2 theta function with 
$z=0$ comes from the complex fermion of the transverse spacetime dimensions, 
while the second level-2 theta function is the one from the 
Liouville fermion. 
$\Theta_{2,2}(\tau,0)\Theta_{0,2}(\tau,z)$ contains a term $q^{\frac12}(y^{+1}+y^{-1})$, 
of which $q^{\frac12}y^{+1}$ has $F^{(\rm NS)}=1$ and 
corresponds to a discrete state.
On the other hand, although
$\Theta_{2,2}(\tau,0)\Theta_{0,2}(\tau,z)$ has $2q^{\frac12}$ in the expansion, they have
$F^{(\rm NS)}=0$ and hence do not correspond to discrete states.
Combined with a similar state in the anti-holomorphic sector, this $h=\frac12$ state 
becomes a real scalar in the four-dimensional spacetime. 
Since $\hat F_{0,0}=\hat F_{\km,2(\km +4)}$,
there is another real scalar. 

Taking account of the Ramond sector again, they constitute a single hyper/vector multiplet 
for typeIIA/IIB strings.  In all, 
for the $A_{\km+1}$ modular invariant with odd $\km$, 
there are $\frac{\km+3} 2$ massless hyper/vector multiplets 
for typeIIA/IIB strings.
$\frac{\km+1} 2$ of them have NS-sector discrete states  
in the $n_{cluster}=0$ cluster,  
whereas one has those in the $n_{cluster}=-1$ cluster. 
R-sector discrete states are all in the $n_{cluster}=-\frac12$ cluster.

In usual ``compact" Gepner models, where the internal $N=2$ CFT 
consists of only the $N=2$ minimal models, the internal $h=0$ state is always 
accompanied by a graviton, an anti-symmetric tensor and a dilaton, with their 
superpartners. In contrast, we have only a massless scalar in the 
spectrum and there is no localized massless graviton due to the constraint 
$F^{(\rm NS)}\geq 1$ for $n_{cluster}=-1$.
\footnote{The author thanks T.Eguchi and Y.Sugawara for 
discussions on this point.} 

We should also note that, although the massless state in the $n_{cluster}=-1$ cluster
comes from the free boson module ${\cal H}_{0,K}$, it does not contain  
the identity representation module of the noncompact $N=2$ CFT 
because that massless state is made of a combination of $|0\rangle$ 
in the free boson module {\em and} an $F^{(\rm NS)}=1$ state in the free 
fermion module. This combination of states has $h=\frac12$ and hence 
is not contained in the $N=2$ identity representation module. This can be 
seen by the fact that the generic (reducible) $N=2$ character with $h=0$, $Q=0$
is decomposed into irreducible characters of the identity representation and 
two discrete series representations with $h=\frac12$, $Q=\pm1$. 
This is consistent with the fact that the identity representation of $SL(2,{\bf R})$ 
does not correspond to a normalizable mode.

\subsection{Massless spectra for even $\km$: A gapless spectrum}
\label{evenkm}
Massless spectra for even $\km$ are similar to those for the odd case, 
but there is a crucial difference.

\begin{figure}
\begin{center}
\includegraphics[width=150mm]{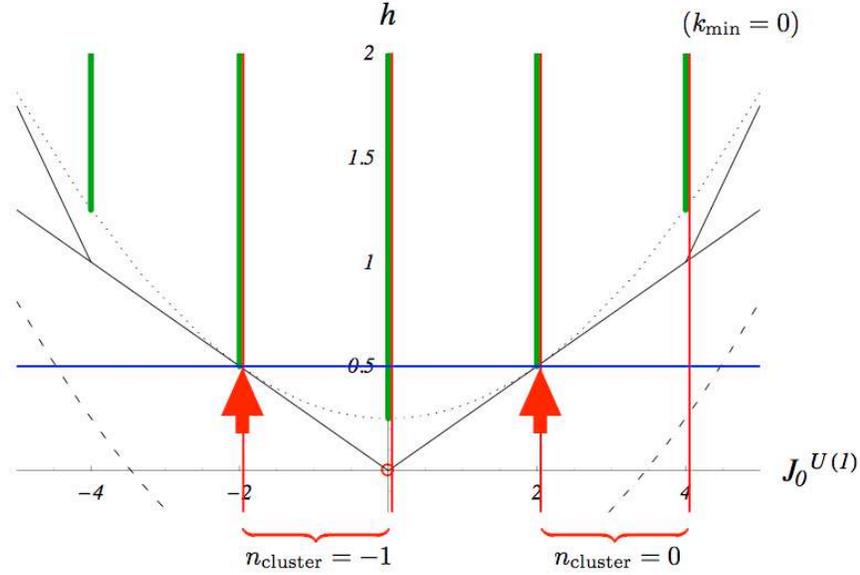}\\
\vskip -3em
(a) The $\km=0$ case. \\
\includegraphics[width=150mm]{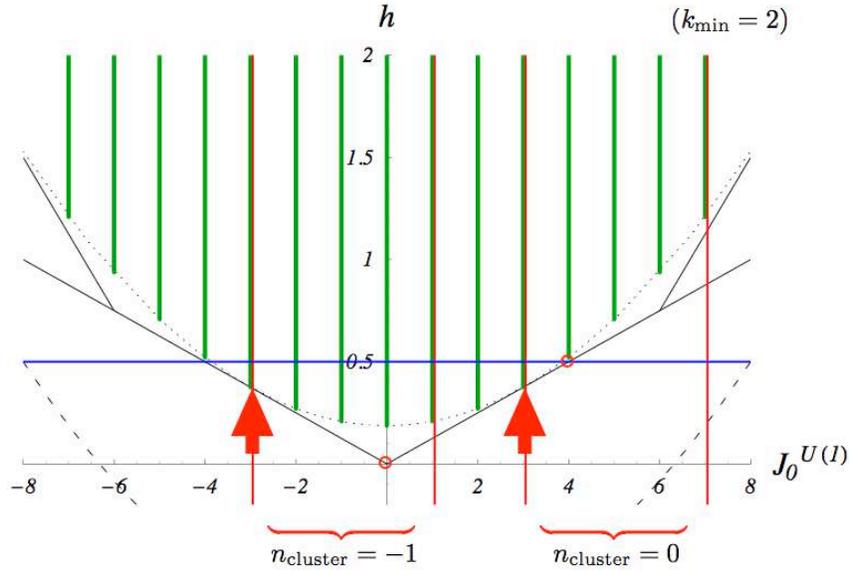}\\
\vskip -3em
(b) The $\km=2$ case.
\end{center}
\caption{Spectra for even $\km$. The small circles show the 
locations of the massless discrete states. Some continuous spectra reach 
the boundary of the unitary region (the arrows). The $N=2$ $U(1)$ 
charge $Q$ is $=\frac{J^{U(1)}_0}{\km+2}$.
\label{kmin_even}
}
\end{figure}

After the contour deformation discussed in 
section \ref{The_discrete_spectrum}, some families of continuous series
``leave behind" discrete series as pole contributions if  
\beqa
\frac{\km+4}2 +(\km+4)n_{cluster}\leq&J^{U(1)}_0&\leq
\left(\frac{\km+4}2 +(\km+4)n_{cluster}\right)+\km +2,
\nonumber\\
\label{J^U(1)_0range_kmineven}
\eeqa
where $n_{cluster}$
is given by (\ref{ncluster}). 
Like in the case of odd $\km$, only the  $n_{cluster}=-1$ and $0$ 
clusters are relevant for the massless spectrum.
In the $n_{cluster}=0$ cluster in the NS sector, 
massless states only come from noncompact $N=2$ representations with 
$h\leq \frac12$, so the upper limit of (\ref{J^U(1)_0range}) is lowered to
\beqa
\frac{\km+4}2 \leq&J^{U(1)}_0&\leq
\km +2.
\label{J^U(1)_0range_kmineven_h<1/2}
\eeqa
Since $\km$ is even, the contours before 
and after the deformation can ``hit" the pole if $J^{U(1)}_0$ is at either of the 
two ends of the domain (\ref{J^U(1)_0range_kmineven}).
For the restricted range  (\ref{J^U(1)_0range_kmineven_h<1/2}), 
the pole can still be located {\em on} the contour {\em after} the deformation if 
$J^{U(1)}_0$ is at the lower limit ($=\frac{\km+4}2$). 
This means that the continuous spectrum already reaches 
the boundary of the unitary region (Figure \ref{kmin_even}). %
Depending on how the contour is deformed to
circumvent this pole,
the residue may or may not contribute to 
the partition function, and irrespective of 
how it is deformed, there exist a continuous spectrum of 
modes arbitrarily close to the discrete mode.

As we discussed in \cite{Mizoguchi}, 
a generic $N=2$ representation becomes reducible at the 
boundary of the unitary region, 
where the generic character is decomposed into a sum of 
characters of discrete (including the identity) representations.  Also,
in that paper we interpreted this massless state as the geometric 
modulus of the conifold. 
In this paper, we regard the geometric moduli of a singular Calabi-Yau 
not as a part  of the continuous spectra, but as pole contributions to 
the partition function. In the conifold ($\km=0$) case, there is 
another massless multiplet from the $n_{cluster}=-1$ sector, 
which has a nonzero mass gap below the lowest end of the continuum 
and hence may be identified as the geometric modulus. 
Since the (deformed) conifold has only one 
modulus (the size of the $S^3$), this would imply that the gapless 
state does not correspond to any topological cycle for general even-$\km$ 
models.

Besides the gapless spectrum at 
$J^{U(1)}_0=\frac{\km+4}2$, there are $\frac\km 2$ possible $J^{U(1)}_0$ 
values 
\beqa
J^{U(1)}_0=\km+2-j~~~(j=0,\ldots,\frac\km2 -1).
\eeqa
which give rise to massless states, similarly to the $\km$ odd case. 
Again $\hat F_{j,j+2}$ and \\ 
$\hat F_{\km-j,\km-j+2}$ 
$(j=0,\ldots,\frac\km2 -1)$ 
correspond to such states. The all reside below the lowest 
limit of the continuous spectrum with a finite mass gap.
The $n_{cluster}=-1$ sector is also similar to that for the $\km$ odd case.

To summarize, for the $A_{\km+1}$ modular invariant with even $\km$, 
there are (excluding the gapless one) 
$\frac\km 2 +1$ massless hyper/vector multiplets 
for typeIIA/IIB strings.
Similarly to the $\km$ odd case, 
$\frac\km 2 $ of them has NS-sector discrete states in the $n_{cluster}=0$ 
cluster, and one has those in the $n_{cluster}=-1$.

We should note that the pattern of the chiral ring structure has already 
been recognized in \cite{ES}. The recognition of the gapless spectrum 
for the even $\km$ case is new, however.

\subsection{Separation of the discrete series for heterotic strings}
\label{separation_heterotic}

Massless discrete spectra for heterotic strings can be similarly 
obtained from the heterotic conversion of (\ref{J0tot_dependence}):
\beqa
&\longrightarrow&\frac 1{4k} 
\sum_{l,\tilde l} N_{l,\tilde l}
\sum_{r\in \Z_{\km +4} +\frac l 2}
\frac{2\hat F^{het}_{l,2r}(\tau,z)(2\hat F_{\tilde l,2r}(\tau,z))^*}
{\left|
y^{\frac{\kappa-1}2}
\tilde\vartheta_1(\tau,z)
\right|^2
\eta^{13}(\tau)(\eta(\tau))^*}.
\label{J0tot_dependencehet}
\eeqa
where $\hat F^{het}_{l,2r}$ is $\hat F^{E_8\times E_8}_{l,2r}$ or 
$\hat F^{SO(32)}_{l,2r}$ given in Appendix. 
In this case we search for $h=1$ states for the left (holomorphic) sector.

\underline{\it $h=1$ states in 
$\hat F^{het}_{j,j+2}$ with odd $\km$}

As we did in the type II case, we first assume that $\km$ is odd.
We have seen in the previous sections that $\hat F^{(+)\rm NS}_{j,j+2}(\tau,z)$ 
$(j=0,\ldots,\frac\km 2 -1)$
has 
$h=\frac12$ discrete states:
\beqa&&
\hat F^{(+)\rm NS}_{j,j+2}(\tau,z)
\cdot q^{\frac14-\frac{\left(
-j+\km+2-\frac{\km +4}2
\right)^2}{2(\km+2)(\km+4)}}
\nonumber\\
&=&q^{\frac14-\frac{\left(
-j+\km+2-\frac{\km +4}2
\right)^2}{2(\km+2)(\km+4)}}
\left(
	\chi^{j,0}_j(\tau,0)
	\Theta_{-2j+2(\km+2),2(\km+2)(\km +4)}
	\left(
	\tau,\frac z{\km +4}
	\right)
\right.\nonumber\\
&&~~~~~~~~~~~~~~~~~~~~~~~~~~~~~
\left.
	+\chi^{\km-j,0}_{-(\km-j)}(\tau,0)
	\Theta_{-2j-4,2(\km+2)(\km +4)}
	\left(
	\tau,\frac z{\km +4}
	\right)+\cdots
\right)\nonumber\\
&&~~~~~~~~~~~~~~~~~~~~~~~~~~~~~\cdot
\left(\Theta_{0,2}(\tau,0)\Theta_{0,2}(\tau,z)+
\Theta_{2,2}(\tau,0)\Theta_{2,2}(\tau,z)\right),
\label{hatFNSjj+2}
\eeqa
where we have written out 
the $m=j+2$ (as well as the $m=j$) term in (\ref{hatFl2r(+)withchi})
because it becomes relevant for the heterotic massless spectrum.
Also, we have already included the factor from the imaginary momentum.
If it is converted to the heterotic versions, the level-2 theta functions 
with argument $(\tau,0)$ change as
\beqa
\Theta_{0,2}=\frac{\vartheta_3+\vartheta_4}{2}
&\rightarrow&
\frac{(\vartheta_3)^{5}-(\vartheta_4)^{5}}{2\eta^{4}}
B^{(E_8)}
,\\
\Theta_{2,2}=\frac{\vartheta_3-\vartheta_4}{2}
&\rightarrow&
\frac{(\vartheta_3)^{5}+(\vartheta_4)^{5}}{2\eta^{4}}
B^{(E_8)},
\eeqa
and also for the $SO(32)$ theory as
\beqa
\Theta_{0,2}=\frac{\vartheta_3+\vartheta_4}{2}
&\rightarrow&
\frac{(\vartheta_3)^{13}-(\vartheta_4)^{13}}{2\eta^{12}}
,\\
\Theta_{2,2}=\frac{\vartheta_3-\vartheta_4}{2}
&\rightarrow&
\frac{(\vartheta_3)^{13}+(\vartheta_4)^{13}}{2\eta^{12}}.
\eeqa
The $h=\frac12$ states come from the lowest term of 
\beqa
\frac{\vartheta_3+\vartheta_4}{2}(\tau,0)
\frac{\vartheta_3+\vartheta_4}{2}(\tau,z)
+
\frac{\vartheta_3-\vartheta_4}{2}(\tau,0)
\frac{\vartheta_3-\vartheta_4}{2}(\tau,z)
&=&1+\cdots,
\eeqa
which is converted to (besides the eta functions)
\beqa
\frac{\vartheta_3^5-\vartheta_4^5}{2}(\tau,0)
\frac{\vartheta_3+\vartheta_4}{2}(\tau,z)
+
\frac{\vartheta_3^5+\vartheta_4^5}{2}(\tau,0)
\frac{\vartheta_3-\vartheta_4}{2}(\tau,z)
&=&10q^{\frac12}+q^{\frac12}(y+y^{-1})+\cdots
\label{E8xE8level2Theta}
\nonumber\\
\eeqa
for the $E_8\times E_8$ case ($B^{(E_8)}=1+\cdots$),
and 
\beqa
\frac{\vartheta_3^{13}-\vartheta_4^{13}}{2}(\tau,0)
\frac{\vartheta_3+\vartheta_4}{2}(\tau,z)
+
\frac{\vartheta_3^{13}+\vartheta_4^{13}}{2}(\tau,0)
\frac{\vartheta_3-\vartheta_4}{2}(\tau,z)
&=&26q^{\frac12}+q^{\frac12}(y+y^{-1})+\cdots
\nonumber\\
\label{SO(32)level2Theta}
\eeqa
for the $SO(32)$ case. 

As we have shown in (\ref{1/2}), 
the first term on the 
right hand side of (\ref{hatFNSjj+2}) starts from
$q^{\frac12} y^{\frac{-j+\km+2}{\km +4}}$. This is in the 
$n_{cluster}=0$ cluster. Either a transverse or a Liouville 
fermion may be excited.
Therefore, (\ref{E8xE8level2Theta}) 
shows that there are $(10+1)$ $h=1$ states in the $E_8\times E_8$ 
theory; the latter singlet comes from $q^{\frac12}y$ which 
has $F^{(\rm NS)}=+1$, while $q^{\frac12}y^{-1}$ does not corresponds 
to a discrete state because  $F^{(\rm NS)}=-1$ is not allowed in the 
$n_{cluster}=0$ cluster. Similarly, we can see from (\ref{SO(32)level2Theta}) that 
there are $(26+1)$ $h=1$ states in the $SO(32)$ 
theory.

On the other hand, the second term of (\ref{hatFNSjj+2}) has also an
expansion $q^{\frac12} y^{\frac{-j-2}{\km +4}}+\cdots$. This is in the $n_{cluster}=-1$
cluster, and therefore it did not produce any massless states in typeII theories.
However, with a Liouville fermion excitation, it is allowed in heterotic theories.
This gives another $h=1$ state for both the $E_8\times E_8$ and $SO(32)$ theories.

Next consider the R-sector terms of $\hat F^{(+)}_{j,j+2}(\tau,z)$:
\beqa&&
\hat F^{(+)\rm R}_{j,j+2}(\tau,z)
\cdot q^{\frac14-\frac{\left(
-j+\km+2-\frac{\km +4}2
\right)^2}{2(\km+2)(\km+4)}}
\nonumber\\
&=&q^{\frac14-\frac{\left(
-j+\km+2-\frac{\km +4}2
\right)^2}{2(\km+2)(\km+4)}}
\left(
	\chi^{j,1}_{j+1}(\tau,0)
	\Theta_{-2j+\km,2(\km+2)(\km +4)}
	\left(
	\tau,\frac z{\km +4}
	\right)+\cdots
\right)\nonumber\\
&&~~~~~~~~~~~~~~~~~~~~~~~~~~~~~\cdot
\left(\Theta_{1,2}(\tau,0)\Theta_{1,2}(\tau,z)+
\Theta_{-1,2}(\tau,0)\Theta_{-1,2}(\tau,z)\right)
\nonumber\\
&=&
(q^\frac14 y^{\frac{-j+\frac\km 2}{\km +4}} +\cdots)
\left(\Theta_{1,2}(\tau,0)\Theta_{1,2}(\tau,z)+
\Theta_{-1,2}(\tau,0)\Theta_{-1,2}(\tau,z)\right).
\label{hatFRjj+2}
\eeqa
This is in the $n_{cluster}=-\frac12$ cluster, and so 
the $y^{+\frac12}$ terms survive. 
Through the heterotic conversion,
\beqa
\frac{\vartheta_2+\tilde\vartheta_1}{2}(\tau,0)
\frac{\vartheta_2+\tilde\vartheta_1}{2}(\tau,z)
+
\frac{\vartheta_2-\tilde\vartheta_1}{2}(\tau,0)
\frac{\vartheta_2-\tilde\vartheta_1}{2}(\tau,z)
&=&q^{\frac14}(y^{\frac12}+y^{-\frac12})+\cdots\\
\left(=~\Theta_{1,2}(\tau,0)\Theta_{1,2}(\tau,z)+
\Theta_{-1,2}(\tau,0)\Theta_{-1,2}(\tau,z)\right)&&\nonumber
\eeqa
is replaced with  
\beqa
\frac{\vartheta_2^5+\tilde\vartheta_1^5}{2}(\tau,0)
\frac{\vartheta_2+\tilde\vartheta_1}{2}(\tau,z)
+
\frac{\vartheta_2^5-\tilde\vartheta_1^5}{2}(\tau,0)
\frac{\vartheta_2-\tilde\vartheta_1}{2}(\tau,z)
&=&16q^{\frac34}(y^{\frac12}+y^{-\frac12})+\cdots
\nonumber\\
\label{E8xE8level2Theta'}
\eeqa
in the $E_8 \times E_8$ case, or
\beqa
\frac{\vartheta_2^{13}+\tilde\vartheta_1^{13}}{2}(\tau,0)
\frac{\vartheta_2+\tilde\vartheta_1}{2}(\tau,z)
+
\frac{\vartheta_2^{13}-\tilde\vartheta_1^{13}}{2}(\tau,0)
\frac{\vartheta_2-\tilde\vartheta_1}{2}(\tau,z)
&=&2^{12}q^{\frac74}(y^{\frac12}+y^{-\frac12})+\cdots
\nonumber\\
\label{SO(32)level2Theta'}
\eeqa
 in the $SO(32)$ case. (\ref{E8xE8level2Theta'}) shows that 
 there are sixteen $h=1$ states in the Ramond sector of the 
 $E_8 \times E_8$ theory,
 while (\ref{SO(32)level2Theta'}) implies no $h=1$ 
 states in the Ramond sector of the 
 $SO(32)$ theory.

Summarizing the $h=1$ states in $\hat F^{het}_{j,j+2}$ $(j=0,\ldots,\frac{\km}2-1)$,
$\hat F^{E_8\times E_8}_{j,j+2}$ has 
\beqa
{\bf 10}\oplus {\bf 1} \oplus  {\bf 1} ~(\mbox{NS sector}),&&
{\bf 16}~(\mbox{R sector})
\nonumber
\eeqa
of $SO(10)$, while $\hat F^{SO(32)}_{j,j+2}$ has
\beqa
{\bf 26} \oplus {\bf 1} \oplus  {\bf 1}~(\mbox{NS sector}),&&
\mbox{no states}~(\mbox{R sector})
\nonumber
\eeqa
of $SO(26)$. Taking into account the right moving part and also the 
symmetry $\hat F^{het}_{j,j+2}=\hat F^{het}_{\km-j,\km-j+2}$,
they become $D=4$, $N=1$ chiral supermultiplets.

\underline{\it $h=1$ states in
$\hat F^{het}_{0,0}$ with odd $\km$}

Just like the type II case, there are also $h=1$ states that contribute to 
$\hat F^{het}_{0,0}$. Before the conversion, the NS-sector terms are
\beqa&&
\hat F^{(-)\rm NS}_{0,0}(\tau,z)
\cdot q^{\frac14-\frac{\km +4}{8(\km+2)}}
\nonumber\\
&=&q^{\frac14-\frac{\km +4}{8(\km+2)}}
\left(
	\chi^{0,0}_0(\tau,0)
	\Theta_{0,2(\km+2)(\km +4)}
	\left(
	\tau,\frac z{\km +4}
	\right)
\right.\nonumber\\
&&~~~~~~~~~~~~~~~
\left.
	+\chi^{0,-2}_{-2}(\tau,0)
	\Theta_{+2(\km+4),2(\km+2)(\km +4)}
	\left(
	\tau,\frac z{\km +4}
	\right)+\cdots
\right)\nonumber\\
&&~~~~~~~~~~~~~~~~~~~~~~~~~~~~~\cdot
\left(\Theta_{0,2}(\tau,0)\Theta_{2,2}(\tau,z)+
\Theta_{2,2}(\tau,0)\Theta_{0,2}(\tau,z)\right).
\label{hatFNS00}
\eeqa
$\hat F^{(+)\rm NS}_{0,0}(\tau,z)$ gives rise to no $h=1$ states 
and hence is not written here. We again included the Liouville energy and 
the imaginary momentum factor. As we saw in the type II analysis, this cancels
the $q^{-\frac\cm{24}}$ of the $N=2$ characters, giving
\beqa
&=&(q^0 y^0 +\cdots + q^1 y^1 +\cdots)
\left(
\frac{\vartheta_3+\vartheta_4}{2}(\tau,0)
\frac{\vartheta_3-\vartheta_4}{2}(\tau,z)
+
\frac{\vartheta_3-\vartheta_4}{2}(\tau,0)
\frac{\vartheta_3+\vartheta_4}{2}(\tau,z)
\right).
\nonumber\\
\label{beforeconversion}
\eeqa
In the $E_8\times E_8$ case, this is converted to 
\beqa
&\rightarrow&(q^0 y^0 +\cdots + q^1 y^1 +\cdots)
\left(
\frac{\vartheta^5_3-\vartheta^5_4}{2}(\tau,0)
\frac{\vartheta_3-\vartheta_4}{2}(\tau,z)
+
\frac{\vartheta^5_3+\vartheta^5_4}{2}(\tau,0)
\frac{\vartheta_3+\vartheta_4}{2}(\tau,z)
\right).
\nonumber\\
&=&(q^0 y^0 +\cdots + q^1 y^1 +\cdots)
\left(
10 q^1(y^{+1}+y^{-1})+\cdots
+1+\cdots
\right).
\eeqa
The $q^0 y^0$ term corresponds to a state in the $n_{cluster}=-1$ cluster.
Therefore, only  terms with $F^{(NS)}\geq 1$ (that is, those containing $y^{+1}$
as a factor in the second parenthesis) are relevant for the discrete 
spectrum, {\em as long as $L^{SL(2,\R)}_0=0$}. This gives ${\bf 10}$.
While states at $L^{SL(2,\R)}_0=0$ in the module 
${\cal H}^{SL(2,{\bf R})}_{\pm,(0,-\frac\kappa 2)}$ have $J^3_0$ charges
\beqa
J^3_0&=&-\frac\kappa 2,~-\frac\kappa 2 -1,~-\frac\kappa 2 -2,\ldots,
\eeqa
those at $L^{SL(2,\R)}_0=1$ have
\beqa
J^3_0&=&-\frac\kappa 2 +1,~~-\frac\kappa 2,~-\frac\kappa 2 -1,\ldots.
\eeqa
Therefore, for states at $L^{SL(2,\R)}_0=1$, the condition $F^{(NS)}\geq -n_{cluster}$ 
is relaxed to $F^{(NS)}\geq -n_{cluster}-1=0$. In this case, $q^0 y^0$ can also 
be paired with $``1"$, with a total conformal weight $h=1$ due to $L^{SL(2,\R)}_0=1$.
This is a singlet.

The $q^1 y^1$ term is in the $n_{cluster}=0$ cluster. This can be paired with $``1"$ 
and gives rise to another singlet of $SO(10)$.

In the $SO(32)$ case, (\ref{beforeconversion}) becomes 
\beqa
&\rightarrow&(q^0 y^0 +\cdots + q^1 y^1 +\cdots)
\left(
\frac{\vartheta^{13}_3-\vartheta^{13}_4}{2}(\tau,0)
\frac{\vartheta_3-\vartheta_4}{2}(\tau,z)
+
\frac{\vartheta^{13}_3+\vartheta^{13}_4}{2}(\tau,0)
\frac{\vartheta_3+\vartheta_4}{2}(\tau,z)
\right).
\nonumber\\
&=&(q^0 y^0 +\cdots + q^1 y^1 +\cdots)
\left(
26 q^1(y^{+1}+y^{-1})+\cdots
+1+\cdots
\right).
\eeqa
A similar analysis shows that there are one ${\bf 26}$
and two ${\bf 1}$ of $SO(26)$.

Finally, we consider the R-sector terms of $\hat F^{het}_{0,0}$:
\beqa&&
\hat F^{(-)\rm R}_{0,0}(\tau,z)
\cdot q^{\frac14-\frac{\km +4}{8(\km+2)}}
\nonumber\\
&=&q^{\frac14-\frac{\km +4}{8(\km+2)}}
\left(
	\chi^{0,-1}_{-1}(\tau,0)
	\Theta_{\km+4,2(\km+2)(\km +4)}
	\left(
	\tau,\frac z{\km +4}
	\right)+\cdots
\right)\nonumber\\
&&~~~~~~~~~~~~~~~~~~~~~~~~~~~~~\cdot
\left(\Theta_{1,2}(\tau,0)\Theta_{-1,2}(\tau,z)+
\Theta_{-1,2}(\tau,0)\Theta_{1,2}(\tau,z)\right).
\nonumber\\
&=&(q^{\frac14}y^{\frac12}+\cdots)
\left(
q^\frac14(y^{+\frac12}+y^{-\frac12})+\cdots
\right).
\label{hatFR00}
\eeqa
In the $E_8\times E_8$ case,
this is converted to
\beqa
&\rightarrow&(q^{\frac14}y^{\frac12}+\cdots)
\left(
16q^\frac34(y^{+\frac12}+y^{-\frac12})+\cdots
\right).
\eeqa
Again, $q^{\frac14}y^{\frac12}$ is in the $n_{cluster}=-\frac12$
cluster and hence chooses only the first 16. In the $SO(32)$ case, 
\beqa
&\rightarrow&(q^{\frac14}y^{\frac12}+\cdots)
\left(
2^{12}q^\frac74(y^{+\frac12}+y^{-\frac12})+\cdots
\right),
\eeqa
and therefore no $h=1$ states.

In all, 
$\hat F^{het}_{0,0}$ has exactly the same set of $h=1$ 
states as an $\hat F^{het}_{j,j+2}$ does.

\underline{\it Massless spectrum for even $\km$}

For the $A_{\km+1}$ modular invariant model with $\km$ odd, 
we have seen that there are $\frac{\km+3}2$ 
sets of massless $N=1$ chiral multiplets, 
in the ${\bf 10}\oplus{\bf 1}\oplus{\bf 1}\oplus{\bf 16}$ representation of $SO(10)$ 
for the $E_8\times E_8$ theory, and 
in the ${\bf 26}\oplus{\bf 1}\oplus{\bf 1}$ representation of $SO(26)$ 
for the $SO(32)$ theory.
For the $A_{\km+1}$ modular invariant model with $\km$ even,
we have similarly $\frac\km2 +1$ 
sets of massless $N=1$ chiral multiplets in the same representations.
In addition, there also exist non-localized ``massless" matter fields corresponding to 
the continuous series representations that reach the boundary of the 
unitary region, as is the case in the type\;II spectrum.

\section{Examples}
\subsection{Type II massless spectrum for $\km=1$}
The central charge of the $N=2$ minimal model is $\cm=1$.
The central charge of the $SL(2,{\bf R})/U(1)$ coset CFT is then 
$9-1=8$, and hence
\beqa
\kappa=\frac{16}5,~~~k=\frac 65.
\eeqa 
The string functions for $\km=1$ are simply
\beqa
c^l_m(\tau)&=&\frac{\delta^{\rm (mod 2)}_{m,l}}{\eta(\tau)},
\eeqa
where $l=0,1$. 
The $\km=1$ minimal characters are
\beqa
\chi^{l,s}_m (\tau,z)&=&\frac{\delta^{\rm (mod 2)}_{m,l+s}}{\eta(\tau)}
\Theta_{2m -3s,6}\left(
\tau, \frac z3
\right).
\eeqa

To find the massless spectrum, it is convenient to use the formulas 
for $\hat F_{l,2r}$ (\ref{}) given in Appendix.
Since 
\beqa
\hat F_{1,2r}=\hat F_{0,5-2r}, 
\label{F1F0symmetry}
\eeqa 
we only consider $l=0$. Then $r$ takes values 
in $\Z_5$. Setting $\km = 1$, we obtain
\beqa
\hat F^{(-)}_{0,2r}(\tau,z)&=&\frac1{\eta(\tau)}\Theta_{-4r,5}\left(
\tau,\frac z5
\right)
\frac12\hat\Lambda_2(\tau,z),\\
\hat F^{(+)}_{0,2r}(\tau,z)&=&\frac1{\eta(\tau)}\Theta_{-4r+5,5}\left(
\tau,\frac z5
\right)
\frac12\hat\Lambda_1(\tau,z).
\eeqa

The NS- and R-sector spectra can be considered separately by writing
\beqa
\hat\Lambda_1(\tau,z)&=&\hat\Lambda_1^{\rm NS}(\tau,z)-\hat\Lambda_1^{\rm R}(\tau,z),\\
\frac12\hat\Lambda_1^{\rm NS}(\tau,z)&\equiv&
\Theta_{1,1}(\tau,z)
\Theta_{(0,0)}(\tau;0,z),
\\
\frac12\hat\Lambda_1^{\rm R}(\tau,z)&\equiv&
\Theta_{0,1}(\tau,z)
\Theta_{(1,1)}(\tau;0,z)
\eeqa
and
\beqa
\hat\Lambda_2(\tau,z)&=&\hat\Lambda_2^{\rm NS}(\tau,z)-\hat\Lambda_2^{\rm R}(\tau,z),\\
\frac12\hat\Lambda_2^{\rm NS}(\tau,z)&\equiv&
\Theta_{0,1}(\tau,z)
\Theta_{(0,2)}(\tau;0,z),\\
\frac12\hat\Lambda_2^{\rm R}(\tau,z)&\equiv&
\Theta_{1,1}(\tau,z)
\Theta_{(1,-1)}(\tau;0,z),
\eeqa
where
\beqa
\Theta_{(s,s')}(\tau;z,z')&\equiv&
\sum_{\nu\in \Z_2}\Theta_{s+2\nu,2}(\tau,z)\Theta_{s'+2\nu,2}(\tau,z').
\eeqa
We define $\hat F_{l,2r}^{(\pm)\rm NS}$ as formulas similar to  
(\ref{hatFl2r(-)}), (\ref{hatFl2r(+)}) but with $\hat\Lambda_2(\tau,z)$,
$\hat\Lambda_1(\tau,z)$ being replaced with $\hat\Lambda_2^{\rm NS}(\tau,z)$,
$\hat\Lambda_1^{\rm NS}(\tau,z)$, respectively, and 
\beqa
\hat F_{l,2r}^{\rm NS}&\equiv&\frac12\left(
\hat F_{l,2r}^{(-)\rm NS}
+\hat F_{l,2r}^{(+)\rm NS}
\right).
\eeqa 
Among them, only $\hat F_{0,0}^{\rm NS}$, $\hat F_{0,-4}^{\rm NS}$ 
and $\hat F_{0,+2}^{\rm NS}(=\hat F_{0,-8}^{\rm NS})$ have theta functions whose $U(1)$ charges 
are in the ranges (\ref{J^U(1)_0range}). 

$\hat F_{0,0}^{\rm NS}(\tau,z)$ has an expansion
\beqa
2\hat F_{0,0}^{\rm NS}(\tau,z)&=&
(\hat F_{0,0}^{(-)\rm NS}
+\hat F_{0,0}^{(+)\rm NS})(\tau,z)\nonumber\\
&=&\frac1{\eta(\tau)}
(\underbrace{
1+q(y+y^{-1})+\cdots}_{\Theta_{0,5}\left(\tau,\frac z5 \right)
\Theta_{0,1}(\tau,z)
})
(\underbrace{
q^{\frac12}(y+y^{-1}+2)+\cdots}_{\Theta_{(0,2)}\left(\tau;0,z \right)
})
\nonumber\\
&&+\frac1{\eta(\tau)}
(\underbrace{
q(y^2+y^{-2}+2)+\cdots}_{\Theta_{5,5}\left(\tau,\frac z5 \right)
\Theta_{1,1}(\tau,z)
})
(\underbrace{
1+2q(y+y^{-1})+\cdots}_{\Theta_{(0,0)}\left(\tau;0,z \right)
}).
\label{hatF0,0NSexpansion}
\eeqa
The cluster number $n_{cluster}$ can be read off from the power of $y$ 
in the expansion of \\
$\Theta_{*,5}(\tau,\frac z5)\Theta_{*,1}(\tau,z)$; if the 
power satisfies 
\beqa
\frac12 + n<(\mbox{The power})<\frac{k+1}2 +n 
\eeqa
for some $n\in \Z$ ($\in\Z +\frac12$) for the NS (R) sector, 
then $n_{cluster}=n$.

The first line of (\ref{hatF0,0NSexpansion}) contains ``$1$" in the first parenthesis;
this is in the $n_{cluster}=-1$ cluster, for which 
$F^{(\rm NS)}\geq 1$.\footnote{To be sure, $F^{(\rm NS)}$ here is the fermion number 
in the NS sector defined in sect.?, which should not be confused with  
$\hat F_{l,2r}^{\rm NS}(\tau,z)$.} Therefore, it can be paired with $q^{\frac12}y$ 
in the second parenthesis, but not with $q^{\frac12}y^{-1}$. 
The second line has no $q^{\frac12}$ terms. Thus we found a single $h=\frac12$ 
state in $\hat F_{0,0}^{\rm NS}(\tau,z)$.

Also $\hat F_{0,2}^{\rm NS}(\tau,z)$ is expanded as
\beqa
2\hat F_{0,2}^{\rm NS}(\tau,z)\times q^{\frac15}&=&
q^{\frac15}(\hat F_{0,2}^{(-)\rm NS}
+\hat F_{0,2}^{(+)\rm NS})(\tau,z)
\nonumber\\
&=&\frac{q^{\frac15}}{\eta(\tau)}
(\underbrace{
q^{\frac45}y^{-\frac25}+\cdots~~~}_{\Theta_{-4,5}\left(\tau,\frac z5 \right)
\Theta_{0,1}(\tau,z)
})
(\underbrace{
q^{\frac12}(y+y^{-1}+2)+\cdots}_{\Theta_{(0,2)}\left(\tau;0,z \right)
})
\nonumber\\
&&+\frac{q^{\frac15}}{\eta(\tau)}
(\underbrace{
q^{\frac1{20}+\frac14}y^{\frac1{10}}(y^{\frac12}+y^{-\frac12})+\cdots}_{\Theta_{+1,5}\left(\tau,\frac z5 \right)
\Theta_{1,1}(\tau,z)
})
(\underbrace{
1+2q(y+y^{-1})+\cdots}_{\Theta_{(0,0)}\left(\tau;0,z \right)
}),
\label{hatF0,2NSexpansion}
\eeqa
where we have included 
the factor\footnote{For $\hat F^{\rm NS}_{0,0}$ such a factor is absent 
because in that case the imaginary momentum factor precisely cancels 
the extra conformal weight of the continuous series.} 
of $q^{\frac15}$ 
coming from the extra weight 
\beqa
\frac14-\frac\cm{24}~=~\frac5{24}
\label{extra_weight}
\eeqa
because of the shortage of the eta functions in the denominator of the partition function, 
minus the ``drop" coming from the imaginary momentum of a possible discrete state
\beqa
\frac1k\left(\Jtot +\frac12 +\frac k2\right)^2
~=~
\frac{\km+4}{2(\km+2)}\left(
\frac{J^{U(1)}_0}{\km+4}-\frac12 -n_{cluster}
\right)^2
~=~\frac1{120}
\eeqa			
 ($\frac5{24}-\frac1{120}=\frac15$).
In this case the term  $q^{\frac15+\frac1{20}+\frac14}y^{\frac1{10}+\frac12}
=q^{\frac12}y^{\frac35}$ in the last line indicates a massless discrete 
state in the $n_{cluster}=0$ cluster, while the other  
$q^{\frac12}y^{\frac1{10}-\frac12}$ 
is in the $n_{cluster}=-1$ cluster, for which $F^{(\rm NS)}$ must be $\geq 1$, and does not 
corresponds to any discrete states.

We can similarly expand $q^{\frac 4{30}}\hat F_{0,-4}^{\rm NS}(\tau,z)$ (where $\frac4{30}$ 
is again $\frac 5{24}$ minus the imaginary momentum contribution $\frac3{40}$),  
but can find no $h=\frac12$ states.
 
Therefore, for the $A_2$ modular invariant (in which $\km=1$),  we find two massless states 
in the NS-NS sector, one from 
$\hat F_{0,0}^{\rm NS}(\hat F_{0,0}^{\rm NS})^*$ and the other from 
$\hat F_{0,2}^{\rm NS}(\hat F_{0,2}^{\rm NS})^*$. Due to (\ref{F1F0symmetry}), 
$\hat F_{1,5}^{\rm NS}(\hat F_{1,5}^{\rm NS})^*$ and 
$\hat F_{1,7}^{\rm NS}(\hat F_{1,7}^{\rm NS})^*$ are also in the summation (\ref{J0tot_dependence}).

We study the Ramond sector in a similar way, and find as many $h=\frac12$
states as in the NS sector due to the supersymmetry. 
All in all, they are two vector multiplets for the typeIIA case and 
two hypermultiplets for the typeIIB case, as we discussed in sect.\ref{oddkm}.

\subsection{$E_8\times E_8$ heterotic massless spectrum for $\km=1$}
Next we turn to the $E_8\times E_8$ heterotic string compactification. Again, we
set $\km=1$.
As we did for $\hat\Lambda_1$ and $\hat\Lambda_2$ in the last subsection, 
we write
\beqa
\hat \Lambda^{E_8\times E_8}_1(\tau,z)
&=&\hat \Lambda^{E_8\times E_8,\rm NS}_1(\tau,z)
+\hat \Lambda^{E_8\times E_8,R}_1(\tau,z),\\
\frac{\frac12\hat \Lambda^{E_8\times E_8,\rm NS}_1(\tau,z)}
{\eta^{14}(\tau)}
&\equiv&
\Theta_{1,1}(\tau,z)\left(
B^{(10)}_v(\tau,0)B^{(2)}_0(\tau,z)
+B^{(10)}_0(\tau,0)B^{(2)}_v(\tau,z)
\right)B^{(E_8)}(\tau,0),
\nonumber
\\
\\
\frac{\frac12\hat \Lambda^{E_8\times E_8,R}_1(\tau,z)}
{\eta^{14}(\tau)}
&\equiv&
\Theta_{0,1}(\tau,z)\left(
B^{(10)}_s(\tau,0)B^{(2)}_s(\tau,z)
+B^{(10)}_{\bar s}(\tau,0)B^{(2)}_{\bar s}(\tau,z)
\right)
B^{(E_8)}(\tau,0),
\nonumber\\
\eeqa
and
\beqa
\hat \Lambda^{E_8\times E_8}_2(\tau,z)
&=&\hat \Lambda^{E_8\times E_8,\rm NS}_2(\tau,z)
+\hat \Lambda^{E_8\times E_8,R}_2(\tau,z),\\
\frac{\frac12\hat \Lambda^{E_8\times E_8,\rm NS}_2(\tau,z)}
{\eta^{14}(\tau)}
&\equiv&
\Theta_{0,1}(\tau,z)\left(
B^{(10)}_0(\tau,0)B^{(2)}_0(\tau,z)
+B^{(10)}_v(\tau,0)B^{(2)}_v(\tau,z)
\right)B^{(E_8)}(\tau,0),
\nonumber\\
\\
\frac{\frac12\hat \Lambda^{E_8\times E_8,R}_2(\tau,z)}
{\eta^{14}(\tau)}
&\equiv&
\Theta_{1,1}(\tau,z)\left(
B^{(10)}_s(\tau,0)B^{(2)}_s(\tau,z)
+B^{(10)}_{\bar s}(\tau,0)B^{(2)}_{\bar s}(\tau,z)
\right)
B^{(E_8)}(\tau,0).
\nonumber\\
\eeqa
Then 
\beqa
\hat F_{l,2r}^{E_8\times E_8}(\tau,z)
&=&
\left(\hat F_{l,2r}^{E_8\times E_8,\rm NS}
+\hat F_{l,2r}^{E_8\times E_8,R}\right)(\tau,z).
\eeqa

Let us first consider $2\hat F_{0,0}^{E_8\times E_8}$:
\beqa
2\hat F_{0,0}^{E_8\times E_8,\rm NS}(\tau,z)
&=&
\left(\hat F_{0,0}^{E_8\times E_8,(-)\rm NS}
+\hat F_{0,0}^{E_8\times E_8,(+)\rm NS}\right)(\tau,z),\nonumber\\
&=&
\frac1{\eta(\tau)}
\left(
\Theta_{0,5}\left(
\tau,\frac z5
\right)
\frac12\hat \Lambda^{E_8\times E_8,\rm NS}_2(\tau,z)
+
\Theta_{5,5}\left(
\tau,\frac z5
\right)
\frac12\hat \Lambda^{E_8\times E_8,\rm NS}_1(\tau,z)
\right)
\nonumber\\
&=&\frac1{\eta(\tau)}
(\underbrace{
1+q(y+y^{-1})+\cdots}_{\Theta_{0,5}\left(\tau,\frac z5 \right)\Theta_{0,1}(\tau,z)}
)
(\underbrace{
1+q(10y+10y^{-1}+40)+\cdots}_{
\mbox{\scriptsize Fermion theta fns.~of } \frac12\hat \Lambda^{E_8\times E_8,\rm NS}_2(\tau,z)
}
)
\nonumber\\
&&+\frac1{\eta(\tau)}
(\underbrace{
q(y^2+y^{-2}+2)+\cdots
}_{\Theta_{5,5}\left(\tau,\frac z5 \right)
\Theta_{1,1}(\tau,z)
})
(\underbrace{
q^{\frac12}(y+y^{-1}+10)+\cdots
}_{
\mbox{\scriptsize Fermion theta fns.~of } \frac12\hat \Lambda^{E_8\times E_8,\rm NS}_1(\tau,z)
}),
\label{2hatF_00E8xE8NS}
\eeqa
\beqa
2\hat F_{0,0}^{E_8\times E_8,R}(\tau,z)
&=&
\left(\hat F_{0,0}^{E_8\times E_8,(-)R}
+\hat F_{0,0}^{E_8\times E_8,(+)R}\right)(\tau,z),\nonumber\\
&=&
\frac1{\eta(\tau)}
\left(
\Theta_{0,5}\left(
\tau,\frac z5
\right)
\frac12\hat \Lambda^{E_8\times E_8,R}_2(\tau,z)
+
\Theta_{5,5}\left(
\tau,\frac z5
\right)
\frac12\hat \Lambda^{E_8\times E_8,R}_1(\tau,z)
\right)
\nonumber\\
&=&
\frac1{\eta(\tau)}
(\underbrace{
q^{\frac14}(y^{\frac12}+y^{-\frac12})+\cdots}_{\Theta_{0,5}\left(\tau,\frac z5 \right)\Theta_{1,1}(\tau,z)}
)
(\underbrace{
16q^{\frac34}(y^{\frac12}+y^{-\frac12})+\cdots}_{
\mbox{\scriptsize Fermion theta fns.~of } \frac12\hat \Lambda^{E_8\times E_8,R}_2(\tau,z)
}
)
\nonumber\\
&&+
\frac1{\eta(\tau)}
(\underbrace{
q^{\frac54}(y^{\frac12}+y^{-\frac12})+\cdots}_{\Theta_{5,5}\left(\tau,\frac z5 \right)\Theta_{0,1}(\tau,z)}
)
(\underbrace{
16q^{\frac34}(y^{\frac12}+y^{-\frac12})+\cdots}_{
\mbox{\scriptsize Fermion theta fns.~of } \frac12\hat \Lambda^{E_8\times E_8,R}_1(\tau,z)
}
).
\label{2hatF_00E8xE8R}
\eeqa
In this case we look for $h=1$ states. Again, we can know which cluster the 
spectrum belongs to by the power of $y$ in the expansion of 
$\Theta_{*,5}(\tau, \frac z5)\Theta_{*,1}(\tau,z)$ because they arose from 
the composition of the $U(1)$ theta function (and the minimal $N=2$ theta 
function which has no $\Jtot$ charge).

In the first line of (\ref{2hatF_00E8xE8NS}), the term ``$1$" in the first parenthesis 
belongs to the $n_{cluster}=-1$ cluster. Therefore, with a lowest $L^{SU(2,\R)}_0(=0)$ state, 
$F^{(\rm NS)}$ must be $\geq 1$ and it can 
be paired with $10 qy$, but not with $10 q y^{-1}$.  This gives a single ${\bf 10}$ 
representation of $SO(10)$.  As we noted in section \ref{separation_heterotic}, 
if $J^+_{-1}|0,-\frac\kappa2\rangle$ (rather than $|0,-\frac\kappa2\rangle$) 
is chosen as the state in the $SL(2,\R)$ module 
${\cal H}_{-,(0,-\frac\kappa 2)}^{SL(2,\R)}$, then $F^{(\rm NS)}$ is relaxed to 
$\geq 0$ and the two ``$1$"s can be paired. This gives a singlet.

The ``$qy$" term is in the $n_{cluster}=0$ 
cluster and can be paired with $1$ in the second parenthesis. This is another singlet. 
On the other hand,  the ``$qy^{-1}$" term is in the $n_{cluster}=-2$ cluster and 
$F^{(\rm NS)}$ must be $\geq 2$. Therefore, it does not give rise to any $h=1$
states. 
Also, no $h=1$ states arise from the second line of (\ref{2hatF_00E8xE8NS}).
A similar analysis can be made for the Ramond sector (\ref{2hatF_00E8xE8R}). 
This confirms sixteen $h=1$ states from the first line. In all, we find a set of 
$10+1+1+16=28$, $h=1$ states in  $2\hat F_{0,0}^{E_8\times E_8}$.

Next we consider $2\hat F_{0,2}^{E_8\times E_8}$:
\beqa
2\hat F_{0,2}^{E_8\times E_8,\rm NS}(\tau,z)
\times q^{\frac15}
&=&
\left(\hat F_{0,2}^{E_8\times E_8,(-)\rm NS}
+\hat F_{0,2}^{E_8\times E_8,(+)\rm NS}\right)(\tau,z)
\times q^{\frac15}
\nonumber\\
&=&
\frac{q^{\frac15}}{\eta(\tau)}
\left(
\Theta_{-4,5}\left(
\tau,\frac z5
\right)
\frac12\hat \Lambda^{E_8\times E_8,\rm NS}_2(\tau,z)
+
\Theta_{1,5}\left(
\tau,\frac z5
\right)
\frac12\hat \Lambda^{E_8\times E_8,\rm NS}_1(\tau,z)
\right)
\nonumber\\
&=&\frac{q^{\frac15}}{\eta(\tau)}
(\underbrace{q^{\frac45}y^{-\frac25}
+\cdots}_{\Theta_{-4,5}\left(\tau,\frac z5 \right)\Theta_{0,1}(\tau,z)}
)
(\underbrace{
1+q(10y+10y^{-1}+40)+\cdots}_{
\mbox{\scriptsize Fermion theta fns.~of } \frac12\hat \Lambda^{E_8\times E_8,\rm NS}_2(\tau,z)
}
)
\nonumber\\
&&+\frac{q^{\frac15}}{\eta(\tau)}
(\underbrace{
q^{\frac3{10}}(y^{\frac35}+y^{-\frac25})+\cdots
}_{\Theta_{1,5}\left(\tau,\frac z5 \right)
\Theta_{1,1}(\tau,z)
})
(\underbrace{
q^{\frac12}(y+y^{-1}+10)+\cdots
}_{
\mbox{\scriptsize Fermion theta fns.~of } \frac12\hat \Lambda^{E_8\times E_8,\rm NS}_1(\tau,z)
}),
\label{2hatF_02E8xE8NS}
\eeqa
\beqa
2\hat F_{0,2}^{E_8\times E_8,R}(\tau,z)
\times q^{\frac15}
&=&
\left(\hat F_{0,2}^{E_8\times E_8,(-)R}
+\hat F_{0,2}^{E_8\times E_8,(+)R}\right)(\tau,z)
\times q^{\frac15}
\nonumber\\
&=&
\frac{q^{\frac15}}{\eta(\tau)}
\left(
\Theta_{0,5}\left(
\tau,\frac z5
\right)
\frac12\hat \Lambda^{E_8\times E_8,R}_2(\tau,z)
+
\Theta_{5,5}\left(
\tau,\frac z5
\right)
\frac12\hat \Lambda^{E_8\times E_8,R}_1(\tau,z)
\right)
\nonumber\\
&=&
\frac{q^{\frac15}}{\eta(\tau)}
(\underbrace{q^{\frac{21}{20}}(y^{\frac1{10}}+y^{-\frac9{10}})
+\cdots}_{\Theta_{-4,5}\left(\tau,\frac z5 \right)\Theta_{1,1}(\tau,z)}
)
(\underbrace{
16q^{\frac34}(y^{\frac12}+y^{-\frac12})+\cdots}_{
\mbox{\scriptsize Fermion theta fns.~of } \frac12\hat \Lambda^{E_8\times E_8,R}_2(\tau,z)
}
)
\nonumber\\
&&+
\frac{q^{\frac15}}{\eta(\tau)}
(\underbrace{q^{\frac1{20}} y^{\frac1{10}}
+\cdots}_{\Theta_{1,5}\left(\tau,\frac z5 \right)\Theta_{0,1}(\tau,z)}
)
(\underbrace{
16q^{\frac34}(y^{\frac12}+y^{-\frac12})+\cdots}_{
\mbox{\scriptsize Fermion theta fns.~of } \frac12\hat \Lambda^{E_8\times E_8,R}_1(\tau,z)
}
).
\label{2hatF_02E8xE8R}
\eeqa
The first line of the NS-sector expansion (\ref{2hatF_02E8xE8NS}) 
contains one $q^1$ term ($=q^{\frac15+\frac45}y^{-\frac25}$), but it 
comes from the $n_{cluster}=-1$ cluster for which $F^{\rm NS}\geq 1$ 
(that is, ``$1$" in the second parenthesis cannot be paired), 
and hence does not give rise to a discrete state. 
The second line has terms proportional to $q^{\frac15+\frac3{10}+\frac12}=q^1$:
\beqa
q\left(
y^{\frac35}+y^{-\frac25}
\right)
\left(
10+y+y^{-1}
\right).
\eeqa
$y^{\frac35}$ is in the $n_{cluster}=0$ cluster, while  $y^{-\frac25}$ 
the $n_{cluster}=-1$ cluster. Therefore, due to the constraint, only $q
y^{\frac35}
\cdot 10
$,
$q
y^{\frac35}
\cdot y
$
and 
$q
y^{-\frac25}
\cdot y
$
correspond to discrete states.
The first is in the ${\bf 10}$ representation, while the latter two are singlets.
The R-sector expansion (\ref{2hatF_02E8xE8R}) similarly gives rise to a 
${\bf 16}$ of $SO(10)$ from the second line of (\ref{2hatF_02E8xE8R}).

Since the left- and right-moving $\hat F_{l,2r}$'s with the same $r$ are paired in 
(\ref{J0tot_dependence}), and since we have seen that $(\hat F_{0,-4})^*$ has no 
$h=\frac 12$ states, we do not need to consider $\hat F^{E_8\times E_8}_{0,-4}$.

To summarize the $\km =1$ $E_8\times E_8$ heterotic massless spectrum, 
we have found two sets of $N=1$ scalar multiplets in 
${\bf 10}\oplus{\bf 1}\oplus{\bf 1}\oplus{\bf 16}$
of $SO(10)$.

\subsection{The three generation model}
Let us consider the $\km=3$,  $A_4$ modular invariant model of the 
$E_8\times E_8$ heterotic string theory.
According to the rule we have found in section \ref{separation_heterotic},
there appear three generations of massless matter multiplets in 
${\bf 10}\oplus{\bf 1}\oplus{\bf 1}\oplus{\bf 16}$
of $SO(10)$, or 
${\bf 27}\oplus{\bf 1}$ of 
$E_6$. They are localized on a four-dimensional spacetime.
It is interesting to note that these three generations are {\em not} 
on an equal footing; for example, 
one generation has the {\bf 10} representation 
from the $n_{cluster}=-1$ cluster,
whereas in the other two generations it comes from the $n_{cluster}=0$ cluster. 
In a more realistic phenomenological application,
this fact might be used as the origin of the differences 
among the generations observed in Nature.

There are no localized gauge fields. Gauge fields correspond to 
the continuous series representations of $SL(2,\R)$ and acquire a mass 
from the Liouville energy. They propagate into the bulk, as is the case for the 
graviton. This situation is analogous to the local GUT in the standard orbifold 
compactification, where the matter fields in the twisted sector constitute
locally a representation of a possibly larger group than the actual unbroken 
gauge symmetry.

\section{Localized modes in six dimensions}

In this section we generalize the analysis in the previous sections to 
six dimensions. The partition function for the internal Calabi-Yau 
is the same except the difference of the level $k$, and it has been shown 
\cite{ES} 
that the discrete spectrum of it for the ALE spaces correctly reflects their 
topological data.

 As we did 
in four dimensions, we couple the internal part to a free superconformal field 
theory describing the six-dimensional flat Minkowski space, perform a GSO 
projection in a suitable way before the continuous and discrete 
representations are separated. This is good because, as we mentioned in 
Introduction, the couplings between the discrete states and the CFT for the 
Minkowski space are automatically consistent with the modular invariance.
A state corresponding to a discrete series representation is always 
associated with some continuous spectrum of states. They arise from 
the same integral with different contours. Therefore, the couplings between 
the discrete states and the Minkowski CFT are not arbitrary but constrained 
by modular invariance of the continuous sector.

In the Calabi-Yau two-fold case, the relation between the levels of the 
$SL(2,{\bf R})$ WZW and the $N=2$ minimal modes is
\beqa
\frac{3\kappa}{\kappa -2} +
\frac{3\km}{\km +2}&=&6,
\eeqa
and hence
\beqa
\kappa-2&=&\km +2 \nonumber\\
&\equiv& k.
\eeqa
Unlike in the three-fold case, $k$ is always an integer for a non-negative 
integer $\km$.

We again consider the internal CFT partition function 
\beqa
Z^{(\nu)}_{CY}(\tau)&=&
\int_0^1 ds_1 \int_0^1 ds_2
\frac{|\Theta_{\nu,2}(\tau,s_1\tau-s_2)|^2}{|\vartheta_1(\tau,s_1\tau-s_2)|^2}
\nonumber\\
&&\cdot
\sqrt{\frac{\tau_2}k}
\sum_{m,\tilde m}
e^{-k \pi \tau_2 s_1^2}
q^{\frac{m^2}k}e^{-2\pi i m (s_1\tau -s_2)}
\bar{q}^{\frac{{\tilde m}^2}k}e^{+2\pi i {\tilde m} (s_1\bar\tau -s_2)}
\eeqa
for $\nu\in\Z_4$,
where $m=\frac{n-kw}2$, ${\tilde m}=-\frac{n+kw}2$ and $n,w\in {\bf Z}$.
This is the same as (\ref{Z_superSL2RoverU1}) with a slight change of notation,
and the Poisson resummation 
(\ref{lattice_decomposition}) has already been done.
Also, we have omitted the overall constant $C$ as before.
We set 
\beqa
m&\equiv&kj +\frac r2,\nonumber\\
\tilde m&\equiv&k\tilde j +\frac{\tilde r}2.
\eeqa
Since both $n$ and $w$ are integers, 
$j$ and $\tilde j$ run independently over $\Z$, whereas $r$ and $\tilde r$ 
take values in $\Z_{2k}$ with a constraint $r+\tilde r=0$ mod $2k$. 
Using this change of variables, $Z^{(\nu)}_{CY}(\tau)$  can be put in the form
\beqa
Z^{(\nu)}_{CY}(\tau)&=&
\int_0^1 ds_1 \int_0^1 ds_2
\frac{|\Theta_{\nu,2}(\tau,s_1\tau-s_2)|^2}{|\vartheta_1(\tau,s_1\tau-s_2)|^2}
\nonumber\\
&&\cdot
\sqrt{\frac{\tau_2}k}
e^{-k \pi \tau_2 s_1^2}
\sum_{r,\tilde r}
\Theta_{r,k}(\tau,s_1\tau -s_2)
(\Theta_{\tilde r,k}(\tau,s_1\tau -s_2))^* .
\label{Z^(nu)}
\eeqa
Since $k=\km +2$, these theta functions are the ones 
appearing in $F_l(\tau,z)$ (\ref{F_l}),
which was introduced in \cite{ES} to construct modular invariants for the 
ADE singularities.
Therefore, generalizing the result in the previous sections, 
we can easily arrive at the supersymmetric 
modular invariant partition function including the discrete series 
contributions
\beqa
Z_{{\cal M}_6\times CY(X_n)}(\tau)&=&
\int_0^1 ds_1
\int_0^1 ds_2
\sqrt{\frac{\tau_2}k}(q\bar q)^{\frac{ks_1^2}4}
\sum_{l,\tilde l}N_{l\tilde l}
\frac{\hat F_l (\tau, s_1\tau -s_2)(\hat F_{\tilde l} (\tau, s_1\tau -s_2))^*}
{\left|\eta^2(\tau)
\vartheta_1(\tau,s_1\tau -s_2)\right|^2},
\nonumber\\
\eeqa
where we have introduced a new set of functions  
\beqa
\hat F_l(\tau,z)&\equiv&
\frac12
\chi^{(\km)}_l (\tau,0)
\left(
\vartheta_3^4-\vartheta_4^4-\vartheta_2^4+\tilde\vartheta_1^4
\right)(\tau,z)
\nonumber\\
&=&
\sum_{\nu \in \Z_4} (-1)^\nu
\sum_{m\in\Z_{2(\km +2)}}
\chi^{l,\nu}_m(\tau,0)
\sum_{
\mbox{\tiny $
							\begin{array}{c}
							\nu_0,\nu_1,\nu_2\in \Z_2\\
							\nu_0+\nu_1+\nu_2 \\
							\equiv 1({\rm mod}2)
							\end{array}$}
							}
							\Theta_{2\nu_0 +\nu,2}(\tau,0)
							\Theta_{2\nu_1 +\nu,2}(\tau,0)
							\nonumber\\
							&&~~~~~~~~~~~~~~~~~~~~~~
							~~~~\cdot
							\Theta_{2\nu_2 +\nu,2}(\tau,z)
							\Theta_{m,\km+2}\left(\tau,z\right)
							\label{hatF_l}
\eeqa
for $l=0,\ldots,\km$.
The $z$-dependences are so chosen that they match 
those of (\ref{Z^(nu)}). 
Again, in going from $Z^{(\nu)}_{CY}(\tau)$ to $Z_{{\cal M}_6\times CY(X_n)}(\tau)$, we have relaxed 
the constraint on $r$ and $\tilde r$ in order to obtain a supersymmetric 
partition function. As before, we can show that $Z_{{\cal M}_6\times CY(X_n)}(\tau)$ is invariant 
under both the modular S- and T-transformations.

In order to separate the discrete series spectrum we define
\beqa
{\cal H}^{(\nu)}_{F_l}&\equiv&\oplus_{\!\!\!\!\!\!\!\!\!\!\!\!\!\!\!\!\!\!\!\!\mbox{\tiny $
							\begin{array}{c}\\
							m \! \in  \! \Z_{2(\km +2)} 
														\end{array}$}
							}~~~
							{\oplus}_{\!\!\!\!\!\!\!\!\!\!\!\!\!\!\!\!\!\!\!\!\mbox{\tiny $
							\begin{array}{c}\\
							\nu_0,\nu_1,\nu_2\in Z_2\\
							\nu_0+\nu_1+\nu_2 \\
							\equiv 1({\rm mod}2)
							\end{array}$}
							}
											\!\!\!\left(
					{\cal H}_{m}^{(\km)l,\nu} \otimes
					{\cal H}_{2\nu_0+\nu,2}\otimes
					{\cal H}_{2\nu_1+\nu,2}
					\otimes {\cal H}_{2\nu_2+\nu,2}
				\otimes 
				{\cal H}
				_{m,\km+2}
				\right),
\nonumber\\
												\end{eqnarray}
where various component modules are defined in section \ref{modules}.
Using this, we can express  
$\hat F_l(\tau,z)$ as
\beqa
\hat F_l(\tau,z)
&=&
i^{-1}\eta^4(\tau)\vartheta_1(\tau,z)\sum_{\nu \in \Z_4} (-1)^\nu~
{\rm Tr}_{{\cal H}^{SL(2,\R)}_{+,(0,0)}\otimes 
{\cal H}^{(\nu)}_{F_l}}
q^{L_0^{SL(2,\R)}+L_0^{N=2}+L_0^{(\nu)}+L_0^{(\nu)}+L_0^{(\nu)}+L_0^{U(1)}-\frac{\cm+7}{24}}
\nonumber\\
&&~~~~~~~~~~~~~~~~~~~~~~~~~~~~~~~~~~~~~~~~~~~~~~~~~~~
\cdot y^{J_0^3+ F^{(\nu)}+J_0^{U(1)}+\frac12},
\eeqa
and 
$Z_{{\cal M}_6\times CY(X_n)}(\tau)$ can be written in this case as
\beqa
Z_{{\cal M}_6\times CY(X_n)}(\tau)&=& \sum_{l,\tilde l}N_{l,\tilde l}
\sqrt{\frac{\tau_2}k}
\int_0^1 d s_1 \int_0^1 d s_2
\sum_{\nu,\tilde \nu\in Z_4}(-1)^{\nu+\tilde \nu}
{\rm Tr}_{
\left({\cal H}^{SL(2,\R)}_{+,(0,0)}\otimes 
{\cal H}^{(\nu)}_{F_l}\right)\otimes
\left({\cal H}^{SL(2,\R)}_{+,(0,0)}\otimes 
{\cal H}^{(\tilde \nu)}_{F_{\tilde l}}\right)
}
\nonumber\\&&
\cdot~
q^{\frac{ks_1^2}4+L_0^{SL(2,\R)}+L_0^{N=2}+L_0^{(\nu)}+L_0^{(\nu)}+L_0^{(\nu)}+L_0^{U(1)}-\frac{\cm+7}{24}
+s_1(J_0^3+ F^{(\nu)}+J_0^{U(1)}+\frac12)}
\nonumber\\&&
\cdot~
\bar q^{\frac{ks_1^2}4+\tilde L_0^{SL(2,\R)}+\tilde L_0^{N=2}
+\tilde L_0^{(\nu)}+\tilde L_0^{(\nu)}+\tilde L_0^{(\nu)}+\tilde L_0^{U(1)}-\frac{\cm+7}{24}
+s_1(\tilde J_0^3+ \tilde F^{(\tilde \nu)}+\tilde J_0^{U(1)}+\frac12)}
\nonumber\\&&
\cdot~e^{-2\pi i s_2(J_0^3+ F^{(\nu)}+J_0^{U(1)}-\tilde J_0^3- \tilde F^{(\tilde \nu)}-\tilde J_0^{U(1)})}
\nonumber\\&&
\cdot~\left|\eta^2(\tau\right|^2.
\eeqa
As before, the $s_2$ integration yields the constraint
\beqa
J_0^3+ F^{(\nu)}+J_0^{U(1)}=\tilde J_0^3+ \tilde F^{(\tilde \nu)}+\tilde J_0^{U(1)},
\eeqa
the left and right hand sides of which we call $J_0^{\rm tot}$ and $\tilde J_0^{\rm tot}$, respectively.

After a similar Fourier transformation and a spectral flow operation, we obtain
\beqa
Z_{{\cal M}_6\times CY(X_n)}(\tau)&=&
\sum_{l,\tilde l}N_{l,\tilde l}
\frac{|\eta^2(\tau)|^2}{-2\pi k}
\sum_{\nu,\tilde \nu\in \Z_4}(-1)^{\nu+\tilde \nu}
\nonumber\\
&&\cdot\left(
{\rm Tr}_{
\left({\cal H}^{SL(2,\R)}_{-,(0,-\frac\kappa 2)}\otimes 
{\cal H}^{(\nu)}_{F_l}\right)\otimes
\left({\cal H}^{SL(2,\R)}_{-,(0,-\frac\kappa 2)}\otimes 
{\cal H}^{(\tilde \nu)}_{F_{\tilde l}}\right)
}
\right.\nonumber\\
&&
~~\int_{-\infty}^\infty
\frac{dp}{ip+ \Jtot +\frac12}
q^{\frac1k\left(p+\frac{ik}2\right)^2
+L_0^{SL(2,\R)}+L_0^{N=2}+L_0^{(\nu)}+L_0^{(\nu)}+L_0^{(\nu)}+L_0^{U(1)}-\frac{\cm+7}{24}}
\nonumber\\&&
~~~~~~~~~~~~~~~~~~~~~~~~\cdot
\bar q^{\frac1k\left(p+\frac{ik}2\right)^2
+\tilde L_0^{SL(2,\R)}+\tilde L_0^{N=2}
+\tilde L_0^{(\nu)}+\tilde L_0^{(\nu)}+\tilde L_0^{(\nu)}+\tilde L_0^{U(1)}-\frac{\cm+7}{24}
}
\nonumber\\
&&
-{\rm Tr}_{
\left({\cal H}^{SL(2,\R)}_{+,(0,0)}\otimes 
{\cal H}^{(\nu)}_{F_l}\right)\otimes
\left({\cal H}^{SL(2,\R)}_{+,(0,0)}\otimes 
{\cal H}^{(\tilde \nu)}_{F_{\tilde l}}\right)
}
\nonumber\\
&&
~~\int_{-\infty}^\infty
\frac{dp}{ip+ \Jtot +\frac12}
q^{\frac{p^2}k
+L_0^{SL(2,\R)}+L_0^{N=2}+L_0^{(\nu)}+L_0^{(\nu)}+L_0^{(\nu)}+L_0^{U(1)}-\frac{\cm+7}{24}}
\nonumber\\&&
\left.\left.
~~~~~~~~~~~~~~~~~~~\cdot
\bar q^{\frac{p^2}k
+\tilde L_0^{SL(2,\R)}+\tilde L_0^{N=2}
+\tilde L_0^{(\nu)}+\tilde L_0^{(\nu)}+\tilde L_0^{(\nu)}+\tilde L_0^{U(1)}-\frac{\cm+7}{24}
}
\rule{0ex}{4ex}
\right)\right|_{\Jtot=\tildeJtot}.
\eeqa

We now deform the contour of the first trace. There is a difference here.
In the present case we have $k=\km+2$, so that $k$ grows linearly as $\km$.
Therefore, if we change the contour of $p$ from $\R$ to $\R+\frac{ik}2$,
then it sweeps across many ($\left[\frac k2\right]$ at most) pole singularities 
through the deformation. (In contrast, $k$ does not exceed two in the threefold case,
and therefore the contour picks up at most a single pole contribution.)

As before, the partition function gets pole contributions from the 
states having
\beqa
-\frac{\km+3}2<\Jtot(=\tildeJtot)<\frac12.
\label{J0totcondition_twofold}
\eeqa
To see the massless spectrum, what we need to do is to find 
conformal weight $\frac12$ NS-sector
states that satisfy (\ref{J0totcondition_twofold}) and 
$\Jtot=\tildeJtot$ in
\beqa
\frac 1k\sum_{l,\tilde l} N_{l,\tilde l}\left|
y^{\frac{1-\kappa}2}
\right|^2
\frac{\hat F_l(\tau,z)(\hat F_{\tilde l}(\tau,z))^*}{\left|
\tilde\vartheta_1(\tau,z)\eta(\tau)
\right|^2}
\label{J0tot_dependence_twofold}
\eeqa 
with taking into account the Liouville energy (the shortage of eta functions)
and the drop of weight due to the imaginary momentum of the discrete states.
The R-sector states follow from supersymmetry.
(\ref{J0tot_dependence_twofold}) is an analogue of (\ref{J0tot_dependence})
and can be similarly derived.

Repeating the previous steps, 
we set
\beqa
J^3_0&\equiv&-\frac\kappa 2-N~~~~~(N=0,1,2,\ldots).
\label{J03=kappa-N}
\eeqa
Then
\beqa
\Jtot&=&-\frac\kappa2-N+F^{(\nu)}+ J^{U(1)}_0
\nonumber\\
&\equiv&-\frac\kappa2-n_{cluster}+ J^{U(1)}_0.
\eeqa
where, again, we defined the number 
\beqa
n_{cluster}&=&N-F^{(\nu)} 
\label{ncluster_twofold}
\eeqa
($\in\Z$ for the NS sector)
to label different 
allowed ranges of $J^{U(1)}_0$. Using this number, we have
\beqa
n_{cluster}+\frac12 ~<~J^{U(1)}_0 ~<~ n_{cluster}+\frac{\km+3}2.
\label{range}
\eeqa
Unlike the threefold case, these ranges overlap with the neighboring ones.
The imaginary momentum factor is (for the holomorphic part) 
\beqa
q^{-\frac1{\km+2}\left(J^{U(1)}_0-\frac12-n_{cluster}\right)^2}.
\label{imaginary_momentum_factor_twofold}
\eeqa
Note that due to (\ref{J03=kappa-N}) and (\ref{ncluster_twofold}) 
a discrete state must satisfy $F^{(\nu)}\geq -n_{cluster}$, the fact 
already used extensively in the threefold analysis.

Let us find weight $\frac12$ state contributions to the NS-sector ($\nu=0,2$) 
terms of $\hat F_l(\tau,z)$ (\ref{hatF_l}). 

If $\nu=0$, at least one of the fermion 
theta must be $\Theta_{2,2}$, and $\chi^{l,0}_m$ is (anti-)chiral primary for $m=\pm l$.

If $m=+l$, we see from 
(\ref{imaginary_momentum_factor_twofold}) that, 
among several choices of $n_{cluster}$, 
a lower $n_{cluster}$ results in a larger drop of conformal weight. On the other hand, 
if $n_{cluster}$ is negatively large,  $F^{(\nu)}\geq -n_{cluster}$ means that 
the Liouville fermion number $F^{(\nu)}$ is also large. It turns out that $n_{cluster}=-1$ 
gives the lowest value of conformal weight. In this case, the power of $q$ is
\beqa
\underbrace{\frac18}_{\rm Liouville~energy}
+\underbrace{\frac l{2(\km +2)}-\frac{\km}{8(\km+2)}}_{\chi^{l,\nu}_m}
+\underbrace{\frac{l^2}{4(\km+2)}}_{\Theta_{m,\km+2}}
+\underbrace{\frac12}_{\Theta_{2,2}}
\underbrace{-\frac{(l+1)^2}{4(\km+2)}}_{\rm imaginary~momentum~factor}
&=&\frac12.\nonumber\\
\eeqa 
If $m=-l$, then $n_{cluster}<-1$ and it does not give any weight $\frac12$ states.

Next we consider $\nu=2$. In this case all the fermion thetas can be $\Theta_{0,2}$
simultaneously. $\chi^{l,2}_m=\chi^{\km-l,0}_{m+\km+2}$ is 
(anti-)chiral primary for $m=\pm(l+2)$.

If $m=l+2$, then the lowest conformal weight arises from $n_{cluster}=0$. The 
counting of the various contributions is
\beqa
\underbrace{\frac18}_{\rm Liouville~energy}
+\underbrace{\frac {\km-l}{2(\km +2)}-\frac{\km}{8(\km+2)}}_{\chi^{l,\nu}_m}
+\underbrace{\frac{(l+2)^2}{4(\km+2)}}_{\Theta_{m,\km+2}}
\underbrace{-\frac{(l+1)^2}{4(\km+2)}}_{\rm imaginary~momentum~factor}
&=&\frac12.\nonumber\\
\eeqa 
On the other hand, if $m=(l+2)$, there are no weight $\frac12$ states.

Therefore, we have seen that there are two NS-sector states of conformal 
weight  $\frac12$ for each $\hat F_l(\tau,z)$. Such states in the R sector
must also be two. As is seen from their imaginary momentum factors, 
these four states have a common $\Jtot$ charge, and therefore 
the $\Jtot=\tildeJtot$ paring can be done as a supermultiplet. 
In the $A_{\km+1}$-type modular invariant theory, in which 
the holomorphic and anti-holomorphic combinations are (fully) diagonal,
there are
\beqa
({\bf 2}_{\rm NS}\oplus{\bf 2}_{\rm R})\otimes
({\bf 2}_{\rm NS}\oplus{\bf 2}'_{\rm R})
&=&{\bf 8}_{\rm bosons}\oplus{\bf 8}_{\rm fermions}
\eeqa 
for each $l=0,\ldots,\km$. If ${\bf 2}_{\rm R}$ and ${\bf 2}'_{\rm R}$ are the 
doublets of the same $SU(2)$ factor of $SO(4)$ (type\;IIB), 
the multiplet contains an anti-selfdual tensor. If, on the other hand, they are 
the different ones (type\;IIA), the multiplet is a vector 
multiplet. 

This spectrum of massless states are precisely the ones expected from 
the geometry of the ALE spaces. This fact has already been anticipated in the
analysis of \cite{ES}. They are opposite to the NS5-branes, and this observation 
is in agreement with the T-duality \cite{OV}.

\section{Summary and discussion}
In this paper, we have considered type\;II and heterotic string compactifications 
on an isolated singularity in the noncompact Gepner model approach. We have 
mainly studied the threefold case, but also briefly discussed the twofold case. 
The conifold-type ADE singular Calabi-Yau threefolds are modeled by conformal 
field theory, which is a tensor product of the $SL(2,\R)/U(1)$ Kazama-Suzuki 
model, an $N=2$ minimal model and a free conformal field theory describing 
the four-dimensional Minkowski space.  We have used the result of \cite{ES} to 
construct new space-time supersymmetric, modular invariant partition functions 
for both type\;II and heterotic string theories, thereby the issue in the earlier 
noncompact Gepner models --- the absence of the localized modes --- has been 
resolved. We have investigated in detail the massless spectra of the localized 
modes. There are differences between when the level of the minimal model $\km$
is odd and when it is even. In particular, we found gapless spectra of continuous 
series representations in the even $\km$ case.  The summary of 
massless spectra for various cases is shown in Table 1.  Among them, 
we have shown that the $\km=3$ compactification of the $E_8\times E_8$ 
heterotic string has three  generations of matter fields in the ${\bf 27}\oplus{\bf 1}$ 
representation of $E_6$. They are not on an equal footing, and we propose that 
this model is worthy of further exploration as a viable alternative string 
model for the $E_6$ unification. 

\begin{table}[htdp]
\caption{A summary of four-dimensional 
massless spectra for the $A_{\km +1}$ modular invariant model.}
\begin{center}
\begin{tabular}{|c|c|c|}
\hline
&&\\
&Odd $\km$& Even $\km$\\
&&\\
\hline
&&\\
Type\;IIA&$\frac{\km +3}2$ hypermultiplets&$\frac{\km}2+1$ hypermultiplets\\
&& (+gapless)\\
&&\\
Type\;IIB&$\frac{\km +3}2$ vector multiplets&$\frac{\km}2+1$ vector multiplets\\
&& (+gapless)\\
&&\\
$E_8\times E_8$ heterotic &$\frac{\km +3}2$ chiral supermultiplets
&$\frac{\km}2 +1$ chiral supermultiplets\\
&in ${\bf 10}\oplus{\bf 1}\oplus{\bf 1}\oplus{\bf 16}$ of $SO(10)$ &
in ${\bf 10}\oplus{\bf 1}\oplus{\bf 1}\oplus{\bf 16}$ of $SO(10)$ \\
&(or ${\bf 27}\oplus{\bf 1}$ of $E_6$)
&(or ${\bf 27}\oplus{\bf 1}$ of $E_6$)
(+gapless)\\
&&\\
$SO(32)$ heterotic&$\frac{\km +3}2$ chiral supermultiplets
&$\frac{\km}2 +1$ chiral supermultiplets\\
&
in ${\bf 26}\oplus{\bf 1}\oplus{\bf 1}$&
in ${\bf 26}\oplus{\bf 1}\oplus{\bf 1}$\\
&of $SO(26)$
&of $SO(26)$ 
(+gapless)\\
&&\\
\hline
\end{tabular}
\end{center}
\label{}
\end{table}%

In the twofold case, we have confirmed in the type\;II case 
that the massless spectra of 
localized modes are consistent with the T-duality between the ALE spaces 
and the systems of NS5-branes. Although the heterotic cases have been 
omitted in this paper, the conversion can straightforwardly be done and 
will be reported in a future publication.

There are no localized gauge fields (nor localized gravity) 
in this model. If we interpret the Virasoro condition as the wave 
equation, as we usually do in critical string theories
on a flat space-time, then the wave operator gets a mass term from 
the Liouville energy.   
However, we should note that, in a curved space,  
one cannot tell whether a field is massless or massive by looking only 
at the wave operator.
A well-known example is the conformal mass in the AdS space \cite{BF}.
Also, in a flat space with a linear-dilaton background, the scalar Laplacian 
in the Einstein frame gets a linear term in the derivative along the linear-dilaton 
direction.  
Therefore, we must be careful when we interpret the Liouville energy 
as the mass of
the gauge fields or gravity. 
%
The decoupling of gravity and gauge fields from the localized modes 
may be regarded as a consequence of the assumption that the  
singularity is isolated. 
It would be interesting to explore the possibility of relaxing somehow this 
assumption so that their couplings may be discussed in the framework of 
conformal field theory.

\section*{Acknowledgments}
The author would like to thank T. Eguchi and Y. Sugawara 
for many useful discussions. 
Special thanks go to Toshiya Kawai for having helpful discussions 
on  singularity theory.
The author also thanks 
H. Fuji,
M. Hatsuda,
Y. Hikida,
Y. Ishimoto,
T. Kikuchi,
T. Kimura,
Y. Kitazawa,
H. Kodama,
Y. Michishita, 
Y. Nagatani,
N. Ohta,
M. Sakaguchi, 
Y. Satoh,
Y. Sekino,
F. Sugino,
and
K. Takahashi
for discussions.

This research was supported in part by 
the Grand-in-Aid for Scientific Research No.16540273
and No.20540287-H20  
from the Ministry of Education,
Culture, Sports, Science and Technology of Japan.

\newpage
\section*{Appendix A Theta functions and $N=2$ minimal characters}
\setcounter{equation}{0}
\renewcommand\theequation{A.\arabic{equation}}
In the Appendices below, we assume that $k$ is a positive integer. 

Theta functions.
\beqa
\Theta_{m,k}(\tau,z)&\equiv&\sum_{n\in {\bf Z}}
q^{k(n+\frac m{2k})^2}
y^{k(n+\frac m{2k})},~~~q=e^{2\pi i \tau},~~~y=e^{2\pi i z},
\eeqa
where the level $k$ is a positive integer, and $m$ is an integer. It satisfies
\beqa
\Theta_{m+2k,k}(\tau,z)&=&\Theta_{m,k}(\tau,z),\\
\Theta_{m,k}(\tau,-z)&=&\Theta_{-m,k}(\tau,z). 
\eeqa

The Jacobi theta functions.
\beqa
\vartheta_3(\tau,z)&\equiv&(\Theta_{0,2}+\Theta_{2,2})(\tau,z),\\
\vartheta_4(\tau,z)&\equiv&(\Theta_{0,2}-\Theta_{2,2})(\tau,z),\\
\vartheta_2(\tau,z)&\equiv&(\Theta_{1,2}+\Theta_{-1,2})(\tau,z),\\
\tilde\vartheta_1(\tau,z)&\equiv&(\Theta_{1,2}-\Theta_{-1,2})(\tau,z).
\eeqa
Here we have introduced the unconventional notation $\tilde\vartheta_1$ 
because it appears in the spectral flow orbit naturally rather than 
$\vartheta_1(\tau,z)=-i\tilde\vartheta_1(\tau,z)$.

The composition formula of theta functions.
\beqa
\Theta_{m,k}(\tau,z)
\Theta_{m',k'}(\tau,z')
&=&
\sum_{r\in \Z_{k+k'}}
\Theta_{2rkk'+km'-k'm,kk'(k+k')}(\tau,u)\Theta_{2rk' + m+m',k+k'}(\tau,v)\\
\mbox{or}&=&
\sum_{r\in \Z_{k+k'}}
\Theta_{2rkk'-km'+k'm,kk'(k+k')}(\tau,-u)\Theta_{2rk+ m+m',k+k'}(\tau,v),
\eeqa
$u=\frac{z'-z}{k+k'}$, $v=\frac{kz +k' z'}{k +k'}$.

The $SU(2)_k$ characters. 
\beqa
\chi^{(k)}_l(\tau,z)&=&\frac{\Theta_{l+1.k+2}-\Theta_{-l-1,k+2}}
{\Theta_{1,2}-\Theta_{-1,2}}(\tau,z)\\
&=&\sum_{m\in\Z_{2k}}c^l_m(\tau)\Theta_{m,k}(\tau,z).
\eeqa
$l=0,1,\ldots,k$. The latter equation defines the string functions $c^l_m(\tau)$.

Symmetries of the level-$k$ string functions.
\beqa
c^l_{m+2k}(\tau)=c^{k-l}_{m+k}(\tau)=c^l_m(\tau).
\eeqa

The $N=2$ minimal characters.
\beqa
\chi^{l,s}_m(\tau,z)=\sum_{r\in \Z_k} c^l_{m+4r-s}(\tau)
\Theta_{2m+(k+2)(4r-s),2k(k+2)}
(\tau,\frac z{k+2}),
\eeqa
$l=0,1,\ldots,k$, $m\in \Z_{2(k+2)}$, $s\in \Z_4$.

\beqa
ch_{l,m}^{\rm (NS)}(\tau,z)&=&(\chi^{l,0}_m+\chi^{l,2}_m)(\tau,z),\\
ch_{l,m}^{(\widetilde{\rm NS})}(\tau,z)&=&(\chi^{l,0}_m-\chi^{l,2}_m)(\tau,z),\\
ch_{l,m}^{\rm (R)}(\tau,z)&=&(\chi^{l,1}_m+\chi^{l,-1}_m)(\tau,z),\\
ch_{l,m}^{(\widetilde{\rm R})}(\tau,z)&=&(\chi^{l,1}_m-\chi^{l,-1}_m)(\tau,z).
\eeqa

Symmetries of the $N=2$ minimal characters.
\beqa
\chi^{l,s}_{m+2(k+2)}(\tau,z)=\chi^{k-l,s+2}_{m+k+2}(\tau,z)=\chi^{l,s}_m(\tau,z).
\eeqa

An identity.
\beqa
\chi^{(k)}_l(\tau,0)\Theta_{s,2}(\tau,-z)
&=&\sum_{m\in \Z_{2(k+2)}}
\Theta_{m,k+2}(\tau, \frac{-2z}{k+2})\chi^{l,s}_m(\tau,z),
\label{character_identity}
\eeqa
which can be proved by using the composition formula:
\beqa
\chi^{(k)}_l(\tau,z+u)\Theta_{s,2}(\tau,u)
&=&\sum_{m\in \Z_{2(k+2)}}
\chi^{l,s}_m(\tau,z)\Theta_{m,k+2}(\tau, u+\frac{kz}{k+2}).
\label{chiTheta=N=2minimalTheta}
%
\eeqa

Modular transformations.
\beqa
\Theta_{m,k}\left(
-\frac 1\tau, \frac z\tau
\right)&=&
\sqrt{\frac\tau{2 i k}} e^{\frac{\pi i k z^2}{2\tau}}
\sum_{m'\in \Z_{2k}}
e^{-\pi i \frac{mm'}k}
\Theta_{m',k}(\tau,z),\\
\Theta_{m,k}
(\tau+1, z)&=&
e^{\frac{\pi i m^2}{2k}}
\Theta_{m,k}
(\tau, z).
\eeqa

\section*{Appendix B Useful expressions of $F_{l,2r} (\tau,z)$ and $\hat F_{l,2r} (\tau,z)$}
\setcounter{equation}{0}
\renewcommand\theequation{B.\arabic{equation}}

\subsection*{$F_{l,2r} (\tau,z)$}
Let us name 
\beqa
\Theta_{(s,s')}(\tau,z)&\equiv&
\sum_{\nu\in \Z_2}\Theta_{s+2\nu,2}(\tau,z)\Theta_{s'+2\nu,2}(\tau,z).
\eeqa
Then
\beqa
&&\Theta_{(0,0)}=\Theta_{(2,2)}=\frac{\vartheta_3^2 + \vartheta_4^2}2,\\
&&\Theta_{(0,2)}=\Theta_{(2,0)}=\frac{\vartheta_3^2 - \vartheta_4^2}2,\\
&&\Theta_{(1,1)}=\Theta_{(-1,-1)}=\frac{\vartheta_2^2 + \tilde\vartheta_1^2}2,\\
&&\Theta_{(1,-1)}=\Theta_{(-1,1)}=\frac{\vartheta_2^2 - \tilde\vartheta_1^2}2.
\eeqa
\beqa
\left(\Theta_{(0,0)}\Theta_{(0,2)}
-\Theta_{(1,1)}\Theta_{(1,-1)}\right)(\tau,z)
&=&
\frac14 \left(
\vartheta_3^4-\vartheta_4^4-\vartheta_2^4+\tilde\vartheta_1^4
\right)(\tau,z)\nonumber\\
&=&0.
\eeqa
Therefore,  we can either write 
\beqa
&&\frac14\chi^{(k)}_l(\tau,0)
\left(\vartheta_3^4-\vartheta_4^4-\vartheta_2^4+\tilde\vartheta_1^4\right)
(\tau,z)\nonumber\\
&=&
\left(\chi^{(k)}_l(\tau,0)\Theta_{(0,0)}(\tau,z)\right)\Theta_{(0,2)}(\tau,z)
-\left(\chi^{(k)}_l(\tau,0)\Theta_{(1,1)}(\tau,z)\right)\Theta_{(1,-1)}(\tau,z)
\label{(chiTheta)Theta(-)}
\eeqa
and use the composition formula for theta functions in the parentheses first, 
or write 
\beqa
&=&
\left(\chi^{(k)}_l(\tau,0)\Theta_{(0,2)}(\tau,z)\right)\Theta_{(0,0)}(\tau,z)
-\left(\chi^{(k)}_l(\tau,0)\Theta_{(1,-1)}(\tau,z)\right)\Theta_{(1,1)}(\tau,z)
\label{(chiTheta)Theta(+)}
\eeqa
and do the same thing in this expression.

Let us compute (\ref{(chiTheta)Theta(-)}) and (\ref{(chiTheta)Theta(+)}) 
in two different ways. We first compute $\chi^{(k)}_l(\tau,0)\Theta_{(s,s')}(\tau,z)$.
Since $\chi^{(k)}_l$ and $\Theta_{s,2}$ are composed into an $N=2$ minimal 
character and a level-($k+2$) theta function as shown in (\ref{chiTheta=N=2minimalTheta}),
we further combine this  level-($k+2$) theta and the remaining level-2 theta in $\Theta_{(s,s')}(\tau,z)$ to find
\beqa
\Theta_{-s',2}(\tau,-u)\Theta_{m,k+2}(\tau, u+\frac{kz}{k+2})&=&
\sum_{r\in\Z_{k+4}}\Theta_{-(k+2)s'-2m +4(k+2)r, 2(k+2)(k+4)}\left(
\tau,-\frac{ 2u +\frac{kz}{k+2}}{k+4}
\right)\nonumber\\
&&\cdot\Theta_{-s'+m+4r, k+4}\left(
\tau,
\frac{k(z+u)}{k+4}
\right),\\
\chi^{(k)}_l(\tau,0)\Theta_{(s,s')}(\tau,z)
&=&
\sum_{r\in\Z_{k+4} +\frac{l+s+s'}2}
\sum_{m\in\Z_{4(k+2)}}
\delta^{\rm (mod 2)}_{m,l+s}
\chi^{l,m-(2r+s-s')}_m (\tau,z)\nonumber\\
&&\cdot
\Theta_{(k+2)2r-(k+4)m,2(k+2)(k+4)}\left(\tau,\frac z{k+2}\right)
\Theta_{2r,k+4}(\tau,0).\nonumber\\
\label{chiTheta(ss')}
\eeqa
Using (\ref{chiTheta(ss')}), we find
\beqa
(\ref{(chiTheta)Theta(-)})
&=&
\sum_{r\in\Z_{k+4} +\frac{l}2}
\Theta_{2r,k+4}(\tau,0)
\sum_{m\in\Z_{4(k+2)}}
\left(
\delta^{\rm (mod 2)}_{m,l}\Theta_{(0,2)}(\tau,z)
-\delta^{\rm (mod 2)}_{m,l+1}\Theta_{(1,-1)}(\tau,z)
\right)\nonumber\\
&&~~~~~~~~~~~~~~~~~~~~~~~~~~\cdot
\chi^{l,m-2r}_m(\tau,z)
\Theta_{(k+2)2r-(k+4)m,2(k+2)(k+4)}\left(\tau,\frac z{k+2}\right)
\nonumber\\
&\equiv&\sum_{r\in\Z_{k+4} +\frac{l}2}
\Theta_{2r,k+4}(\tau,0)
F_{l,2r}^{(-)}(\tau,z),
\\
(\ref{(chiTheta)Theta(+)})
&=&
\sum_{r\in\Z_{k+4} +\frac{l}2}
\Theta_{2r,k+4}(\tau,0)
\sum_{m\in\Z_{4(k+2)}}
\left(
\delta^{\rm (mod 2)}_{m,l}\Theta_{(0,0)}(\tau,z)
-\delta^{\rm (mod 2)}_{m,l+1}\Theta_{(1,1)}(\tau,z)
\right)\nonumber\\
&&~~~~~~~~~~~~~~~~~~~~~~~\cdot
\chi^{l,m-2r+2}_m(\tau,z)
\Theta_{(k+2)2r-(k+4)m,2(k+2)(k+4)}\left(\tau,\frac z{k+2}\right)\nonumber\\
&\equiv&\sum_{r\in\Z_{k+4} +\frac{l}2}
\Theta_{2r,k+4}(\tau,0)
F_{l,2r}^{(+)}(\tau,z).
\eeqa

On the other hand, if we apply the composition formula to the two theta 
functions in $\Theta_{(s,s')}$, we find 
\beqa
\Theta_{(s,s')}(\tau,z)&=&\sum_{\nu\in\Z_2}
\sum_{t\in\Z_4}
\Theta_{8t+8\nu+2s+2s',16}\left(
\tau,\frac z2\right)
\Theta_{4t-s+s',4}(\tau,0),
\eeqa
\beqa
\chi^{(k)}_l(\tau,0)\Theta_{(s,s')}(\tau,z)
&=&
\sum_{m\in \Z_{2k}}c^l_m(\tau)\Theta_{m,k}(\tau,0)
\sum_{\nu\in\Z_2}
\sum_{t\in\Z_4}
\Theta_{8t+8\nu+2s+2s',16}\left(
\tau,\frac z2\right)
\Theta_{4t-s+s',4}(\tau,0)
\nonumber\\
&=&
\sum_{r'\in \Z_{2(k+4)}}\Theta_{r' k+4}(\tau,0)
\sum_{m\in \Z_{2k}}
c^l_{r'+4m+s-s'}(\tau)\Theta_{\frac{s+s'}2,1}(\tau,2z)
\nonumber\\
&&~~~~~~~~~~~~~~~~~~~~~~~\cdot
\Theta_{-4r'+(k+4)(-4m-s+s'),4k(k+4)}(\tau,0).
\eeqa
This way of composition of theta functions leads to different expressions of 
(\ref{(chiTheta)Theta(-)})
and (\ref{(chiTheta)Theta(+)}):
\beqa
(\ref{(chiTheta)Theta(-)})
&=&\sum_{\Z_{k+4}+\frac l2}
\Theta_{2r,k+4}(\tau,0)
\sum_{m\in \Z_{2k}}
c^l_{2r+4m}(\tau)
\Theta_{-8r-(k+4)4m,4k(k+4)}(\tau,0)\nonumber\\
&&
\cdot(\Theta_{0,1}(\tau,2z)\Theta_{(0,2)}(\tau,z)
-\Theta_{1,1}(\tau,2z)\Theta_{(1,-1)}(\tau,z)),\\
&&\nonumber\\
(\ref{(chiTheta)Theta(+)})
&=&\sum_{\Z_{k+4}+\frac l2}
\Theta_{2r,k+4}(\tau,0)
\sum_{m\in \Z_{2k}}
c^l_{2r+4m-2}(\tau)
\Theta_{-8r-(k+4)(4m-2),4k(k+4)}(\tau,0)\nonumber\\
&&
\cdot(\Theta_{1,1}(\tau,2z)\Theta_{(0,0)}(\tau,z)
-\Theta_{0,1}(\tau,2z)\Theta_{(1,1)}(\tau,z)).
\eeqa
Thus we find, for $r\in \Z_{k+4}+\frac l2$,
\beqa
F^{(-)}_{l,2r}(\tau,z)&=&
\sum_{m\in\Z_{4(k+2)}}
\chi^{l,m-2r}_m(\tau,z)
\Theta_{(k+2)2r-(k+4)m,2(k+2)(k+4)}\left(\tau,\frac z{k+2}\right)
\nonumber\\
&&~~~~~~~~~~\cdot
\left(
\delta^{\rm (mod 2)}_{m,l}\Theta_{(0,2)}(\tau,z)
-\delta^{\rm (mod 2)}_{m,l+1}\Theta_{(1,-1)}(\tau,z)
\right)
\label{Fl2r(-)1}
\\
&=&
\sum_{m\in \Z_{2k}}
c^l_{2r+4m}(\tau)
\Theta_{-8r-(k+4)4m,4k(k+4)}(\tau,0)\nonumber\\
&&
~~~~~~~~~~\cdot(\Theta_{0,1}(\tau,2z)\Theta_{(0,2)}(\tau,z)
-\Theta_{1,1}(\tau,2z)\Theta_{(1,-1)}(\tau,z))
\nonumber\\
&=&\frac12
\sum_{m\in \Z_{2k}}
c^l_{2r+4m}(\tau)
\Theta_{-8r-(k+4)4m,4k(k+4)}(\tau,0)
\Lambda_2(\tau,z),
\\
&&\nonumber\\
F^{(+)}_{l,2r}(\tau,z)&=&
\sum_{m\in\Z_{4(k+2)}}
\chi^{l,m-2r+2}_m(\tau,z)
\Theta_{(k+2)2r-(k+4)m,2(k+2)(k+4)}\left(\tau,\frac z{k+2}\right)
\nonumber\\
&&~~~~~~~~~~\cdot
\left(
\delta^{\rm (mod 2)}_{m,l}\Theta_{(0,0)}(\tau,z)
-\delta^{\rm (mod 2)}_{m,l+1}\Theta_{(1,1)}(\tau,z)
\right)
\label{Fl2r(+)1}
\\
&=&
\sum_{m\in \Z_{2k}}
c^l_{2r+4m-2}(\tau)
\Theta_{-8r-(k+4)(4m-2),4k(k+4)}(\tau,0)\nonumber\\
&&
~~~~~~~~~~\cdot(\Theta_{1,1}(\tau,2z)\Theta_{(0,0)}(\tau,z)
-\Theta_{0,1}(\tau,2z)\Theta_{(1,1)}(\tau,z))\nonumber\\
&=&
\frac12
\sum_{m\in \Z_{2k}}
c^l_{2r+4m-2}(\tau)
\Theta_{-8r-(k+4)(4m-2),4k(k+4)}(\tau,0)
\Lambda_1(\tau,z),
\eeqa
where 
 \beqa
\Lambda_1(\tau,z)&\equiv&2\left(
\Theta_{1,1}(\tau,2z)
\Theta_{(0,0)}(\tau,z)
-\Theta_{0,1}(\tau,2z)
\Theta_{(1,1)}(\tau,z)
\right),
\\
\Lambda_2(\tau,z)&\equiv&
2\left(
\Theta_{0,1}(\tau,2z)
\Theta_{(0,2)}(\tau,z)
-\Theta_{1,1}(\tau,2z)
\Theta_{(1,-1)}(\tau,z)
\right)
\eeqa
are the same as (\ref{Lambda1(z)}), (\ref{Lambda2(z)}) in the text. 
(The definition of $\Theta_{(s,s')}(\tau,z)$ is given at the beginning of 
this Appendix.)
In particular, even if $k=0$, 
the equation (\ref{chiTheta=N=2minimalTheta}) still holds if we define
\beqa
\chi^{l=0,s}_m(\tau,z)&\equiv&\delta^{{\rm (mod 4)}}_{m,s}
~~~(m, s\in\Z_4),
\eeqa
then we have
\beqa
F^{(-)}_{l,2r}(\tau,z)&=&\left\{
\begin{array}{ll}
\frac12\Lambda_2(\tau,z)&\mbox{if $r=0,2$},
\\
0&\mbox{if $r=1,3$},\end{array}
\right.\\
F^{(+)}_{l,2r}(\tau,z)&=&\left\{
\begin{array}{ll}
0&\mbox{if $r=0,2$},
\\
\frac12\Lambda_1(\tau,z)&\mbox{if $r=1,3$}.
\end{array}
\right.
\eeqa
The total $F_{l,2r}(\tau,z)$ function (\ref{F}) is given by 
\beqa
F_{l,2r}(\tau,z)&=&\frac12\left( F^{(-)}_{l,2r}(\tau,z)+F^{(+)}_{l,2r}(\tau,z)\right)
\nonumber\\
&=&
\frac12\sum_{\nu\in\Z_{4(\km +2)}}
\sum_{
\mbox{\tiny $
							\begin{array}{c}
							\nu_0,\nu_1,\nu_2\in \Z_2\\
							\nu_0+\nu_1+\nu_2 \\
							\equiv 1({\rm mod}2)
							\end{array}$}
							}
(-1)^{\nu}
							\chi^{l,l-2r+2\nu_0+\nu}_{l+\nu}(\tau,z)
							\Theta_{2\nu_1+\nu,2}(\tau,z)
							\Theta_{2\nu_2+\nu,2}(\tau,z)
							\nonumber\\
&&~~~~~~~~~~~~~~
\cdot
\Theta_{(\km +2)2r-(\km+4)(l+\nu),2(\km+2)(\km +4)}
\left(
\tau,\frac z{\km +2}
\right).\label{Fl2r}
\eeqa

$F_{l,2r}(\tau,z)$ satisfies 
\beqa
F_{l,2r+2(k+4)}(\tau,z)&=&F_{l,2r}(\tau,z)
\eeqa
which is obvious due to the periodicity of theta functions. Also it is 
easy to show that \cite{ES1}
\beqa
F_{k-l,2r+k+4}(\tau,z)&=&F_{l,2r}(\tau,z).
\eeqa

\subsection*{$\hat F_{l,2r} (\tau,z)$}
$\hat F_{l,2r}(\tau,z)$ functions (\ref{hatF}) can be obtained by 
modifying the $z$-dependences of various theta functions as 
\beqa
\chi^{l,s}_{m}(\tau,z)&\rightarrow&\chi^{l,s}_{m}(\tau,0),\\
(\Theta_{(s,s')}(\tau,z)\equiv~)
\sum_{\nu\in \Z_2}\Theta_{s+2\nu,2}(\tau,z)\Theta_{s'+2\nu,2}(\tau,z)
&\rightarrow&
\sum_{\nu\in \Z_2}\Theta_{s+2\nu,2}(\tau,0)\Theta_{s'+2\nu,2}(\tau,z),
\\
\Theta_{(k+2)2r-(k+4)m,2(k+2)(k+4)}\left(\tau,\frac z{k+2}\right)
&\rightarrow&
\Theta_{(k+2)2r-(k+4)m,2(k+2)(k+4)}\left(\tau,\frac z{k+4}\right)\nonumber\\
\eeqa
in 
$F^{(\pm)}_{l,2r}(\tau,z)$.
The following formulas are useful:
\beqa
\hat{F}^{(-)}_{l,2r}(\tau,z)&\equiv&
\sum_{m\in\Z_{4(k+2)}}
\chi^{l,m-2r}_m(\tau,0)
\Theta_{(k+2)2r-(k+4)m,2(k+2)(k+4)}\left(\tau,\frac z{k+4}\right)
\nonumber\\
&&~~~~~~~~~~\cdot
\left(
\delta^{\rm (mod 2)}_{m,l}\Theta_{(0,2)}(\tau;0,z)
-\delta^{\rm (mod 2)}_{m,l+1}\Theta_{(1,-1)}(\tau;0,z)
\right)
\label{hatFl2r(-)withchi}
\\
&=&
\sum_{m\in \Z_{2k}}
c^l_{2r+4m}(\tau)
\Theta_{-8r-(k+4)4m,4k(k+4)}\left(\tau,\frac{z}{2(k+4)}\right)\nonumber\\
&&
~~~~~~~~~~\cdot(\Theta_{0,1}(\tau,z)\Theta_{(0,2)}(\tau;0,z)
-\Theta_{1,1}(\tau,z)\Theta_{(1,-1)}(\tau;0,z))
\nonumber\\
&=&\frac12
\sum_{m\in \Z_{2k}}
c^l_{2r+4m}(\tau)
\Theta_{-8r-(k+4)4m,4k(k+4)}\left(\tau,\frac{z}{2(k+4)}\right)
\hat\Lambda_2(\tau,z),
\label{hatFl2r(-)}
\\
&&\nonumber\\
\hat{F}^{(+)}_{l,2r}(\tau,z)&\equiv&
\sum_{m\in\Z_{4(k+2)}}
\chi^{l,m-2r+2}_m(\tau,0)
\Theta_{(k+2)2r-(k+4)m,2(k+2)(k+4)}\left(\tau,\frac z{k+4}\right)
\nonumber\\
&&~~~~~~~~~~\cdot
\left(
\delta^{\rm (mod 2)}_{m,l}\Theta_{(0,0)}(\tau;0,z)
-\delta^{\rm (mod 2)}_{m,l+1}\Theta_{(1,1)}(\tau;0,z)
\right)
\label{hatFl2r(+)withchi}
\\
&=&
\sum_{m\in \Z_{2k}}
c^l_{2r+4m-2}(\tau)
\Theta_{-8r-(k+4)(4m-2),4k(k+4)}\left(\tau,\frac{z}{2(k+4)}\right)\nonumber\\
&&
~~~~~~~~~~\cdot(\Theta_{1,1}(\tau,z)\Theta_{(0,0)}(\tau;0,z)
-\Theta_{0,1}(\tau,z)\Theta_{(1,1)}(\tau;0,z))\nonumber\\
&=&\frac12
\sum_{m\in \Z_{2k}}
c^l_{2r+4m-2}(\tau)
\Theta_{-8r-(k+4)(4m-2),4k(k+4)}\left(\tau,\frac{z}{2(k+4)}\right)
\hat\Lambda_1(\tau,z),
\label{hatFl2r(+)}
\nonumber\\
\\
\hat{F}_{l,2r}(\tau,z)&=&\frac12\left(\hat{F}^{(-)}_{l,2r}(\tau,z)+\hat{F}^{(+)}_{l,2r}(\tau,z)\right),
\eeqa
where 
 \beqa
\hat\Lambda_1(\tau,z)&=&2\left(
\Theta_{1,1}(\tau,z)
\Theta_{(0,0)}(\tau;0,z)
-\Theta_{0,1}(\tau,z)
\Theta_{(1,1)}(\tau;0,z)
\right),
\label{hatLambda1'}
\\
\hat\Lambda_2(\tau,z)&=&
2\left(
\Theta_{0,1}(\tau,z)
\Theta_{(0,2)}(\tau;0,z)
-\Theta_{1,1}(\tau,z)
\Theta_{(1,-1)}(\tau;0,z)
\right),
\label{hatLambda2'}
\\
\Theta_{(s,s')}(\tau;z,z')&\equiv&
\sum_{\nu\in \Z_2}\Theta_{s+2\nu,2}(\tau,z)\Theta_{s'+2\nu,2}(\tau,z').
\eeqa
The expressions (\ref{hatLambda1'}) and (\ref{hatLambda2'})
are equivalent to the definitions (\ref{hatLambda1}) and (\ref{hatLambda2}) 
in the text.
(Note, again, that $k$ here is $\km$ in (\ref{hatF}).)

$\hat F_{l,2r}(\tau,z)$ also satisfies 
\beqa
\hat F_{l,2r+2(k+4)}(\tau,z)&=&\hat F_{l,2r}(\tau,z),\\
\hat F_{k-l,2r+k+4}(\tau,z)&=&\hat F_{l,2r}(\tau,z),\\
\hat F_{l,-2r}(\tau,z)&=&\hat F_{l,2r}(\tau,-z).
\label{Fhatsymmetry}
\eeqa

\section*{Appendix C Heterotic conversion}
\setcounter{equation}{0}
\renewcommand\theequation{C.\arabic{equation}}
In Gepner models, any modular invariant partition function for type\;II 
strings can be converted to that for heterotic strings by a straightforward 
procedure \cite{Gepner}, which we review in this Appendix.

Let us denote level-1 affine $SO(2n)$ characters by
\beqa
B^{(2n)}_0(\tau,z)&\equiv&\frac{(\vartheta_3(\tau,z))^n+(\vartheta_4(\tau,z))^n}{2(\eta(\tau))^n},\\
B^{(2n)}_v(\tau,z)&\equiv&\frac{(\vartheta_3(\tau,z))^n-(\vartheta_4(\tau,z))^n}{2(\eta(\tau))^n},\\
B^{(2n)}_s(\tau,z)&\equiv&\frac{(\vartheta_2(\tau,z))^n+(\tilde\vartheta_1(\tau,z))^n}{2(\eta(\tau))^n},\\
B^{(2n)}_{\bar s}(\tau,z)&\equiv&\frac{(\vartheta_2(\tau,z))^n-(\tilde\vartheta_1(\tau,z))^n}{2(\eta(\tau))^n}.
\eeqa
Writing them as a column vector ${\bf B}^{(2n)}(\tau,z)$, their modular $S$- and $T$-transformations 
are given in the matrix notation 
\beqa
\left.{\bf B}^{(2n)}(\tau,z)\right|_S&=&e^{\frac{n\pi i z^2}\tau}S^{(2n)}{\bf B}^{(2n)}(\tau,z),\\
S^{(2n)}&=&\frac12\left(
\begin{array}{cccc}
1&1&1&1\\
1&1&-1&-1\\
1&-1&i^{-n}&-i^{-n}\\
1&-1&-i^{-n}&i^{-n}
\end{array}
\right)
\eeqa
and
\beqa
\left.{\bf B}^{(2n)}(\tau,z)\right|_T&=&T^{(2n)}{\bf B}^{(2n)}(\tau,z),\\
T^{(2n)}&=&\left(
\begin{array}{cccc}
e^{-\frac{n\pi i}{12}}&&&\\
&-e^{-\frac{n\pi i}{12}}&&\\
&&e^{+\frac{n\pi i}{6}}&\\
&&&e^{+\frac{n\pi i}{6}}
\end{array}
\right).
\eeqa

We also define
\beqa
B^{(E_8)}(\tau,z)&\equiv&\frac{(\vartheta_3(\tau,z))^8+(\vartheta_4(\tau,z))^8
+(\vartheta_2(\tau,z))^8+(\tilde\vartheta_1(\tau,z))^8}{2(\eta(\tau))^8},\\
\eeqa
then
\beqa
\left.B^{(E_8)}(\tau,z)\right|_S&=&e^{\frac{8\pi i z^2}\tau}B^{(E_8)}(\tau,z),
\\
\left.B^{(E_8)}(\tau,z)\right|_T&=&e^{-\frac{8\pi i}{12}}B^{(E_8)}(\tau,z).
\label{T^(E8)}
\eeqa

If we set $z=0$,  then we find
\begin{alignat}{4}
&S^{(d+8)}&=\quad&S^{(d+24)}&=\quad&S^{(d)}&=\quad&M^T S^{(d)} M,\\
&e^{-\frac8{12}\pi i}~T^{(d+8)}&=\quad&T^{(d+24)}&=\quad&
\left(
\begin{array}{cccc}
-1&&&\\
&-1&&\\
&&1&\\
&&&1
\end{array}
\right)
T^{(d)}&=\quad&M^T T^{(d)} M,
\label{T^(d)}
\end{alignat}
where 
\beqa
M=M^T=M^{-1}=
\left(
\begin{array}{cccc}
&1&&\\
1&&&\\
&&-1&\\
&&&-1
\end{array}
\right).
\eeqa
$d$ is the transverse space dimensions (that is, $d=2$ 
for a four-dimensional flat Minkowski spacetime with a Calabi-Yau three-fold, 
and $d=4$ for six-dimensional one with a Calabi-Yau two-fold).
Therefore, $MB^{(E_8)}{\bf B}^{(d+8)}(\tau,0)$ and $M{\bf B}^{(d+24)}(\tau,0)$ 
transform exactly in the same manner as ${\bf B}^{(d)}(\tau,0)$ does under the 
modular $S$- and $T$-transformations.  
This means that starting from any modular invariant partition function for 
type\;II strings, we can obtain one for the $E_8\times E_8$ heterotic string theory 
by replacing the left-moving fermion theta functions as (with all $z$'s
being equal to zero)
\beqa
\frac{(\vartheta_3)^{\frac d2}+(\vartheta_4)^{\frac d2}}{2\eta^{\frac d2}}
~(=B^{(d)}_0)
&\rightarrow&
\frac{(\vartheta_3)^{\frac{d+8}2}-(\vartheta_4)^{\frac{d+8}2}}{2\eta^{\frac{d+8}2}}
B^{(E_8)}
~(=B^{(d+8)}_v B^{(E_8)}),\\
\frac{(\vartheta_3)^{\frac d2}-(\vartheta_4)^{\frac d2}}{2\eta^{\frac d2}}
~(=B^{(d)}_v)
&\rightarrow&
\frac{(\vartheta_3)^{\frac{d+8}2}+(\vartheta_4)^{\frac{d+8}2}}{2\eta^{\frac{d+8}2}}
B^{(E_8)},
~(=B^{(d+8)}_0 B^{(E_8)}),\\
\frac{(\vartheta_2)^{\frac d2}+(\tilde\vartheta_1)^{\frac d2}}{2\eta^{\frac d2}}
~(=B^{(d)}_s)
&\rightarrow&
-\frac{(\vartheta_2)^{\frac{d+8}2}+(\tilde\vartheta_1)^{\frac{d+8}2}}{2\eta^{\frac{d+8}2}}
B^{(E_8)},
~(=-B^{(d+8)}_s B^{(E_8)}),\\
\frac{(\vartheta_2)^{\frac d2}-(\tilde\vartheta_1)^{\frac d2}}{2\eta^{\frac d2}}
~(=B^{(d)}_{\bar s})
&\rightarrow&
-\frac{(\vartheta_2)^{\frac{d+8}2}-(\tilde\vartheta_1)^{\frac{d+8}2}}{2\eta^{\frac{d+8}2}}
B^{(E_8)}
~(=-B^{(d+8)}_{\bar s} B^{(E_8)})
\eeqa
and also for the $SO(32)$ theory as
\beqa
\frac{(\vartheta_3)^{\frac d2}+(\vartheta_4)^{\frac d2}}{2\eta^{\frac d2}}
~(=B^{(d)}_0)
&\rightarrow&
\frac{(\vartheta_3)^{\frac{d+24}2}-(\vartheta_4)^{\frac{d+24}2}}{2\eta^{\frac{d+24}2}},
~(=B^{(d+24)}_v)\\
\frac{(\vartheta_3)^{\frac d2}-(\vartheta_4)^{\frac d2}}{2\eta^{\frac d2}}
~(=B^{(d)}_v)
&\rightarrow&
\frac{(\vartheta_3)^{\frac{d+24}2}+(\vartheta_4)^{\frac{d+24}2}}{2\eta^{\frac{d+24}2}},
~(=B^{(d+24)}_0)\\
\frac{(\vartheta_2)^{\frac d2}+(\tilde\vartheta_1)^{\frac d2}}{2\eta^{\frac d2}}
~(=B^{(d)}_s)
&\rightarrow&
-\frac{(\vartheta_2)^{\frac{d+24}2}+(\tilde\vartheta_1)^{\frac{d+24}2}}{2\eta^{\frac{d+24}2}},
~(=-B^{(d+24)}_s)\\
\frac{(\vartheta_2)^{\frac d2}-(\tilde\vartheta_1)^{\frac d2}}{2\eta^{\frac d2}}
~(=B^{(d)}_{\bar s})
&\rightarrow&
-\frac{(\vartheta_2)^{\frac{d+24}2}-(\tilde\vartheta_1)^{\frac{d+24}2}}{2\eta^{\frac{d+24}2}}
~(=-B^{(d+24)}_{\bar s}).
\eeqa

Applying these rules in $\hat F_{l,2r}(\tau,z)$, we obtain
\beqa
\hat F^{E_8\times E_8}_{l,2r}(\tau,z)&=&
\frac12\left(
\hat F^{E_8\times E_8(-)}_{l,2r}(\tau,z)
+\hat F^{E_8\times E_8(+)}_{l,2r}(\tau,z)
\right),
\label{F^E_8xE_8}
\\
\hat F^{E_8\times E_8(-)}_{l,2r}(\tau,z)&\equiv&
\sum_{m\in \Z_{2k}}
c^l_{2r+4m}(\tau)
\Theta_{-8r-(k+4)4m,4k(k+4)}\left(\tau,\frac{z}{2(k+4)}\right)
\frac12\hat \Lambda^{E_8\times E_8}_2(\tau,z),\nonumber\\
\\
\hat F^{E_8\times E_8(+)}_{l,2r}(\tau,z)
&\equiv&
\sum_{m\in \Z_{2k}}
c^l_{2r+4m-2}(\tau)
\Theta_{-8r-(k+4)(4m-2),4k(k+4)}\left(\tau,\frac{z}{2(k+4)}\right)
\frac12\hat \Lambda^{E_8\times E_8}_1(\tau,z),\nonumber\\
\\
\frac{\frac12\hat \Lambda^{E_8\times E_8}_1(\tau,z)}
{\eta^{14}(\tau)}
&\equiv&\left(
\Theta_{1,1}(\tau,z)\left(
B^{(10)}_v(\tau,0)B^{(2)}_0(\tau,z)
+B^{(10)}_0(\tau,0)B^{(2)}_v(\tau,z)
\right)
\right.
\nonumber\\
&&
\left.
+
\Theta_{0,1}(\tau,z)\left(
B^{(10)}_s(\tau,0)B^{(2)}_s(\tau,z)
+B^{(10)}_{\bar s}(\tau,0)B^{(2)}_{\bar s}(\tau,z)
\right)\right)B^{(E_8)}(\tau,0),
\nonumber\\
\\
\frac{\frac12\hat \Lambda^{E_8\times E_8}_2(\tau,z)}
{\eta^{14}(\tau)}
&\equiv&\left(
\Theta_{0,1}(\tau,z)\left(
B^{(10)}_0(\tau,0)B^{(2)}_0(\tau,z)
+B^{(10)}_v(\tau,0)B^{(2)}_v(\tau,z)
\right)
\right.
\nonumber\\
&&
\left.
+
\Theta_{1,1}(\tau,z)\left(
B^{(10)}_s(\tau,0)B^{(2)}_{\bar s}(\tau,z)
+B^{(10)}_{\bar s}(\tau,0)B^{(2)}_s(\tau,z)
\right)\right)B^{(E_8)}(\tau,0),
\nonumber\\
\label{Lambda2^E_8xE_8}
\eeqa
and
\beqa
\hat F^{SO(32)}_{l,2r}(\tau,z)&=&
\frac12\left(
\hat F^{SO(32)(-)}_{l,2r}(\tau,z)
+\hat F^{SO(32)(+)}_{l,2r}(\tau,z)
\right),
\label{F^SO(32)}
\\
\hat F^{SO(32)(-)}_{l,2r}(\tau,z)&\equiv&
\sum_{m\in \Z_{2k}}
c^l_{2r+4m}(\tau)
\Theta_{-8r-(k+4)4m,4k(k+4)}\left(\tau,\frac{z}{2(k+4)}\right)
\frac12\hat \Lambda^{SO(32)}_2(\tau,z),\nonumber\\
\\
\hat F^{SO(32)(+)}_{l,2r}(\tau,z)
&\equiv&
\sum_{m\in \Z_{2k}}
c^l_{2r+4m-2}(\tau)
\Theta_{-8r-(k+4)(4m-2),4k(k+4)}\left(\tau,\frac{z}{2(k+4)}\right)
\frac12\hat \Lambda^{SO(32)}_1(\tau,z),
\nonumber\\
\\
\frac{\frac12\hat \Lambda^{SO(32)}_1(\tau,z)}
{\eta^{14}(\tau)}
&\equiv&
\Theta_{1,1}(\tau,z)\left(
B^{(26)}_v(\tau,0)B^{(2)}_0(\tau,z)
+B^{(26)}_0(\tau,0)B^{(2)}_v(\tau,z)
\right)
\nonumber\\
&&
+
\Theta_{0,1}(\tau,z)\left(
B^{(26)}_s(\tau,0)B^{(2)}_s(\tau,z)
+B^{(26)}_{\bar s}(\tau,0)B^{(2)}_{\bar s}(\tau,z)
\right),
\nonumber\\
\\
\frac{\frac12\hat \Lambda^{SO(32)}_2(\tau,z)}
{\eta^{14}(\tau)}
&\equiv&
\Theta_{0,1}(\tau,z)\left(
B^{(26)}_0(\tau,0)B^{(2)}_0(\tau,z)
+B^{(26)}_v(\tau,0)B^{(2)}_v(\tau,z)
\right)
\nonumber\\
&&
+
\Theta_{1,1}(\tau,z)\left(
B^{(26)}_s(\tau,0)B^{(2)}_{\bar s}(\tau,z)
+B^{(26)}_{\bar s}(\tau,0)B^{(2)}_s(\tau,z)
\right).
\label{Lambda2^SO(32)}
\eeqa

\section*{Appendix D Proof of the regularization formula}
\setcounter{equation}{0}
\renewcommand\theequation{D.\arabic{equation}}
In this Appendix we prove the regularization formula.
Let 
\beqa
f(z,\epsilon)&\equiv&-\sum_{n=0}^\infty
\frac{e^{-n\epsilon}}{z-n},
\eeqa
then $f(z,\epsilon)$ has simple poles at $z=n$, $(n=0,1,2,\ldots)$ with residue 
$-e^{-n\epsilon}$. On the other hand, the gamma function $\Gamma(-z)$ 
has also simple poles at $z=n$, $(n=0,1,2,\ldots)$, and so do 
$\frac{\partial}{\partial z}\log(\Gamma(-z))$ at the same locations with residue $-1$.
Therefore, comparing the singularities, we may write
\beqa
\frac{\partial^2}{\partial z^2}\log(\Gamma(-z))&=&\sum_{n=0}^\infty
\frac{+1}{(z-n)^2}+{\rm const}.
\eeqa
Subtracting
\beqa
\frac{\partial}{\partial z}f(z,\epsilon)&=&\sum_{n=0}^\infty \frac{e^{-n\epsilon}}{(z-n)^2}
\eeqa
from both sides and integrating with respect to $z$, we find
\beqa
\frac\partial{\partial z}\log\Gamma(-z) - f(z,\epsilon)&=&-\sum_{n=0}^\infty
\frac{1-e^{-n\epsilon}}{z-n}+az+b
\eeqa
for some constants $a$ and $b$. The first term is $O(\epsilon)$.

To determine $a$ and $b$, we set $z=-1$ and $z=2$:
\beqa
f(-1,\epsilon)&=& - e^\epsilon \log(1-e^{-\epsilon}),\\
f(-2,\epsilon)&=& - e^{2\epsilon} \log(1-e^{-\epsilon})- e^\epsilon,
\eeqa 
and therefore
\beqa
- e^\epsilon \log(1-e^{-\epsilon})&=&
\left.\frac\partial{\partial z}\log\Gamma(-z)\right|_{z=-1}+O(\epsilon)-(-a+b),\\
- e^{2\epsilon} \log(1-e^{-\epsilon})- e^\epsilon&=&
\left.\frac\partial{\partial z}\log\Gamma(-z)\right|_{z=-2}+O(\epsilon)-(-2a+b).
\eeqa

\beqa
\psi(z)&\equiv&\frac\partial{\partial z}\log \Gamma(z)
\eeqa
is known as the psi-function (or the $d\Gamma$-function), and 
\beqa
\frac\partial{\partial z}\log\Gamma(-z)&=&-\psi(-z).
\eeqa 
The psi-function satisfies the recursion relation
\beqa
\psi(z+1)&=&\frac1 z + \psi(z),
\eeqa
and hence 
\beqa
\psi(2)&=&\psi(1)+1\nonumber\\
&=&-{\cal C}+1,
\eeqa
where 
\beqa
-\psi(1)&=&\left.\frac\partial{\partial z}\log\Gamma(-z)\right|_{z=-1}
~=~{\cal C}
\eeqa
is known as Euler's constant. Using these data, we find
\beqa
a&=&O(\epsilon),\\
b&=&\log\epsilon + {\cal C} +O(\epsilon)+O(\epsilon\log\epsilon),
\eeqa
and obtain the final form of the regularization formula 
\beqa
-\sum_{n=0}^\infty\frac{e^{-n\epsilon}}{z-n}
&=&-\log\epsilon+\frac\partial{\partial z}\log\Gamma(-z)
-{\cal C}+O(\epsilon)+O(\epsilon\log\epsilon).
\eeqa
That this formula is correct can also be confirmed numerically by Mathematica. 

\newpage
%

 \end{document}